\documentclass{aa}
\usepackage{graphicx}
\usepackage{txfonts}
\usepackage{lscape}
\usepackage{natbib,twoopt}
\usepackage[breaklinks=true]{hyperref} 
\bibpunct{(}{)}{;}{a}{}{,} 
\makeatletter
\newcommandtwoopt{\citeads}[3][][]{\href{https://adsabs.harvard.edu/abs/#3}%
{\def\hyper@linkstart##1##2{}%
\let\hyper@linkend\@empty\citealp[#1][#2]{#3}}}
\newcommandtwoopt{\citetads}[3][][]{\href{https://adsabs.harvard.edu/abs/#3}%
{\def\hyper@linkstart##1##2{}%
\let\hyper@linkend\@empty\citet[#1][#2]{#3}}}
\newcommandtwoopt{\citepads}[3][][]{\href{https://adsabs.harvard.edu/abs/#3}%
{\def\hyper@linkstart##1##2{}%
\let\hyper@linkend\@empty\citep[#1][#2]{#3}}}
\newcommandtwoopt{\citeyearads}[3][][]%
{\href{http://adsabs.harvard.edu/abs/#3}
{\def\hyper@linkstart##1##2{}%
\let\hyper@linkend\@empty\citeyear[#1][#2]{#3}}}
\newcommandtwoopt{\citetalads}[3][][]{\href{https://adsabs.harvard.edu/abs/#3}%
{\def\hyper@linkstart##1##2{}%
\let\hyper@linkend\@empty#1\citetalias{#3}#2}}
\newcommandtwoopt{\citepalads}[3][][]{\href{https://adsabs.harvard.edu/abs/#3}%
{\def\hyper@linkstart##1##2{}%
\let\hyper@linkend\@empty\citepalias[#1][#2]{#3}}}
\makeatother

\defcitealias{2009A&A...493..339W}{V}
\defcitealias{2016A&A...590A...1R}{VII}

\begin{document}

   \title{The XMM-Newton serendipitous survey}

   \subtitle{VIII: The first XMM-Newton serendipitous source catalogue from
     overlapping observations\thanks{Based on observations obtained with
       XMM-Newton, an ESA science mission with instruments and contributions
       directly funded by ESA Member States and NASA.}\fnmsep\thanks{The
       catalogue is available in FITS format via the SSC web pages at
       \url{https://xmmssc.irap.omp.eu} and searchable via XCatDB
       \url{https://xcatdb.unistra.fr/3xmmdr7s}, XSA
       \url{https://www.cosmos.esa.int/web/xmm-newton/xsa}, and CDS \url{https://vizier.u-strasbg.fr/viz-bin/VizieR?-source=IX/56}.}}

   \titlerunning{The XMM-Newton serendipitous survey. VIII: The first source
     catalogue from overlapping observations}

   \author{I.\ Traulsen\inst{1}
          \and A.\ D.\ Schwope\inst{1}
          \and G.\ Lamer\inst{1}
          \and J.\ Ballet\inst{2}
          \and F.\ Carrera\inst{3}
          \and M.\ Coriat\inst{4}
          \and M.\ J.\ Freyberg\inst{5}
          \and L.\ Michel\inst{6}
          \and C.\ Motch\inst{6}
          \and S.\ R.\ Rosen\inst{7}
          \and N.\ Webb\inst{4}
          \and M.\ T.\ Ceballos\inst{3}
          \and F.\ Koliopanos\inst{4}
          \and J.\ Kurpas\inst{1}
          \and M.\ J.\ Page\inst{8}
          \and M.\ G.\ Watson\inst{7}
          }

   \institute{Leibniz-Institut f\"ur Astrophysik Potsdam (AIP), An der
             Sternwarte 16, 14482 Potsdam, Germany\\
              \email{itraulsen@aip.de}
         \and    
             AIM, CEA, CNRS, Universit\'e Paris-Saclay, Universit\'e Paris
             Diderot, Sorbonne Paris Cit\'e, F-91191 Gif-sur-Yvette, France
         \and    
             Instituto de F\'isica de Cantabria (CSIC-UC), Avenida de los
             Castros, 39005 Santander, Spain
         \and    
             IRAP, Universit\'e de Toulouse, CNRS, UPS, CNES, 9 Avenue du
             Colonel Roche, BP 44346, 31028 Toulouse Cedex 4, France
         \and    
             Max-Planck-Institut f\"ur extraterrestrische Physik,
             Giessenbachstra{\ss}e 1, 85748 Garching, Germany
         \and    
             Observatoire astronomique, Universit\'e de Strasbourg, CNRS, UMR
             7550, 11 rue de l’Universit\'e, 67000 Strasbourg, France
         \and    
             Department of Physics \& Astronomy, University of Leicester,
             Leicester, LE1 7RH, UK
         \and    
              Mullard Space Science Laboratory, University College London,
              Holbury St Mary, Dorking, Surrey RH5 6NT, UK
             }

   \date{Received July 24, 2018; accepted February 7, 2019}


\abstract
  {XMM-Newton has observed the X-ray sky since early 2000. The XMM-Newton
    Survey Science Centre Consortium has published catalogues of X-ray and
    ultraviolet sources found serendipitously in the individual
    observations. This series is now augmented by a catalogue dedicated to
    X-ray sources detected in spatially overlapping XMM-Newton observations.}
  {The aim of this catalogue is to explore repeatedly observed sky regions. It
    thus makes use of the long(er) effective exposure time per sky area and
    offers the opportunity to investigate long-term flux variability directly
    through the source-detection process.}
  {A new standardised strategy for simultaneous source detection on multiple
    observations was introduced, including an adaptive-smoothing method to
    describe the image background. It was coded as a new task within the
    XMM-Newton Science Analysis System and used to compile a catalogue of
    sources from 434 stacks comprising 1\,789 overlapping XMM-Newton
    observations that entered the 3XMM-DR7 catalogue, have a low background
    and full-frame readout of all EPIC cameras.}
  {The first stacked catalogue is called 3XMM-DR7s. It contains 71\,951 unique
    sources with positions and parameters such as fluxes, hardness ratios,
    quality estimates, and information on inter-observation variability,
    directly derived from a simultaneous fit. Source parameters are calculated
    for the stack and for each contributing observation. About 15\,\% of the
    sources are new with respect to 3XMM-DR7. Through stacked source
    detection, the parameters of repeatedly observed sources are determined
    with higher accuracy than in the individual observations. The method is
    more sensitive to faint sources and tends to produce fewer spurious
    detections.}
  {With this first stacked catalogue we demonstrate the feasibility and
    benefit of the approach. It supplements the large data base of XMM-Newton
    detections with additional, in particular faint, sources and adds
    variability information. In the future, the catalogue will be expanded to
    larger samples and continued within the series of serendipitous XMM-Newton
    source catalogues.}


\maketitle


\section{Introduction}
\label{sec:introduction}

  \begin{figure}
    \centering
    \includegraphics[width=87mm]{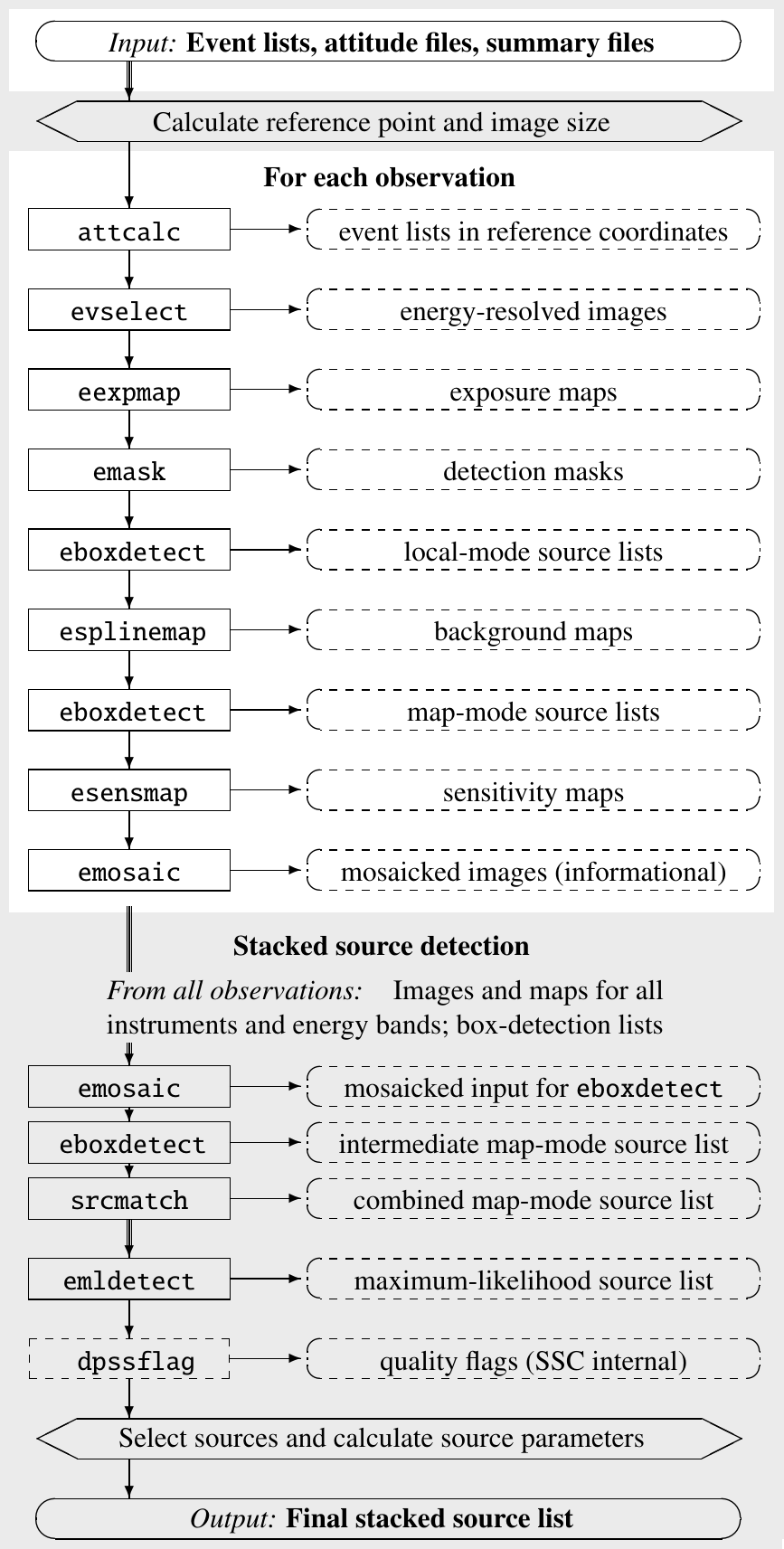}
    \caption{Structure of the task \texttt{edetect\_stack}. Internal steps are
      shown in hexagonal boxes, calls to external SAS tasks in rectangles, and
      their data products in dashed rounded boxes. Stages dealing with all
      observations simultaneously are highlighted by a grey background. In
      `local mode', \texttt{eboxdetect} uses an internally determined local
      background value, while in `map mode' an external background map is
      applied. This is produced by \texttt{esplinemap}, which is named after
      its first functionality and now run in its new adaptive-smoothing mode
      (Sect.~\ref{sec:bkgfitting}).}
    \label{fig:tasks}
  \end{figure}

  ESA's X-ray mission, XMM-Newton \citepads{2001A&A...365L...1J}, launched in
  December 1999, is dedicated to pointed X-ray and ultraviolet to optical
  observations. Its large field of view and effective area also make it
  suitable for survey-like searches for serendipitous X-ray detections.  Up to
  one hundred (or more) sources are found in addition to the main target in
  each XMM-Newton observation with the EPIC CCD instruments pn\linebreak
  \citepads{2001A&A...365L..18S}, MOS1, and MOS2
  \citepads{2001A&A...365L..27T}. The XMM-Newton Survey Science Centre
  Consortium \citepads[SSC, ][]{2001A&A...365L..51W} has been generating
  catalogues of individual detections, merged into unique sources, from public
  XMM-Newton observations since the beginning of the mission. The series of
  XMM-Newton Serendipitous Source Catalogues are produced from pointed
  observations with the EPIC instruments. The most recent data release
  3XMM-DR8 of the third generation catalogue was published on May 16th,
  2018. The catalogue series and the underlying software are described by
  \citetads[][hereafter: Paper\,V]{2009A&A...493..339W} and
  \citetads[][hereafter: Paper\,VII]{2016A&A...590A...1R}. Complementary
  source catalogues are the Slew Survey Source Catalogue
  \citepads{2008A&A...480..611S} from EPIC-pn data taken during telescope
  slews and the XMM-Newton OM Serendipitous Ultraviolet Source Survey
  Catalogue \citepads{2012MNRAS.426..903P} from data taken with the Optical
  Monitor. The software to reduce and analyse XMM-Newton data and to compile
  the catalogues has been developed by the SSC and the XMM-Newton Science
  Operations Centre (SOC) and is released regularly by the SOC.

  After seventeen years in orbit, XMM-Newton has re-observed many patches of
  the sky. Overall, almost a third of the XMM-Newton sky has been visited more
  than once. This may occur from planned repeated observations of variable
  objects or calibration targets, mosaic observations of large regions, or
  unplanned overlaps of independent observations. To properly exploit the
  survey potential of the growing body of multiply imaged sky areas in the
  XMM-Newton archive, we (members of the SSC) have now developed a new
  standardised approach to source detection in multiple observations.
  Previous work on overlapping observations includes the ROSAT catalogues
  \citepads{1999A&A...349..389V,2016A&A...588A.103B}, for which the photons of
  all exposures covering a sky region are merged, the SwiftFT
  \citepads{2011A&A...528A.122P} and 1SXPS \citepads{2014ApJS..210....8E}
  catalogues, for which overlapping images are merged, and the upcoming second
  release of the Chandra Source Catalogue \citepads{2015ASPC..495..297E}, for
  which the photons of observations with aim-points within 1\arcmin\ are
  merged. For the XMM-Newton EPIC data with a strongly position-dependent
  point spread function (PSF), we perform simultaneous multi-band PSF fitting
  in all individual images without merging them. A maximum-likelihood
  algorithm is employed in the five standard energy bands \emph{(1)}
  0.2$-$0.5\,keV, \emph{(2)} 0.5$-$1.0\,keV, \emph{(3)} 1.0$-$2.0\,keV,
  \emph{(4)} 2.0$-$4.5\,keV, and \emph{(5)} 4.5$-$12.0\,keV. This is similar
  to the method used to produce the other XMM-Newton source
  catalogues. Parameters of each source are derived from overlays of the
  empirical PSFs for the respective instrument, energy band, and off-axis
  position. The full procedure from the input event lists to the final stacked
  source list has been made available to all users within the XMM-Newton
  Science Analysis System \citepads[SAS, ][]{2004ASPC..314..759G}.

  This paper, number VIII in the series of publications dedicated to the
  catalogues of serendipitous detections in XMM-Newton pointing-mode
  observations, introduces the first catalogue of X-ray sources from spatially
  overlapping EPIC observations. Being the first release using stacked source
  detection, it also serves as a method validation and as a feasibility
  study. It has been compiled from a selection of good-quality data, namely
  overlapping 3XMM-DR7 observations with large usable chip area and reasonably
  low background. All sources in the groups of selected observations are
  included in the catalogue, whether detected in overlapping or
  non-overlapping parts of their fields of view. Within the series of
  XMM-Newton serendipitous source catalogues, it is named 3XMM-DR7s.

  The following Section~\ref{sec:processing} describes the data processing and
  source detection on multiple observations, an implementation of an adaptive
  smoothing technique to model the background in the images, and the detection
  efficiency and sensitivity for overlapping
  observations. Section~\ref{sec:observations} contains the selection criteria
  of the observations that enter the first stacked catalogue and a new
  automated strategy to identify and reject observations with a high
  background throughout the whole observation.  Section~\ref{sec:catalogue}
  covers the compilation of the catalogue and describes its properties and the
  access it and to the auxiliary products. Section~\ref{sec:summary} gives
  information on planned future catalogue versions and a summary.

\section{Data processing and source detection}
\label{sec:processing}

  The new catalogue 3XMM-DR7s is based on archival XMM-Newton data that
  entered 3XMM-DR7. Throughout the paper, we refer to it as the stacked
  catalogue and to the other releases from source detection on single
  observations as the 3XMM catalogues. The term `stack' is used for a group
  of overlapping observations for which simultaneous source detection is
  performed. In the context of XMM-Newton observations, `exposure' stands
  for the measurement by one of its instruments within an
  observation. `Images' are created for each observation, instrument, and
  energy band separately, if not noted otherwise. If several images are merged
  into a single file, it is called a `mosaic'.

  3XMM-DR7s is processed with the SAS software version 16 and calibration
  files as of July 2017. We follow the data handling outlined in
  \citetalads[Papers\,][]{2009A&A...493..339W},
  \citetalads{2016A&A...590A...1R}, and the 3XMM-DR4 online
  documentation\footnote{\url{https://xmmssc-www.star.le.ac.uk/Catalogue/3XMM-DR4/UserGuide_xmmcat.html}},
  using the same parameters as in the 3XMM pipeline wherever applicable. The
  tasks are adjusted to the needs of source detection on multiple
  observations, including the handling of many input files and large image
  sizes, runtime improvements, wider ranges of allowed parameter values than
  in single observations, for example the minimum detection likelihood, and
  additional output used to create the final stacked source list. The
  standardised approach to perform stacked source detection on multiple
  observations has entered the SAS as a new task \texttt{edetect\_stack}
  together with the updates to the existing source-detection tasks. Its
  structure is illustrated in Fig.~\ref{fig:tasks}. It is a combination of
  newly written Perl code and up to eleven other SAS tasks, comprising three
  major steps: \emph{(i)} Input data to source detection are prepared for each
  observation individually (described in the next two sub-sections). All input
  images are created with the same binning, reference coordinates, and size,
  large enough to cover the sky areas of all observations in the
  stack. \emph{(ii)} Source detection is run on all input data simultaneously
  (described in Sect.~\ref{sec:srcdet}) and the results per input image are
  stored in an intermediate source list. In both steps,
  \texttt{edetect\_stack} determines the appropriate parameter values for the
  other SAS tasks and calls them. \emph{(iii)} Sources which enter the final
  source list are selected and their source parameters calculated from the
  results of step (ii). For source detection on a single observation, this
  step is part of the task \texttt{emldetect}. For multiple observations,
  modifications are needed and a module of \texttt{edetect\_stack} refines
  this functionality of \texttt{emldetect} (described at the end of
  Sect.~\ref{sec:srcdet}).

  \subsection{Preparation of the input data for maximum-likelihood source
    detection}
  \label{sec:inputdata}

  Event lists and attitude files to produce the new catalogue are taken from
  the set of files used to produce the XMM-Newton Serendipitous Source
  Catalogues 3XMM-DR5 to DR7. Within the pipeline processing, the event lists
  are filtered for good time intervals (GTIs) per CCD with a minimum GTI
  length of 10\,s, cleaned of bad pixels and merged per instrument. They are
  publicly available via the XMM-Newton Science Archive
  (XSA\footnote{\url{https://www.cosmos.esa.int/web/xmm-newton/xsa}}). For the
  3XMM catalogues, time intervals of background flares are identified in the
  merged event lists for each instrument using an optimised flare filtering
  method. Observations in mosaic mode have been split into sub-pointings and
  attributed individual observation identifiers. More details on the pipeline
  can be found in \citetalads[Paper\,]{2016A&A...590A...1R}. For the stacked
  catalogue, the XSA event lists are filtered with the 3XMM GTIs. If two event
  lists per instrument are available with the same observation identifier,
  they are combined using the task \texttt{merge}. Within
  \texttt{edetect\_stack}, information about the telescope boresight during
  the exposure is obtained from the attitude files. Therefore, they are also
  filtered with the combined GTIs of all EPIC instruments for the stacked
  catalogue to eliminate erroneously recorded coordinate shifts.

  The filtered event lists and attitude files of a stack of observations are
  passed to the task \texttt{edetect\_stack}. It establishes a common
  coordinate system for the stack from the pointing coordinates in the
  attitude files, which is used for all subsequent source-detection steps.
  The events are projected onto reference coordinates in the local tangent
  plane using the task \texttt{attcalc}\footnote{The maximum fractional area
    distortion introduced by tangential projection in the images used for the
    stacked catalogue with side lengths up to 4\degr\ is smaller than $4\times
    10^{-3}$ and thus negligible in source detection.}. The reference point of
  the projection is calculated as the average of the minimum and maximum
  coordinates of all overlapping observations. The size of the sky area
  covered by them is derived from their pointing coordinates and position
  angles. Using the projected event lists, the input files for source
  detection are prepared for each contributing observation individually,
  namely images and corresponding exposure maps, detection masks, and
  background maps for the three EPIC instruments and the five 3XMM energy
  bands over the full sky area of the stack. The images are created in bins of
  4\arcsec\ $\times$ 4\arcsec\ by the task \texttt{evselect}. Exposure maps
  are created by \texttt{eexpmap} and give the exposure time per instrument,
  taking invalid pixels and relative detector efficiency into account. They
  serve as input to the detection masks and background maps. For the
  source-detection tasks, a second set of vignetting-corrected exposure maps
  is produced. Detection masks are created by \texttt{emask} for each
  instrument and give the valid pixels per image. They are derived from the
  lowest energy band, which defines the most conservative mask. Background
  maps are created by \texttt{esplinemap} and give the modelled background in
  counts per pixel. Its new adaptive-smoothing mode is described in more
  detail in the next sub-section. In addition to these mandatory input files
  for source detection, two sets of products are created for purely
  informational purposes: All input images and those per energy band are
  combined into mosaics by \texttt{emosaic} to illustrate the
  stacks. Sensitivity maps are calculated by \texttt{esensmap} per instrument
  and energy band.

  \subsection{Modelling the EPIC background by an adaptive smoothing technique}
  \label{sec:bkgfitting}

  \begin{figure}
    \centering
    \includegraphics[width=88mm]{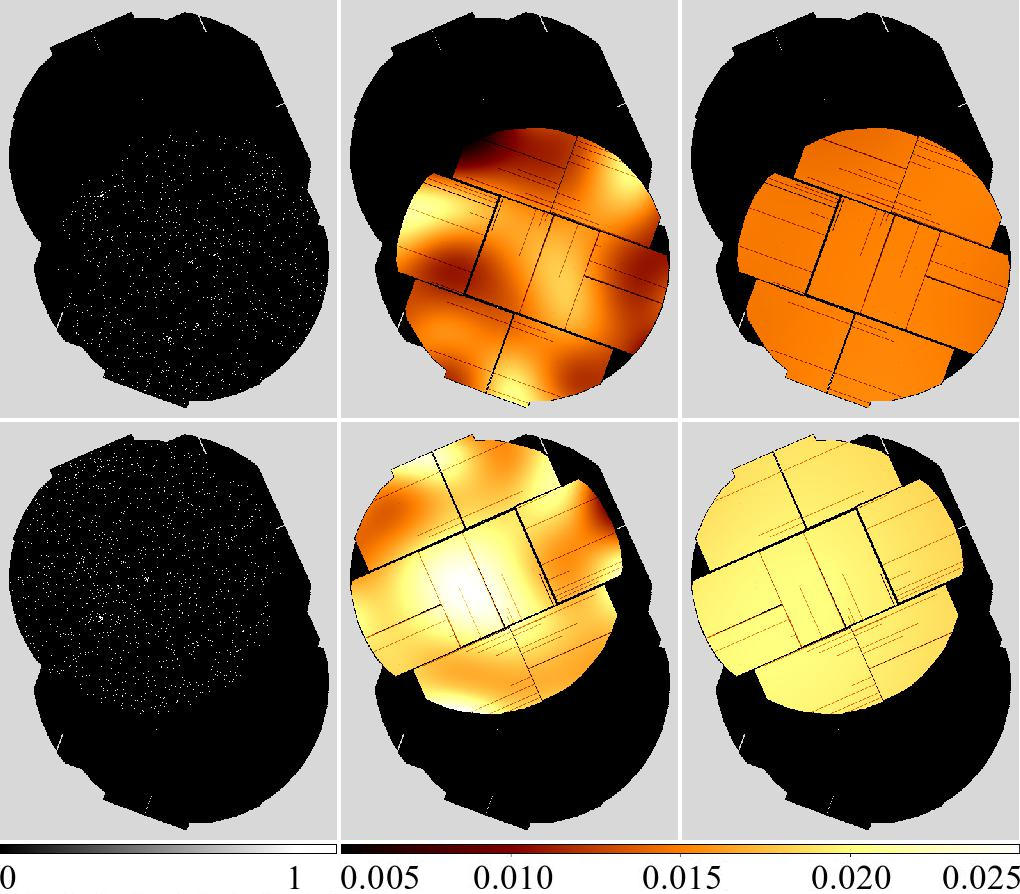}
    \caption{Example of low-amplitude brightness fluctuations in the
      background maps produced by spline fits: images (left), spline fits
      (middle), and adaptive smoothing (right) of MOS2 in the 2.0$-$4.5\,keV
      band of a stack of two observations (identifiers 0741033401 and
      0741033501). The sky region covered by all instruments is shown in
      black.}
    \label{fig:splineexample}
  \end{figure}

  \begin{figure*}
    \centering
    \parbox[b][82mm][b]{152mm}{
    \includegraphics[height=40.5mm]{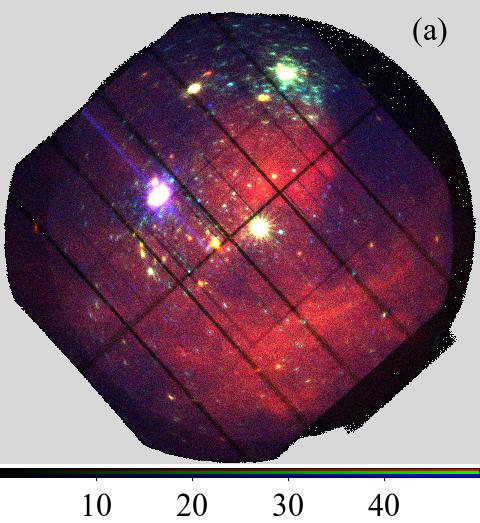}\hskip.5mm%
    \includegraphics[height=40.5mm]{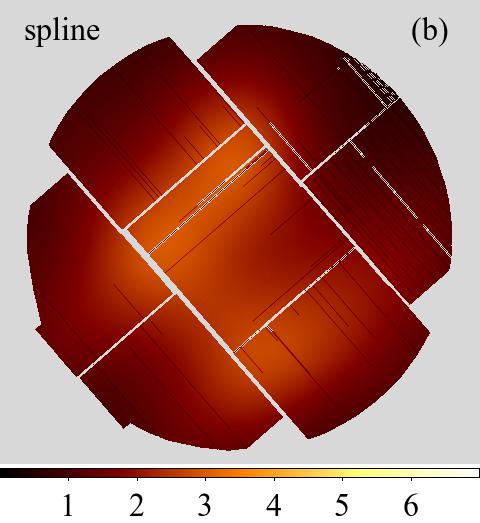}\hskip.5mm%
    \includegraphics[height=40.5mm]{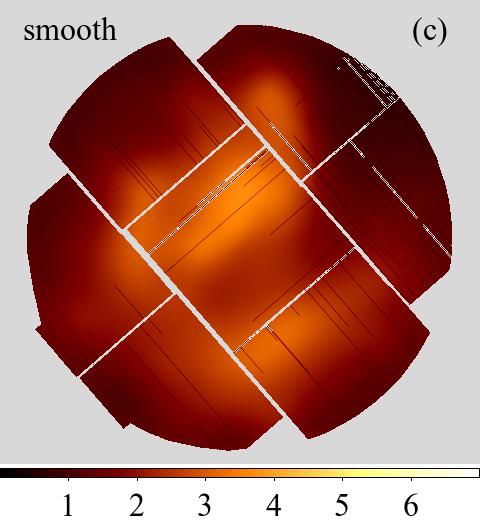}\hskip2mm%
    \includegraphics[height=40.5mm]{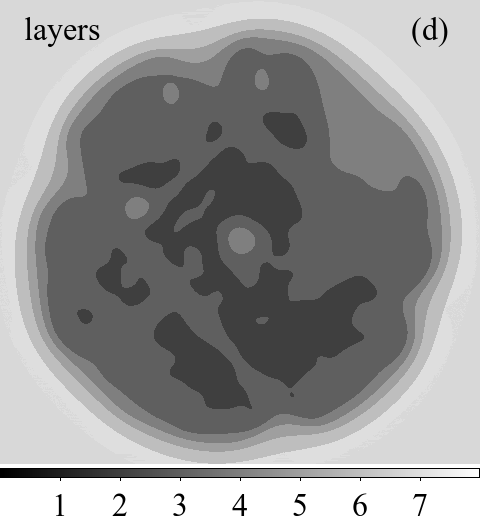}\vfill
    \includegraphics[height=40.5mm]{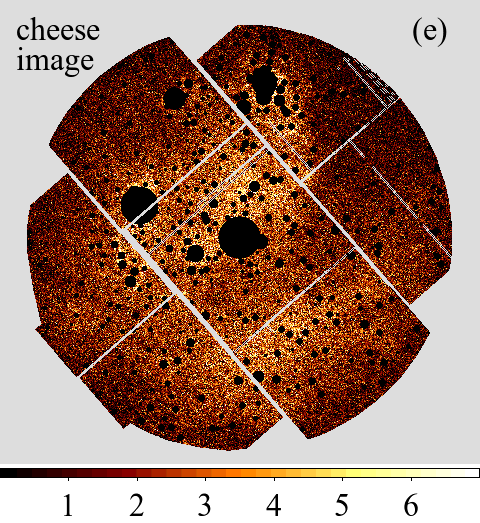}\hskip.5mm%
    \includegraphics[height=40.5mm]{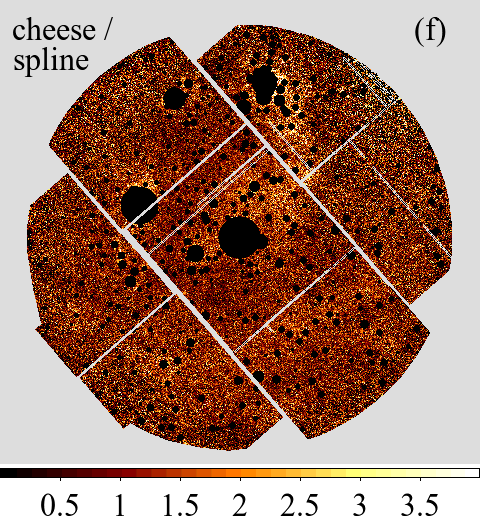}\hskip.5mm%
    \includegraphics[height=40.5mm]{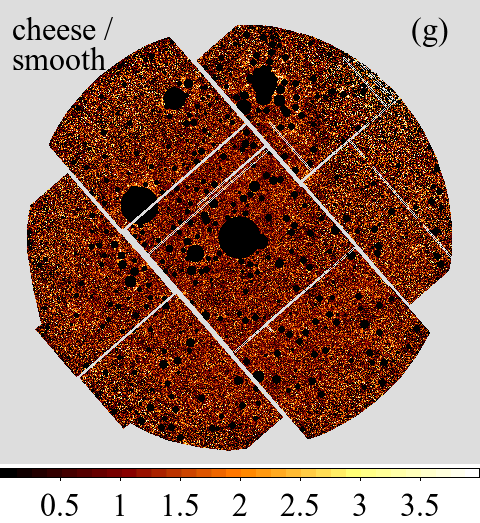}\hskip2mm%
    \includegraphics[height=40.5mm]{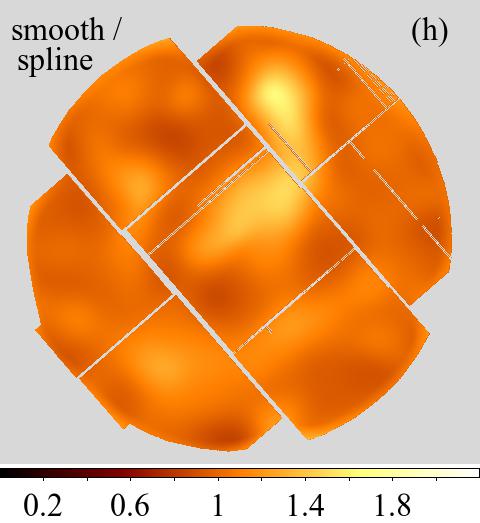}
    }\hskip2mm%
    \includegraphics[height=82mm]{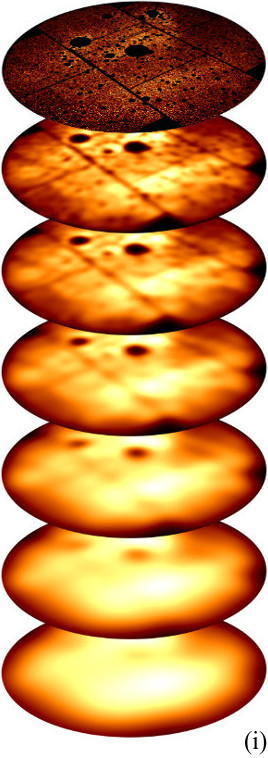}
    \caption{Different methods of background modelling, illustrated in the
      example of an observation of the $\eta$ Carinae region
      (obs.\ id.\ 0112560101). The panels include a three-band false-colour
      image of the EPIC observation (0.2$-$1.0\,keV, 1.0$-$2.0\,keV,
      2.0$-$12.0\,keV), showing \emph{(a)} the complex background structure of
      the field, \emph{(b)} the MOS1 background map derived from a spline fit
      and \emph{(c)} from adaptive smoothing, \emph{(e)} the source-excised
      image, \emph{(f)} and \emph{(g)} its ratio to the two background maps
      and \emph{(h)} the ratio between them. The source-excised image is
      smoothed with a Gaussian kernel of increasing width \emph{(i)}. All
      images have a linear intensity scale. The smoothed layers which are
      chosen per image pixel to construct the background map according to
      their signal-to-noise ratio are shown in grey-scale \emph{(d)}.}
    \label{fig:smoothing}
  \end{figure*}

  The EPIC background includes an internal instrumental background and
  external components such as the cosmic X-ray background together with a
  time-variable local particle background linked to the complex interaction of
  solar activity with the Earth’s magnetosphere
  \citepads[e.g.][]{2003A&A...409..395R}. For source detection, time intervals
  dominated by high and variable background are filtered from the 3XMM-DR7
  event lists\linebreak \citepalads[see Sect.\ 3.2.3 of
    Paper\,]{2016A&A...590A...1R}. The remaining background is modelled based
  on source-excised images by \texttt{esplinemap} and used within the
  source-detection tasks. To construct the source-excised images, sliding-box
  source detection is performed on the input images by \texttt{eboxdetect},
  run in the so-called local mode, in which a local background level is
  directly estimated from the image, using a frame around the search box. The
  resulting list of tentative source positions is passed to the task
  \texttt{esplinemap}, which excludes circular regions centred at the listed
  positions within a brightness-dependent radius from each input image.

  \begin{figure*}
    \centering
  \includegraphics[width=60mm]{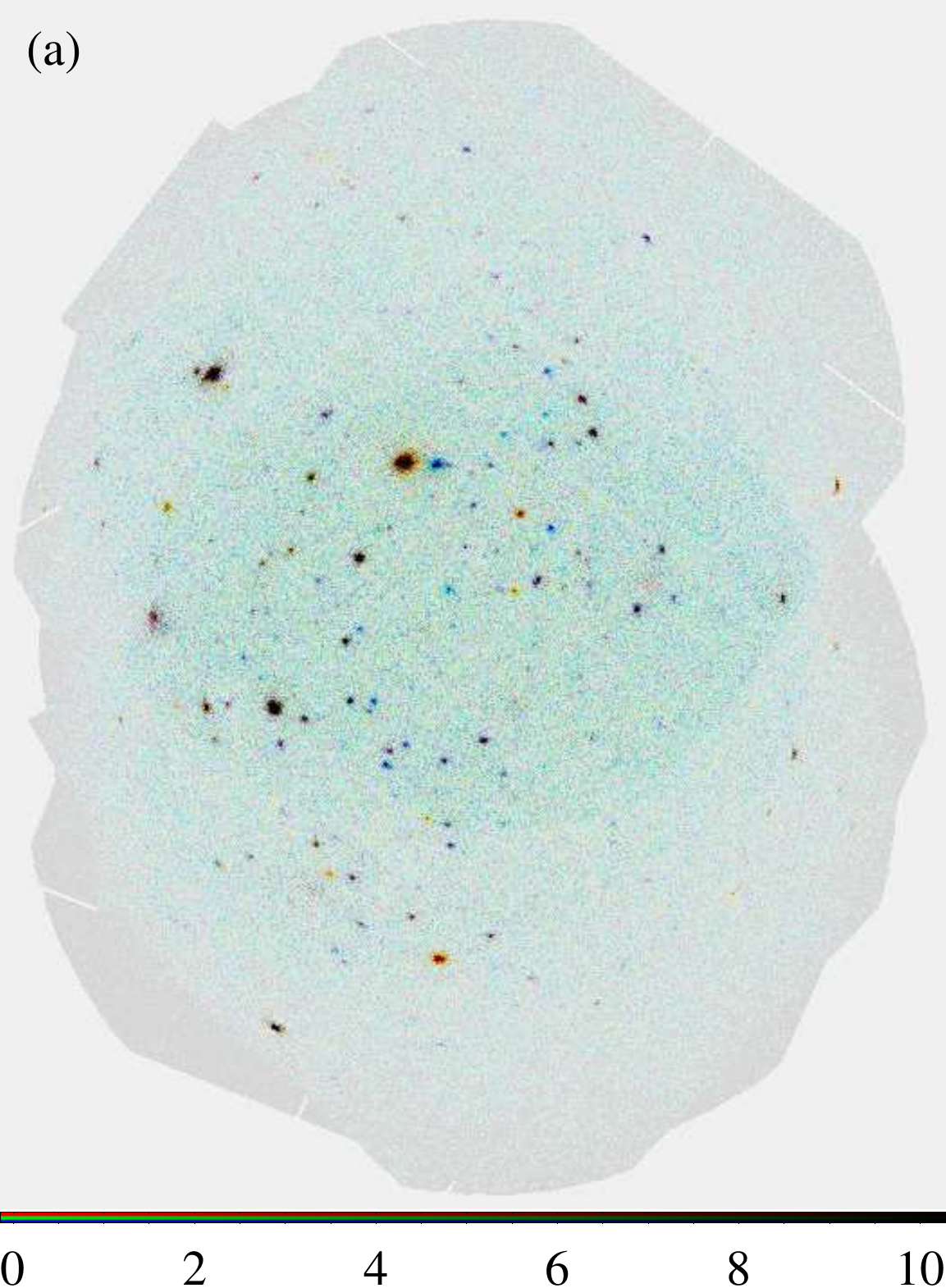}\hskip2pt%
  \includegraphics[width=60mm]{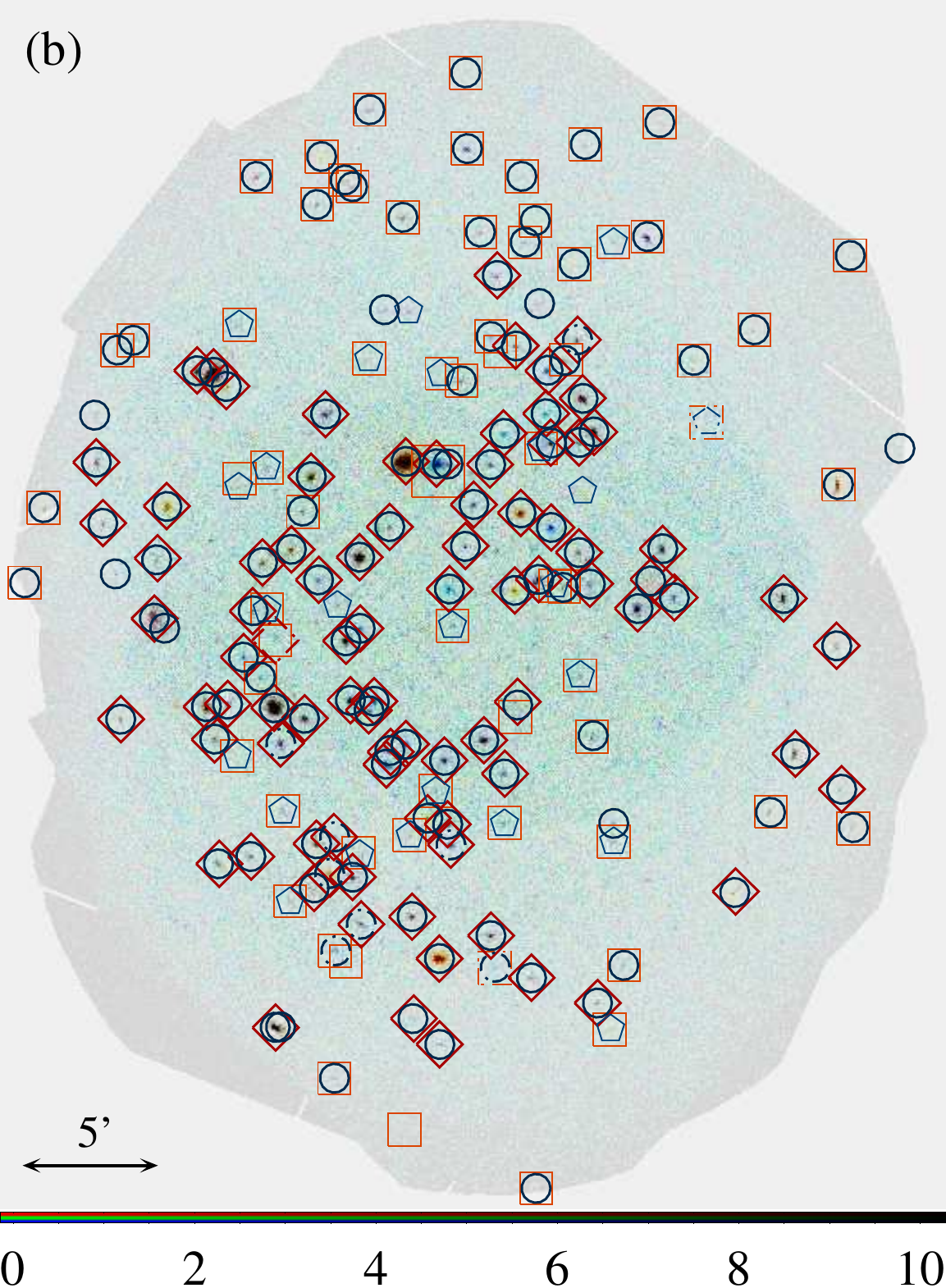}\hskip2pt%
  \includegraphics[width=60mm]{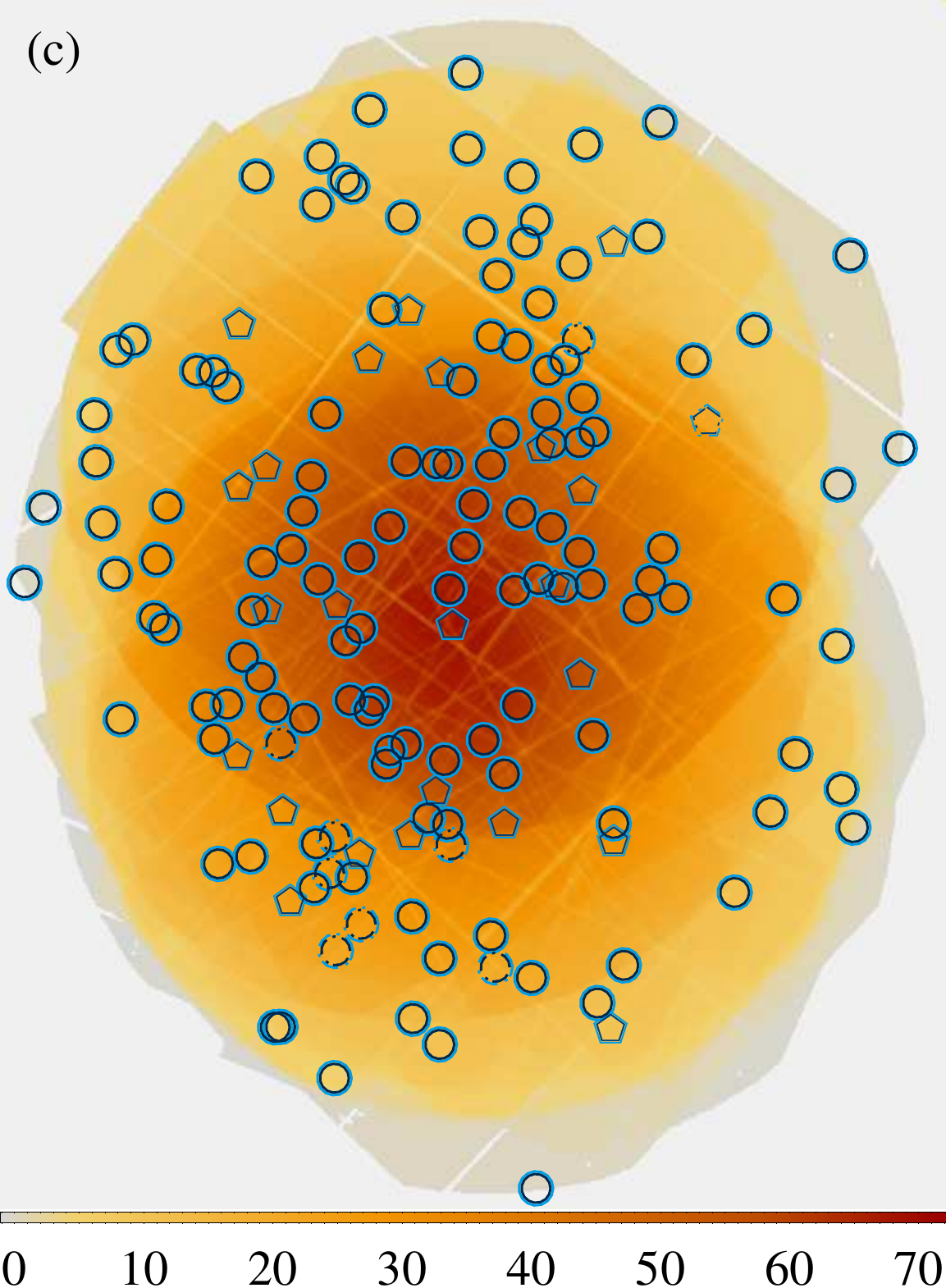}
  \caption{Example of stacked source detection: six overlapping observations
    within the Magellanic Bridge. \emph{(a)} Mosaic of all images. The three
    colour-coded energy bands are: 0.2$-$1.0\,keV (red), 1.0$-$2.0\,keV
    (green), 2.0$-$12.0\,keV (blue). Colour intensity scales linearly with the
    number of counts. \emph{(b)} The same mosaic image with source
    identifications overlaid. \emph{Blue circles and pentagons:} 158 sources
    detected by stacked source detection. Circles mark sources that exceed the
    likelihood threshold in total or in more than one contributing
    observation. \emph{Red diamonds and boxes:} 152 sources detected in the
    individual observations. Diamonds mark sources that exceed the likelihood
    threshold in more than one contributing observation.  Dashed symbols mark
    sources that have been flagged by the task \texttt{dpssflag}
    (cf.\ Sect.~\ref{sec:columns}). \emph{(c)} Mosaic of the vignetted
    exposure maps with the identifications of the sources in the stacked
    catalogue, using the same symbols as in panel \emph{(b)}. The exposure
    time has been averaged over the instruments and energy bands and is given
    in units of kiloseconds in the colour bar.}
    \label{fig:detimages}
  \end{figure*}

  A spline fit has been the standard method to model the background and
  extrapolate it to the source positions in single observations; this is also
  employed for the 3XMM catalogues. It gives a reasonably good description of
  the background behaviour in most images of standard size from single
  pointings. Test runs, however, have revealed that its current SAS
  implementation, which was designed for single observations, can result in
  undesired overshoot or ringing effects for images that are larger than a
  single XMM-Newton EPIC field of view as needed for stacked source detection
  (two examples are shown in Fig.~\ref{fig:splineexample}). The artefacts
  occur in particular close to the sharp transition between the exposed and
  the unexposed image area within and outside a single field of
  view. Furthermore, the splines may smooth out small-scale variations in very
  complex background structures. Thus, an adaptive filtering method to model
  the background emission has been introduced in
  \texttt{esplinemap}\footnote{Although the task is now capable of three
    different methods of background modelling including spline fits and
    smoothing, its initial name \texttt{esplinemap} is retained to be
    consistent with former SAS versions.} as an alternative to the spline
  fitting. The source-excised images, normalised by the exposure maps, and the
  corresponding masks are convolved with a Gaussian kernel. The resulting
  smoothed images are divided by the smoothed masks, compensating for the
  unknown background flux in the masked source regions. To account for
  different background structures in individual image areas, an optimum
  smoothing radius is determined pixel by pixel such that the final adaptively
  smoothed background map has a uniform signal-to-noise ratio, which limits
  the allowed noise fluctuations. Therefore, the initial width of the Gaussian
  kernel is increased by a factor of $\sqrt{2}$ in eight steps. The counts per
  pixel in the smoothed images are the weighted average over the kernel extent
  centred at the pixel position. Their Poissonian signal-to-noise ratio is
  calculated as the square root of the counts under the kernel. For each
  pixel, the two smoothed images with the signal-to-noise ratios closest to
  the pre-defined (user-supplied) optimum are selected. The background value
  with the desired signal-to-noise ratio is linearly interpolated between
  them. Small-scale structures are thus covered by the images with the
  narrowest smoothing radii, while the cut-out regions around the sources are
  filled by values from those with a broad Gaussian kernel. The new default
  parameters of this method in \texttt{esplinemap} have been chosen
  empirically as a brightness level of $5\times
  10^{-4}$\,cts\,arcsec$^{-2}\,\textrm{s}^{-1}$ to cut out sources, a minimum
  smoothing radius of 10\,px, corresponding to 40\arcsec\ when using standard
  image binning, and a signal-to-noise ratio of 30. For the catalogue images,
  these values result in a reasonable compromise between minimising the
  remaining photon noise in the background map and retaining the resolution
  for true spatial background variations.

  The smoothed background maps are generally in good agreement with the input
  images. For the 26\,835 catalogue images, the median deviation between the
  total counts of the source-excised background maps and images is below
  2\,\%. Figure~\ref{fig:smoothing} provides an example comparing the
  spline-fit background and the results of adaptive smoothing for a single
  observation of the region of $\eta$~Carinae. The large-scale variation of
  its complex background structure (Fig.~\ref{fig:smoothing}a) is well
  described by the spline fit (Fig.~\ref{fig:smoothing}b), while small-scale
  structure becomes additionally visible in the adaptive smoothing fit
  (Fig.~\ref{fig:smoothing}c). The differences between the two methods are
  most obvious in a comparison of the ratios between the source-excised image
  (Fig.~\ref{fig:smoothing}e) and the source-excised background maps
  (Fig.~\ref{fig:smoothing}f and g) and in a direct comparison of the
  background maps (Fig.~\ref{fig:smoothing}h). Figure~\ref{fig:smoothing}i
  shows six of the eight layers with increasing smoothing radii, from which
  the smoothed background map has been constructed, and
  Fig.~\ref{fig:smoothing}d the layer chosen for each image pixel. Tests on
  selected fields with complex background and of large images processed with
  both methods confirm a more robust approximation of the observed background
  by adaptive smoothing in these cases. However, it may be less sensitive to
  extended low surface-brightness sources, in particular if small cut-out
  radii are chosen for the source-excised images. Adaptive smoothing has been
  chosen as the standard approach for the new catalogue, whose first version
  is restricted to fields without large extended emission (see
  Sect.~\ref{sec:observations}).

  \subsection{Source detection on stacked images}
  \label{sec:srcdet}

  All data products described in Sect.~\ref{sec:inputdata} are used in
  parallel by the source-detection tasks, which couple images, exposure maps,
  and background maps for each observation, instrument, and energy band, and
  detection masks for each observation and instrument. Simultaneous source
  detection is performed by means of the usual two-step process used for
  XMM-Newton data: sliding-box source detection followed by maximum-likelihood
  fitting. This was described originally in
  \citetalads[Paper\,][]{2009A&A...493..339W}. In the following paragraphs,
  essentials common to source detection on single and on multiple observations
  are summarised, followed by the modifications introduced for the stacked
  catalogue. Both detection steps test the null hypothesis that all counts
  collected arise from random background fluctuations and no source is
  present. The null-hypothesis probability $P_\mathrm{null}$ is converted into
  a measure for detection significance by the logarithmic likelihood $L=-\ln
  P_\mathrm{null}$, which is given in the XMM-Newton source lists.

  First, all images are searched for tentative sources by a sliding-box source
  detection using the task \texttt{eboxdetect}. The initial run is made with a
  20\arcsec\ box size. Two subsequent runs increase the box size by a factor
  two each to facilitate searches for extended sources. Detections from
  previous runs are overwritten if one is found at the same position with a
  higher signal-to-noise ratio. For each image $i$, a logarithmic likelihood
  \begin{equation}\label{eq:ebox_li}
    L_i(c_i,c_b)=-\ln P_\Gamma(c_i,c_b)
  \end{equation}
  is calculated such that the measured counts $c_i=c_s+c_b$ within the
  detection box exceed the level of pure Poissonian noise. $c_s$ are the
  source and $c_b$ the background counts in the detection region. $P_\Gamma$
  is the regularised incomplete gamma function
  \begin{equation}\label{eq:igamma}
      P_\Gamma(a,x) = \frac{\int_0^x e^{-t}t^{a-1}dt}{\int_0^\infty
        e^{-t}t^{a-1}dt} ,
  \end{equation}
  used here as the cumulative distribution function of a Poisson
  distribution. According to \citet{1932fisher}, the natural logarithms of
  probabilities $P_i$ from $n$ independent tests of the same null hypothesis
  can be combined as $= -2 \sum_{i=1}^n{\ln{P_i}}$, which follows a $\chi^2$
  distribution with $2n$ degrees of freedom.  The detection likelihoods of a
  source in $n$ individual images is hence calculated as
  \begin{equation}\label{eq:ebox_ltot}
    \textrm{LIKE}_\textrm{eboxdetect}=-\ln\left(1-P_\Gamma(n,
    \sum_{i=1}^n{L_i})\right) ,
  \end{equation}
  making use of $P_\Gamma$ as the $\chi^2$ cumulative distribution
  function. The combined EPIC detection likelihoods are also called
  `equivalent likelihoods', referring to \citet{1932fisher}.  All images are
  considered for which the source position lies within the detection
  mask. Their number can thus vary from source to source within one detection
  run. Sources are selected if their equivalent likelihood exceeds a
  pre-defined minimum, and passed to the task \texttt{emldetect} to calculate
  their parameters by maximum-likelihood fitting. A good likelihood cut
  represents a compromise between being as complete as possible with respect
  to real sources and as strict as possible with respect to spurious
  detections.

  The equivalent likelihood depends on the number of photons in the detection
  box and on the number of images over which they are distributed, because the
  large number of images in multiple observations leads to large corrections
  when combining their individual detection likelihoods according to
  Eq.~\ref{eq:ebox_ltot}. In particular, the sensitivity of the sliding-box
  detection decreases if few counts are distributed over an increasing number
  of images (cf.\ Sect.~\ref{sec:artstack}). To avoid the loss of real sources
  solely because of the number of images of multiple observations, the stacked
  box detection step was hence reduced to the same number as used for a single
  observation: one image for each EPIC instrument and energy band, limiting
  the number of images $n$ in Eq.~\ref{eq:ebox_ltot} to fifteen. Therefore,
  the corresponding images of all contributing observations are summed per
  instrument and energy band by the task \texttt{emosaic} within
  \texttt{edetect\_stack}; likewise the corresponding exposure maps,
  background maps, and detection masks. These mosaics are exclusively used in
  the sliding-box run. However, transient sources that are significant in a
  subset of the observations may disappear from the pre-selection if box
  detection is restricted to the mosaics. Thus, \texttt{eboxdetect} is also
  called for each observation separately. For the stacked catalogue, a
  likelihood cut of five is used in all \texttt{eboxdetect} runs. The source
  lists of all observations and the one based on the mosaics are merged by
  \texttt{srcmatch} within a fixed radius of $2\sqrt{2}$ times the pixel size,
  chosen to cover the area of two by two pixels. The matching radius for
  standard images with a default binning of 4\arcsec\ thus becomes
  11.3\arcsec. The likelihood column of the merged source list holds the
  maximum detection likelihood of a source.

  Next, the task \texttt{emldetect} determines the parameters of all sources
  in the merged box-detection source list in all images per observation,
  instrument, and energy band simultaneously by means of maximum-likelihood
  fitting. Details on the approach and the parameters chosen for the catalogue
  processing are given in Sect.~4.4.3 of
  \citetalads[Paper\,][]{2009A&A...493..339W}. All input images are combined
  with their respective background image, exposure map, and detection mask. In
  each image, the appropriate PSF is chosen at the tentative source position
  for the instrument configuration. The common source position and extent and
  the counts per image are fitted within an area of
  1\arcmin$\times$1\arcmin\ in all images for which the PSF overlaps with the
  field of view as defined in the detection mask. \texttt{emldetect} scales
  each PSF with the counts measured in the image. Thus, it does not need to
  merge PSFs a priori and to make assumptions about the source spectrum. The
  detection sensitivity is then approximately the same for all incident source
  spectra \citepads{2009A&A...495..989S} and nearly independent of the
  accuracy of the instrument cross-calibration.  To choose the sources that
  are considered real and to minimise the spurious content, a significance
  level needs to be defined. For each source, the detection likelihood in the
  given fitting setup is derived using the best-fit C-statistic
  \citepads{1976A&A....52..307C,1979ApJ...228..939C}, minimising the sum of
  the deviations
  \begin{equation}\label{eq:cash}
    C_i(c_i) = 2 \sum_{k=1}^N (m_k - c_k \ln(m_k))
  \end{equation}
  between measured counts $c$ and the model prediction $m$ in a region of $N$
  pixels, where $c_i$ stands for the sum of source counts $c_s$ and background
  counts $c_b$ in the detection region as before. It is compared to the null
  hypothesis that the signal purely arises from background counts $c_b$,
  resulting in the logarithmic likelihood ratios $\Delta C_i = C_i(c_i) -
  C_i(c_b)$. According to \citetads{1979ApJ...228..939C}, the $\Delta C$
  values follow a $\chi^2$ distribution with $\nu$ degrees of freedom, which
  is the number of varied parameters. The $\Delta C_i$ of the $n$ images
  involved are combined into the equivalent likelihood
  \begin{equation}\label{eq:eml_ltot}
    \textrm{DET\_ML}_\textrm{emldetect} =
    -\ln\left(1-P_\Gamma(\frac{\nu}{2},\sum_{i=1}^n\frac{\Delta C_i}{2})\right) ,
    \end{equation}
  using the regularised incomplete gamma function $P_\Gamma$
  (Eq.~\ref{eq:igamma}). The likelihood values are then a measure for
  detection significance that the collected counts exceed random background
  fluctuations. The $\nu$ free parameters are the coordinates of the source,
  its extent, and the counts per image in which the source lies within the
  instrumental detection mask. If the likelihood of the source being extended
  falls below a threshold of four or its extent radius below
  6\arcsec\ \citepalads[see Paper\,][]{2009A&A...493..339W}, the source extent
  is set to zero and $\nu$ is reduced by one to $n+2$. Using these
  definitions, the degradation of the detection sensitivity with the number of
  images for faint sources is less prominent than for \texttt{eboxdetect}
  (cf.\ Sect.~\ref{sec:artstack}), and \texttt{emldetect} is applied to all
  images of the stack simultaneously. Deviating from the standard procedure
  for individual observations, \texttt{emldetect} is called by
  \texttt{edetect\_stack} with a minimum detection likelihood of zero to store
  the parameters of each box-detection source and each image in an
  intermediate source list without \mbox{(de-)}selecting sources.

  A separate module of the task \texttt{edetect\_stack} is dedicated to the
  calculation of the final source parameters, to performing a quality
  assessment, and to source filtering. In particular, the total equivalent
  likelihood over all observations and the likelihoods for each individual
  observation are calculated for each detection. Sources are included in the
  final source list if at least one of these equivalent likelihoods exceeds a
  user-defined minimum. As in the 3XMM catalogues, a likelihood of at least
  six is required in the stacked catalogue. An example of stacked source
  detection on archival observations of the Magellanic Bridge region is shown
  in Fig.~\ref{fig:detimages}. For comparison, \texttt{emldetect} was also run
  for each observation separately.
  The resulting detections are joined within a matching radius of 15\arcsec,
  the radius used to create the 3XMM catalogues of unique sources, and shown
  in red in Fig.~\ref{fig:detimages}b. A comparison between source lists from
  stacks and individual observations is given in Sect.~\ref{sec:detqual}.

  The results of \texttt{edetect\_stack} are provided in two FITS-format
  source lists with different structure: one \texttt{emldetect}-like list and
  one in catalogue-like format. The first is described in the task
  documentation of
  \texttt{emldetect}\footnote{\url{http://xmm-tools.cosmos.esa.int/external/sas/current/doc/emldetect/}}. The
  second list includes an all-observation all-EPIC summary row for each
  detected source plus one additional row for each individual contributing
  observation of this particular source.
   These latter catalogue-like source lists are the basis of the new stacked
   catalogue. Details on their columns are found in Sect.~\ref{sec:columns}
   and Table~\ref{tab:columns}.

  \subsection{Testing detection efficiency and sensitivity with artificial stacks}
  \label{sec:artstack}

  The efficiency of the new stacked source detection was investigated in
  several tests using long archival observations. Stacks were constructed by
  dividing their event lists into shorter ones. Source detection was performed
  following the recipes given above and a reference source list created from
  the full exposure.

  \begin{figure}
    \centering
    \includegraphics[width=88mm]{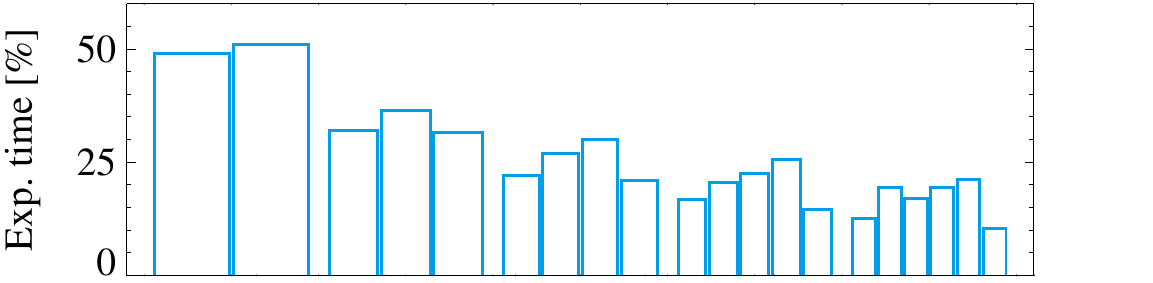}\vskip-1pt
    \includegraphics[width=88mm]{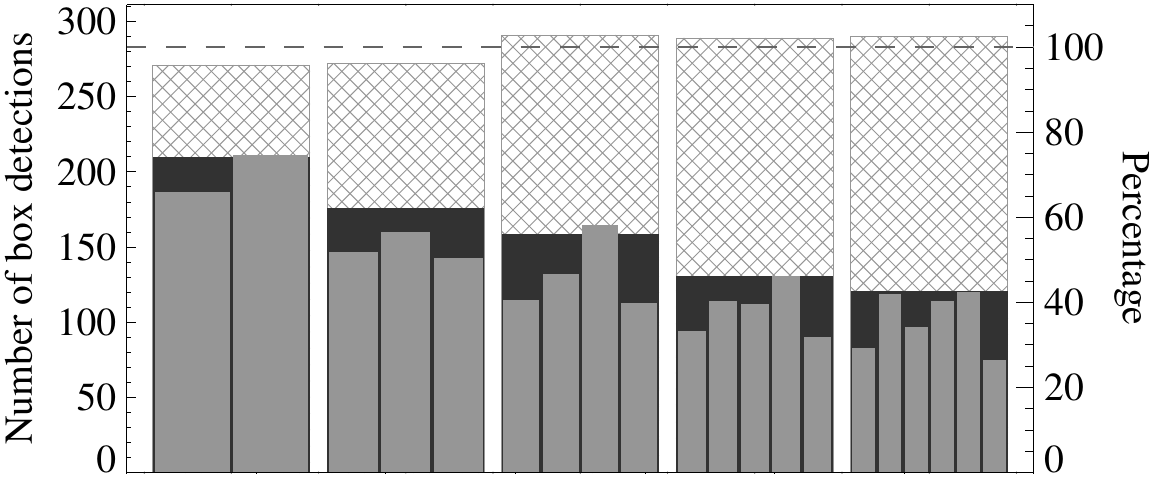}\vskip-1pt
    \includegraphics[width=88mm]{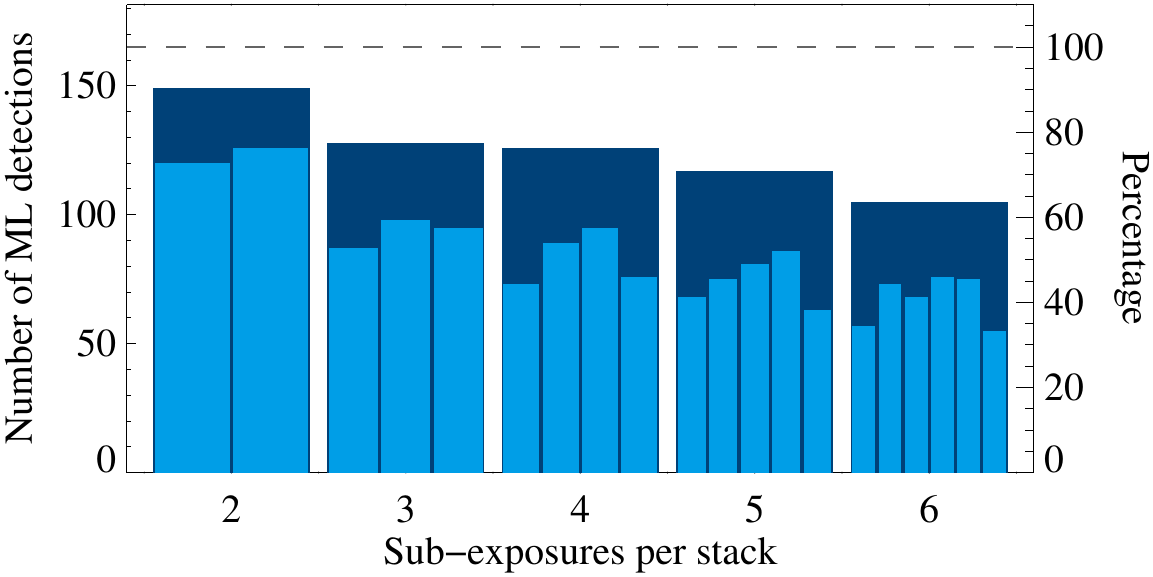}
    \caption{Stacked source detection on an observation split into several
      nearly equally long sub-exposures (obs.\ id.\ 0555780201). \emph{Upper
        panel:} Percentage of exposure time. \emph{Middle panel:} Sliding box
      detections that are submitted to \texttt{emldetect}. Cross-hatched bars
      mark those found in the fifteen mosaics of all sub-exposures, dark
      filled bars the detections found when running \texttt{eboxdetect} on all
      individual images simultaneously, and light filled bars the box
      detections in each individual sub-exposure. \emph{Lower panel:} Final
      maximum-likelihood detections with a minimum total detection likelihood
      of at least six in the stack (dark blue) and in the individual
      sub-exposures (light blue). The dashed horizontal line marks the result
      of source detection on the full, unsplit observation.}
    \label{fig:pseudomosaics}
  \end{figure}

  \begin{figure}
    \centering
    \includegraphics[width=88mm]{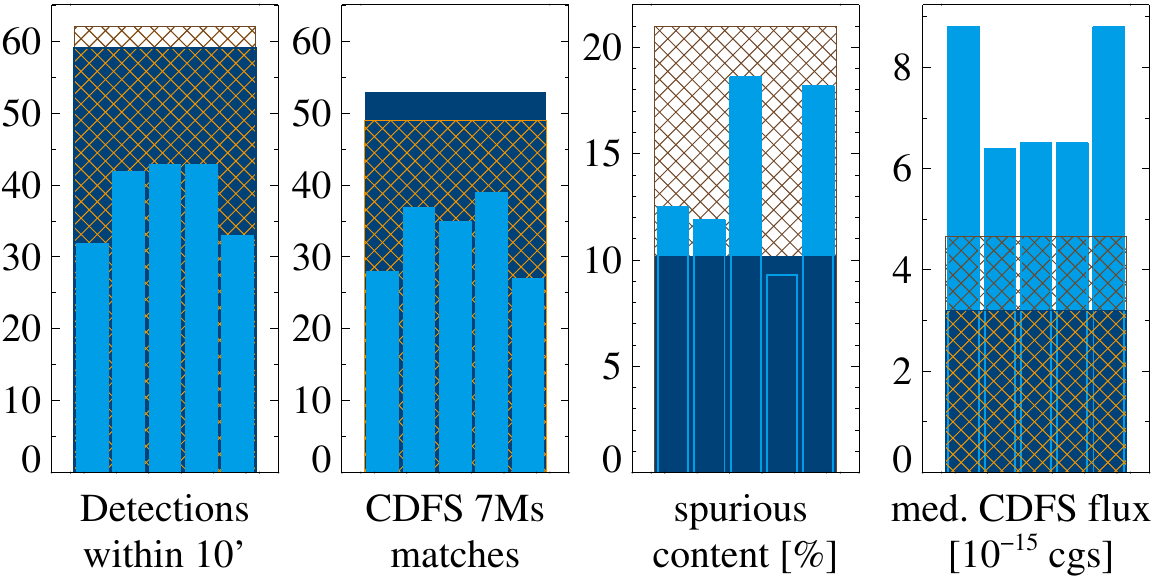}
    \caption{The five-component artificial stack from
      Fig.~\ref{fig:pseudomosaics} compared to the CDFS 7\,Ms catalogue within
      a 10\arcmin\ circle. \emph{From left to right:} All XMM-Newton
      detections, those with a Chandra match within 5\arcsec, fraction of
      detections without Chandra counterpart, and Chandra full-band fluxes of
      the matches. Dark blue bars denote the results from stacked source
      detection, light blue bars those from source detection on the individual
      sub-exposures, and orange cross-hatched bars their joined source lists.}
    \label{fig:CDFSmatch}
  \end{figure}

  In the first experiment, the detection efficiency and its dependence on the
  number of overlapping observations was investigated. Selected observations
  with an exposure time of at least 100\,ks were split into two to six
  sub-exposures of similar duration. The results of source detection on the
  various stacks were compared to those for the full
  observation. Figure~\ref{fig:pseudomosaics} shows an observation of the
  Chandra Deep Field South, a deep extragalactic survey field
  (obs.\ id.\ 0555780201). As expected, the number of sliding-box detected
  sources decreases drastically if all input images are used in parallel but
  remains approximately constant for the corresponding mosaics. A slight
  increase in box detections with the number of sub-exposures indicates more
  false positives. The number of maximum-likelihood detected sources also
  tends to decrease close to the detection limit when the number of
  sub-exposures increases. The source counts are distributed among more
  images, resulting in lower detection likelihoods per image, and the fit has
  more degrees of freedom, resulting in larger corrections when calculating
  the total equivalent likelihood. The overall sensitivity, hence the number
  of reliably detected sources is reduced with an increasing number of short
  sub-exposures. A given source will thus have different likelihood values in
  a stack or one long observation of the same length despite the correction
  scheme applied (see below for a quantitative assessment).

  \begin{figure}
    \centering
    \includegraphics[width=77mm]{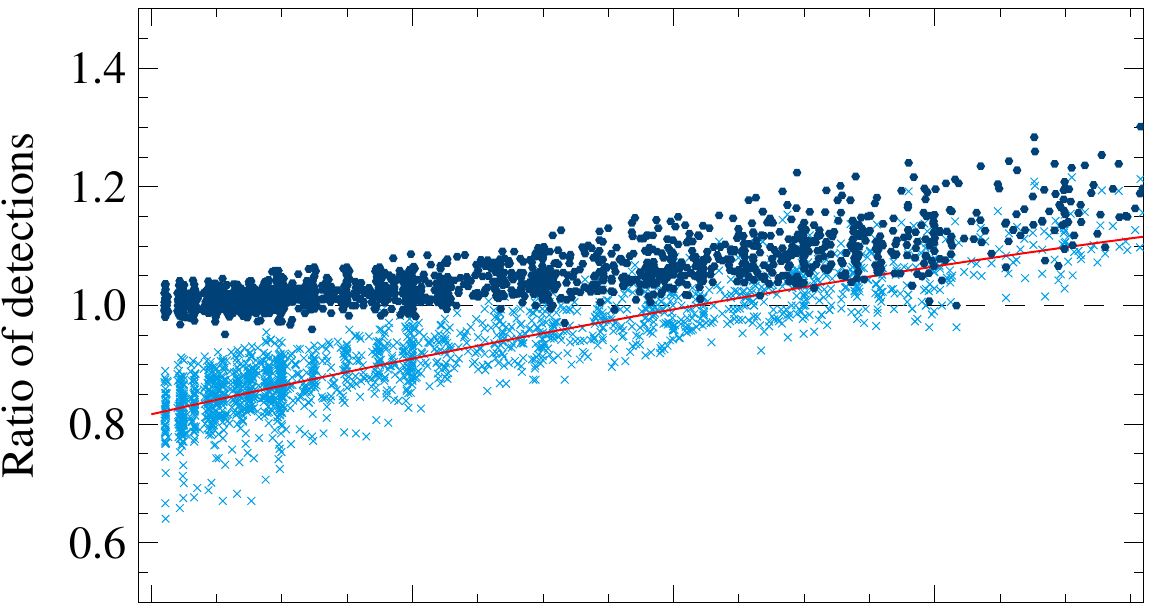}\vskip1pt
    \includegraphics[width=77mm]{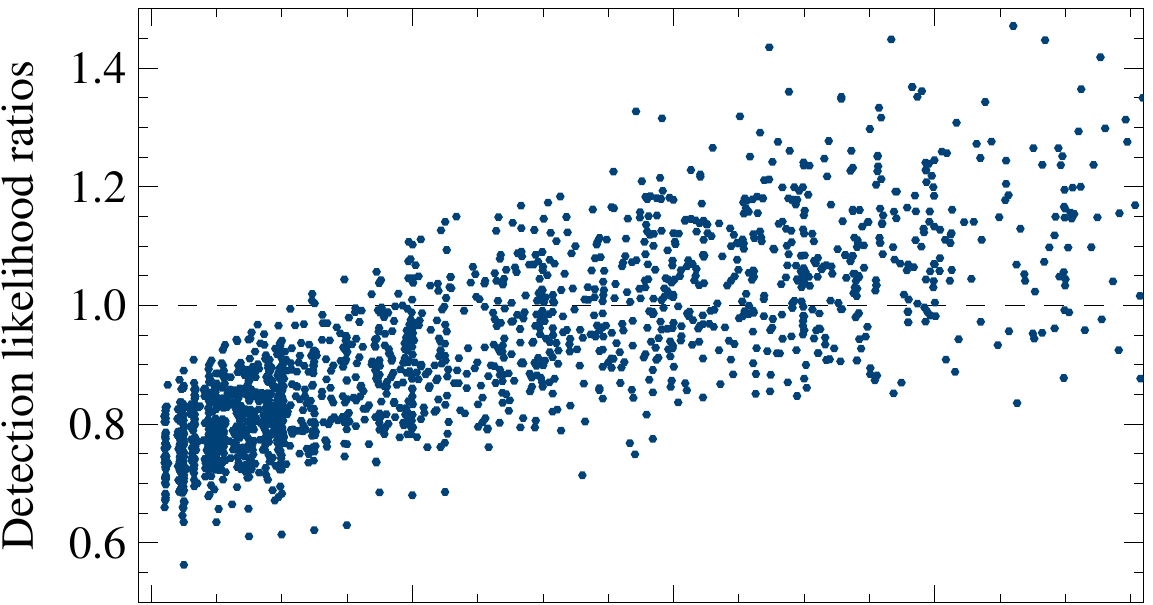}\vskip1pt
    \includegraphics[width=77mm]{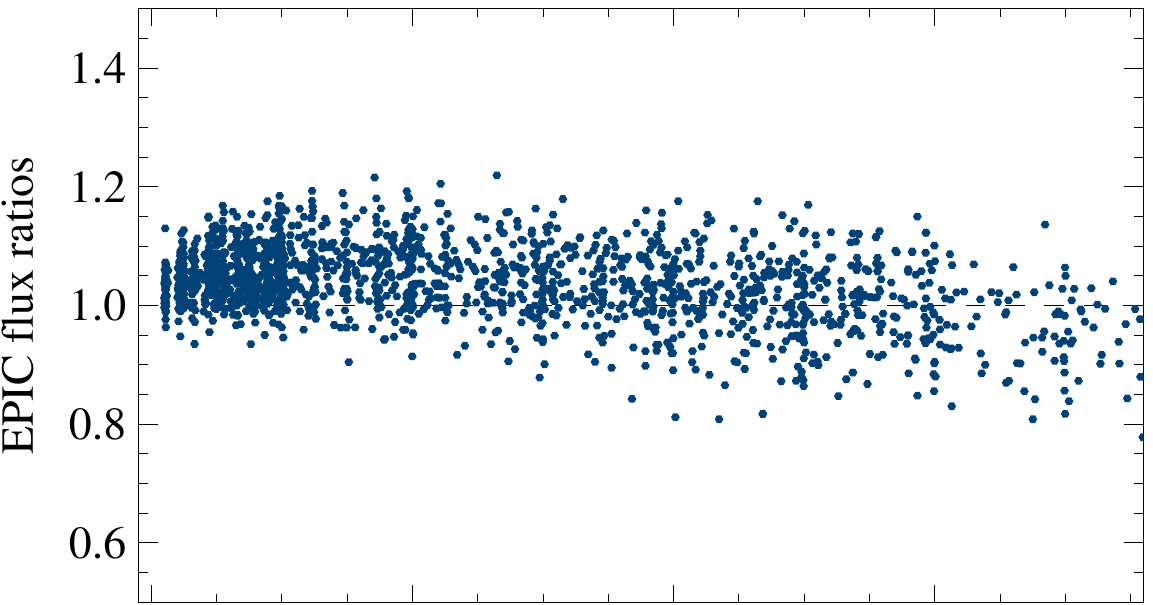}\vskip1pt
    \includegraphics[width=77mm]{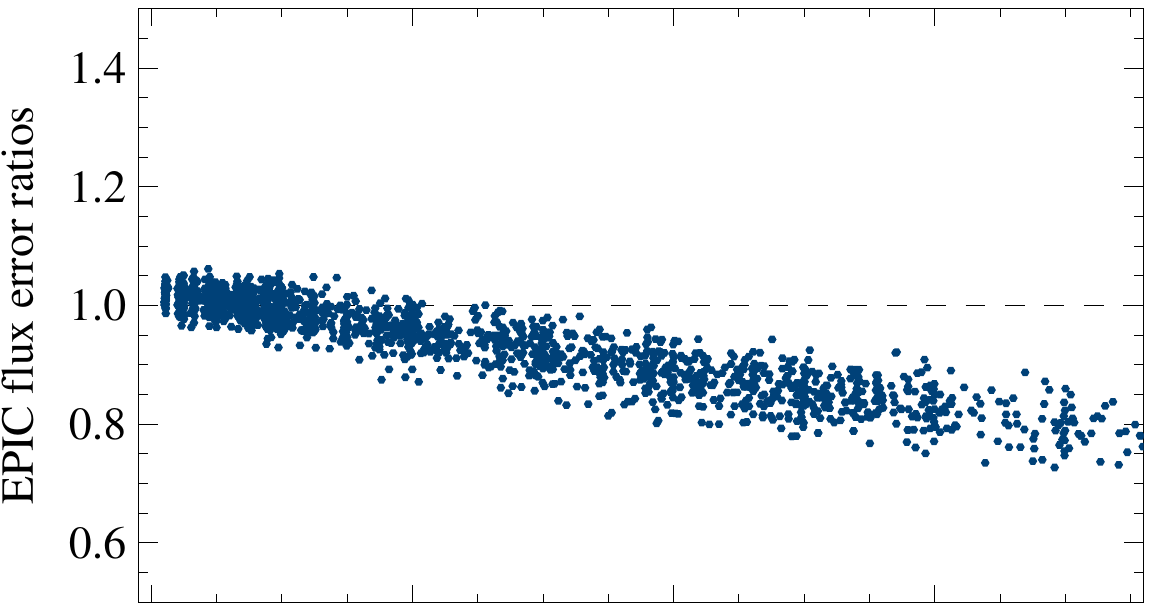}\vskip1pt
    \includegraphics[width=77mm]{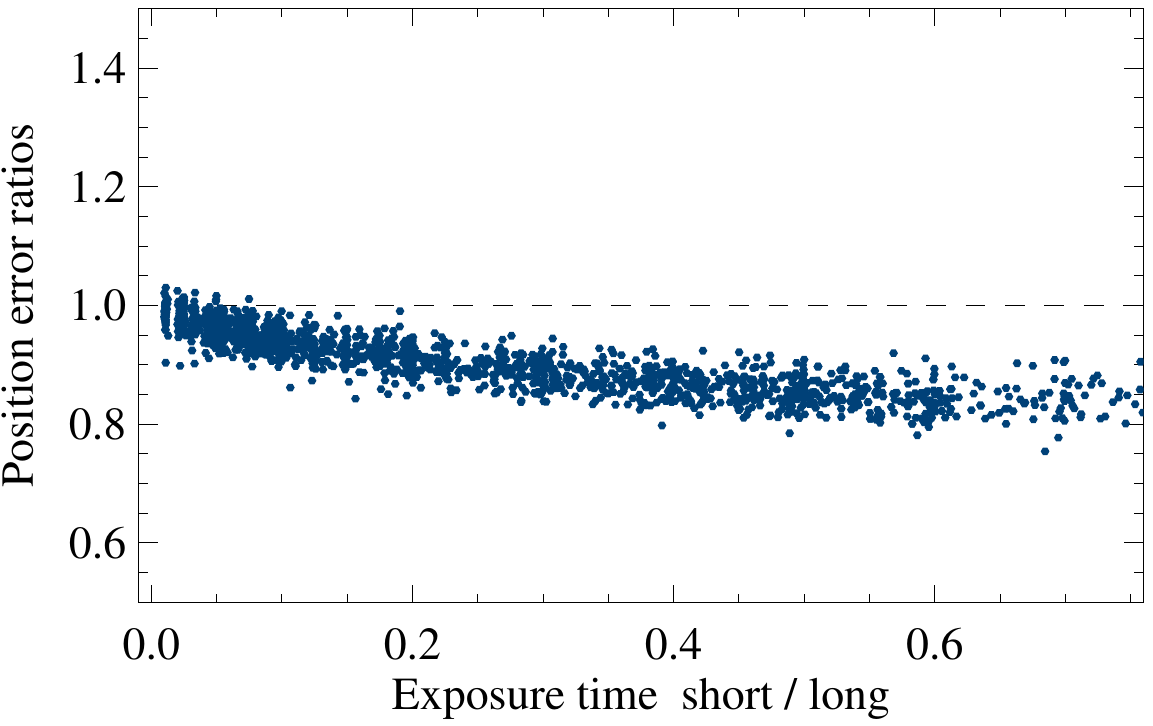}
    \caption{Source parameters derived from stacked source detection in a
      longer and a shorter part of long observations, compared to source
      detection in the longer part only. Each dot represents the ratio of the
      median values of the sources detected in one stack to the median values
      of the sources detected in the long sub-exposure alone. Sources with an
      equivalent detection likelihood above six in at least one exposure are
      included in the stacked source list. The light blue crosses in the
      uppermost panel mark the ratio of sources with a total likelihood above
      six. The red curve is a 2nd-order polynomial fit to guide the eye.}
    \label{fig:pseudomostwo}
  \end{figure}

  \begin{figure}
    \centering
    \includegraphics[width=76mm]{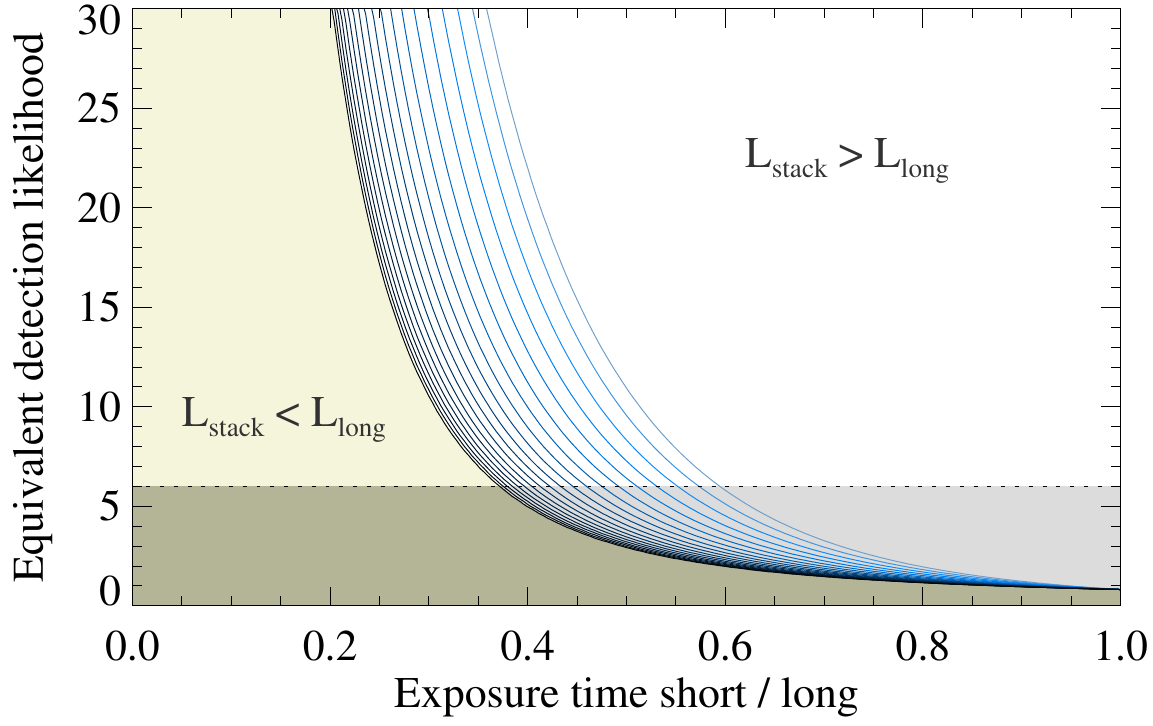}
    \caption{Numerically calculated limiting detection likelihood in stacks of
      a long and a short observation. For given counts (source plus
      background) and exposure time ratios, the detection likelihood
      $L_\mathrm{long}$ in the long observation and $L_\mathrm{stack}$ in the
      stack are calculated. The curves show equal likelihoods
      $L_\mathrm{stack}=L_\mathrm{long}$, each for a fixed count
      number. Counts increase from right to left from 15 to 5\,000 in 18 steps
      of 0.14\,dex.}
    \label{fig:detlimits}
  \end{figure}

  To further investigate the reliability and spurious content of the stacked
  detections, the artificial five-component stack of
  Fig.~\ref{fig:pseudomosaics} was compared with the 7\,Ms catalogue of the
  Chandra Deep Field South survey \citepads{2017ApJS..228....2L}, which is
  expected to include all detectable sources of the much shorter single
  XMM-Newton observation. The comparison was restricted to the innermost
  10\arcmin\ of the Chandra field, corresponding to a Chandra flux limit of
  about $4\times 10^{-16}\,\textrm{erg\,cm}^{-2}\,\textrm{s}^{-1}$. From the
  sub-exposures of the artificial EPIC stacks, a joint source list was created
  by merging the individual lists. Its flux limit is about $2\times
  10^{-15}\,\textrm{erg\,cm}^{-2}\,\textrm{s}^{-1}$. Detections were merged
  within a radius of 15\arcsec\ (the radius used to create the 3XMM catalogues
  of unique sources). The EPIC and the Chandra detections were then matched
  within a radius of 5\arcsec, taking the higher source density of the Chandra
  catalogue into account. Each match is considered a true source and each EPIC
  detection without a Chandra counterpart is considered spurious, including
  the considerable fraction of long-term variable sources that were
  undetectable during the Chandra observation
  \citepads[see][]{2009A&A...497..423M}. Figure~\ref{fig:CDFSmatch} shows the
  number of sources and the median Chandra full-band fluxes for stacked source
  detection, for source detection in the individual sub-exposures, and for
  their combined source list. The flux sensitivity and the number of reliable
  detections are higher in the stack than in the sub-exposures alone, and the
  spurious content decreases significantly, in this example by about 50\%.

  In a second experiment, the detection efficiency for combinations of two
  observations with different exposure times was investigated. As described in
  Sect.~\ref{sec:srcdet}, the combined detection likelihood of a source
  depends not only on the number of photons collected, but also on the number
  of images used in the fit. The number of images and thus the number of free
  parameters in Eqs.~\ref{eq:ebox_ltot} and \ref{eq:eml_ltot} increases by the
  number of energy bands times the number of active instruments in each
  observation that is added to the stack. For faint sources close to the
  detection limit, the combined likelihoods decrease if an observation with
  low likelihood is added to an observation with high likelihood. To quantify
  the effect, 54 long observations with common properties (full-frame mode,
  $\geq$99\,\% chip area usable for serendipitous science, clean exposure time
  above 75\,ks in all instruments) were selected. They were divided into two
  parts to construct artificial stacks. The longer exposure has a fixed
  length, while the shorter one is increased in uniform time steps. Four
  setups are chosen. The first combines a long sub-exposure that covers 50\,\%
  of the total effective exposure time and a short sub-exposure that covers
  5\,\%, 10\,\%, 15\,\%, \dots of it. The second combines a 65\,\% part and
  multiples of 2.5\,\% exposure time, the third an 80\,\% part and multiples
  of 2\,\%, and the fourth a 90\,\% part and multiples of 1\,\% exposure
  time. For the resulting more than 1\,800 pairs of a long and a short
  exposure, stacked source detection is run to compare the results to single
  detection on the longer alone.

  Figure~\ref{fig:pseudomostwo} shows how the detection likelihoods and source
  parameters depend on the exposure time ratios between short and long part
  (see Table~\ref{tab:columns} for the definitions of the stacked source
  parameters). In general, the detection likelihood and thus the number of
  sources increase with exposure time, while the statistical errors on the
  source parameters decrease. For two sub-exposures with an exposure time
  ratio of at least about 40\,\%, more and fainter sources are reliably
  detected in the stacks than in the individual sub-exposures. For lower
  exposure time ratios, the median detection likelihood and the number of
  sources above the detection limit decrease for purely statistical reasons,
  because more degrees of freedom of the fit enter Eq.~\ref{eq:eml_ltot}. The
  limiting exposure time ratio above which the total detection likelihood
  increases with respect to the single detection depends on the
  signal-to-noise ratio and on the detection likelihood itself. The dependence
  can be estimated by a simplified simulation using the \texttt{eboxdetect}
  definition of detection likelihoods given in Eqs.~\ref{eq:ebox_li} and
  \ref{eq:ebox_ltot}. For a fixed number of counts in the long observation
  with 15 images, the equivalent detection likelihood is calculated and
  compared to the combined likelihood of this long and a short
  observation. Counts are assumed to scale linearly with exposure time and to
  be the same in each of the fifteen images of an observation, while in real
  observations, counts depend on energy band and instrument
  characteristics. The source counts among the chosen total counts are derived
  for which the detection likelihood in the long observation $L_\mathrm{long}$
  equals the likelihood in the stack $L_\mathrm{stack}$. Equal detection
  likelihoods $L_\mathrm{stack}=L_\mathrm{long}$ are shown in
  Fig.~\ref{fig:detlimits} for different numbers of counts as a function of
  the exposure time ratio.  Sources whose likelihood in the long observation
  lies above the curve are recovered in the stack with a higher detection
  likelihood. Sources below the curve have a lower likelihood in the stack and
  may be lost if they fall below the detection limit of six (dotted horizontal
  line). The effect is less prominent for the \texttt{emldetect} likelihoods
  which are based on $C$ statistic but still depend on the number of degrees
  of freedom of the fit. The simulation confirms the empirical finding that
  higher detection sensitivity is reached for exposure-time ratios above
  0.35$-$0.60, depending on the count number.

  The stacked catalogue thus includes all sources which reach the minimum
  detection likelihood in at least one observation (dark blue dots in the
  uppermost panel of Fig.~\ref{fig:pseudomostwo}) or in total. This approach
  preserves strongly variable sources. It is possible, however, that some of
  the additional sources with total detection likelihood below the threshold
  of six are spurious. A simple filtering expression may be applied to the
  source list to extract sources with total detection likelihood above six
  only.

\section{Field selection for the catalogue}
\label{sec:observations}

  The catalogue of sources in overlapping observations is based on the data
  used to compile 3XMM-DR7 and its selection criteria: Per observation, each
  EPIC exposure enters 3XMM if it has a minimum net exposure time of 1\,ks,
  which is the sum of good-time intervals after filtering the event list, and
  non-empty images in all five energy bands. This first release of a stacked
  catalogue comprises good-quality observations on which additional
  requirements regarding observational setup and usability are imposed. These
  are introduced in the following sub-sections.

  \subsection{Determining continuously high background}
  \label{sec:highbkg}

\begin{figure}
    \centering
    \includegraphics[width=44mm]{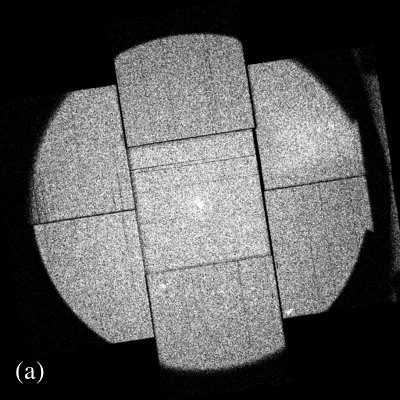}\hskip2pt%
    \includegraphics[width=44mm]{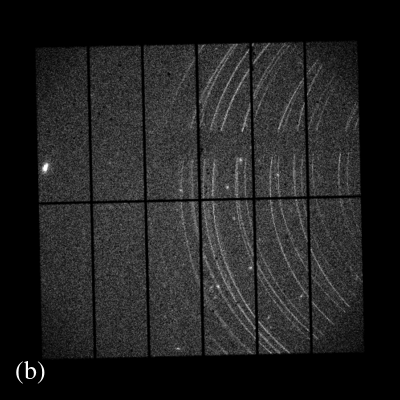}\vskip1pt
    \includegraphics[width=44mm]{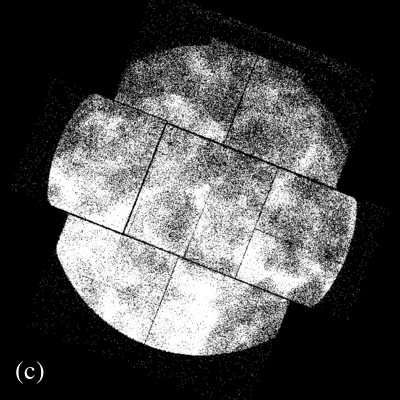}\hskip2pt%
    \includegraphics[width=44mm]{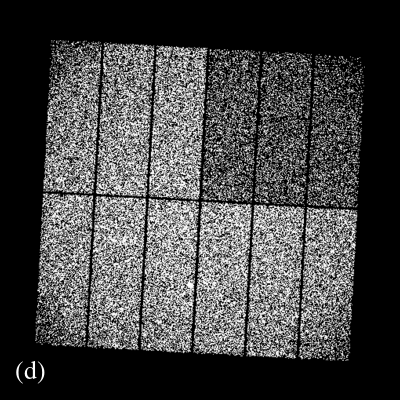}
    \caption{Examples of different types of increased background intensity in
      EPIC observations which have been assigned a Cauchy probability above
      the limit of $87\,\%$ and do not have a HIGH\_BACKGROUND warning flag in
      3XMM-DR7: \emph{(a)} continuously high background, exceeded by few
      sources only (obs.\ id.\ 0200171401 MOS1), \emph{(b)} single reflection
      patterns, caused by a bright X-ray source outside the field of view, but
      close to it (obs.\ id.\ 0604820101 pn), \emph{(c)} extended diffuse
      emission (obs.\ id.\ 0650220201 MOS2), \emph{(d)} different brightness
      levels of the EPIC-pn quadrants owing to continuous counting mode
      (obs.\ id.\ 0406752601). The images are created with a linear brightness
      scale ranging from zero to half their exposure time in kiloseconds.}
    \label{fig:bkgimages}
  \end{figure}

  Observations with very high particle-induced background need to be
  identified before performing source detection for the stacked catalogue
  since their low signal-to-noise can lower the overall detection likelihoods
  of sources in the field and cause loss of sources. For the 3rd generation of
  the Serendipitous Source Catalogues 3XMM, an optimised flare filtering
  technique was introduced, described in Sect.\ 3.2.3 of
  \citetalads[Paper\,]{2016A&A...590A...1R}. The count-rate threshold of the
  background light curve above which time intervals are rejected is
  automatically determined from its signal-to-noise ratio. This method
  efficiently excludes intervals of high flaring background which are shorter
  than the total exposure, but is less capable of identifying images with
  persistently high background or features not resolved by source detection
  and thus regarded as part of the background, examples of which are given in
  Fig.~\ref{fig:bkgimages}. We employ a new standardised approach to determine
  the mean background level of an observation from broad-band background
  images and use it to find remaining high background emission after applying
  the good-time intervals from the 3XMM flare filtering. The method is
  described in Appendix~\ref{sec:hbmethod} and applied to all 3XMM-DR7
  exposures taken in full frame, extended full frame, or large window mode to
  establish a high-background cut. For each instrument, probabilities are
  derived from their median background rate per unit area that measure the
  background level of the full observation. From trial runs of source
  detection on combinations of high- and low-background fields, we choose a
  probability threshold of 87\,\% to exclude observations from the
  pre-selection for the stacked catalogue, reducing the risk of loss of
  detections because of background contamination. Using this cut, the majority
  of the observations flagged by the DR7 screeners are also discarded by the
  automatic procedure and 537 additional observations (overlapping or not) are
  newly defined as affected by high-background, like the examples shown in
  Fig.~\ref{fig:bkgimages}.

  \subsection{Selection criteria and grouping of observations}
  \label{sec:stacks}

  Observations are selected for the first stacked catalogue if they fulfil the
  following criteria (the number of the 9\,710 DR7 observations remaining
  after each filtering step given in brackets):

  \begin{enumerate}

    \item All three EPIC instruments were active (8\,022) and

    \item each EPIC instrument was operated in full-frame mode, including
      Extended Full-Frame Mode for EPIC-pn (6\,937).

    \item At least 99\,\% of the chip area are usable according to a
      classification of OBS\_CLASS$\leq2$ in 3XMM-DR7 (4\,741).

    \item The mean background level of each instrument (pn: quadrant) lies
      below the threshold defined in Sect.~\ref{sec:highbkg} (4\,370).

    \item The observation overlaps with another one by at least 20\,\% in
      area, approximated as an angular separation of up to 20\arcmin\ between
      the aim points (2\,207).

  \end{enumerate}

  OBS\_CLASSes indicate the fraction of the usable chip area and are adopted
  from 3XMM-DR7 without further revision. The assignment of an OBS\_CLASS
  depends on a combination of automatic flagging, manual flagging, and
  background properties within a partly subjective screening process. By using
  a maximum OBS\_CLASS of two, we are aiming at excluding complex background
  structures and large extended objects, which are not the main interest of
  serendipitous source detection. The fractional area may be slightly
  different for similar observations of the same field, possibly resulting in
  different OBS\_CLASSes.

  The resulting list of stacks includes three well-studied survey fields that
  cannot simply be supplied to \texttt{edetect\_stack} as a black box, namely
  M31 and the extra-galactic surveys XXL North and South. Numerous source
  candidates in the bright core of M31 and the large extent of the XXL surveys
  prevent them from being processed within a reasonable runtime on standard
  PCs, which were employed to compile the catalogue. Observations of the M31
  core are thus manually de-selected, and 28 observations of its outer parts
  remain in the catalogue. The large associations comprising the XXL surveys
  are composed of more than a hundred members each and are completely
  discarded.

  All adjacent overlapping observations are sorted into one group or
  `stack'. The final sample includes 1\,789 observations in 434 stacks, the
  majority of them having two or three members. The number of observations per
  stack size is given in Table~\ref{tab:stacks}.

\section{Catalogue construction and properties}
\label{sec:catalogue}

  \subsection{Organisation of the catalogue}
  \label{sec:columns}

  For each of the observation groups described in
  Sect.~\ref{sec:observations}, stacked source detection is run using the new
  task \texttt{edetect\_stack}.  The stacked catalogue is constructed from the
  unique source lists of the 434 stacks and comprises 71\,951 sources. It
  lists the parameters from the combined fit for each source and, in addition,
  one row for each observation that was involved in this fit. All source
  parameters are directly derived from the results of the simultaneous fit to
  all observations in a stack. Values per observation refer to the subset of
  images taken during this observation. The catalogue can be reduced to the
  one-source-one-row layout of the 3XMM slim source catalogues using a
  selection expression on the identifier columns given below, such as
  N\_CONTRIB. Its columns are mostly organised in the style of the 3XMM
  catalogues with the same definitions of their values wherever applicable and
  fully listed in Table~\ref{tab:columns} of the Appendix. In this section, we
  describe the most relevant parameters, modifications to the 3XMM column
  definitions, and newly introduced columns.

  \emph{Source identifier.} The unique source identifier SRCID in the stacked
  catalogue is a 16-digit number, composed of a preceding `3', linking it to
  the convention of the 3XMM catalogues that the detection identifier of
  individual detections starts with a `1' and the source identifier of unique
  matches between them starts with a `2', followed by the lowest OBS\_ID of
  the contributing observations (10 digits), and the identifier within the
  \texttt{emldetect} source list (5 digits), for example 3020624020100030 for
  the thirtieth detection in a stack with 0206240201 being the lowest
  identifier of all the observations for which the detection was in the field
  of view. The five-digits identifiers are not continuous, because the
  temporary \texttt{emldetect} source list comprises all input detections, and
  only the significant ones among them are transferred to the final source
  list.

  Each source is attributed an IAU name of the form
  3XMMs\,J\emph{hhmmss.s}$\pm$\emph{ddmmss}, including the truncated
  sexagesimal right ascension and declination of the source. It is given in
  the column IAUNAME.

  \emph{Observations included.} N\_OBS gives the total number of observations
  per stack and N\_CONTRIB the number of contributing observations for which
  the source position is inside the field of view. Both column values are set
  to null (undefined) in the observation-specific rows and can thus be used to
  select the summary rows per source.

  \begin{table}
  \caption{Stacks from which the new catalogue is compiled. \emph{(a)} Number
    of observations per stack. \emph{(b)} Number of stacks.}
  \label{tab:stacks}
  \centering
  \begin{tabular}{rr@{\qquad\qquad}rr@{\qquad\qquad}rr}
  \hline\hline\noalign{\smallskip}
  (a) & (b) &    (a) & (b) &    (a) & (b) \\
  \hline\noalign{\smallskip}
   2 &  269 &     11 &   2 &     22 &   1 \\
   3 &   74 &     12 &   3 &     23 &   1 \\
   4 &   16 &     13 &   1 &     24 &   2 \\
   5 &   14 &     15 &   2 &     25 &   2 \\
   6 &   15 &     16 &   1 &     28 &   1 \\
   7 &    4 &     18 &   2 &     32 &   1 \\
   8 &    5 &     19 &   4 &     49 &   1 \\
   9 &    4 &     20 &   1 &     52 &   1 \\
  10 &    4 &     21 &   2 &     66 &   1 \\
  \hline
  \end{tabular}
  \end{table}

  \emph{Source coordinates.} The position of the source is considered to be
  the same in all contributing observations and images in the simultaneous
  fit, while the source counts are determined separately per image (see
  Sect.~\ref{sec:astrometry} for a discussion of the astrometric accuracy). It
  is given in equatorial, galactic and image coordinate systems in the RA,
  DEC, LII, BII, and X\_IMA, Y\_IMA columns. Image coordinates refer to the
  common coordinate system of each stack (Sect.~\ref{sec:srcdet}) and are
  listed together with their individual errors $\sigma_\textrm{X\_IMA}$,
  $\sigma_\textrm{Y\_IMA}$. The combined position error RADEC\_ERR is
  calculated from them as
  $(\sigma_\textrm{X\_IMA}^2+\sigma_\textrm{Y\_IMA}^2)^{0.5}$, converted to
  arcseconds. For symmetric errors in both dimensions, RADEC\_ERR/$\sqrt{2}$
  is the one-dimensional $1\sigma$ position error, giving the interval that
  includes 68\,\% of normally distributed coordinate values.
  $\sqrt{2.3/2}\times$RADEC\_ERR is the two-dimensional error, giving the
  radius of a circularised ellipse that includes 68\,\% of normally
  distributed pairs of coordinates.

  \emph{Equivalent detection likelihoods.} Maximum detection likelihoods are
  determined per input image, summed, and converted from the total number of
  degrees of freedom to the mathematical equivalent of a two-parameter fit
  (see Sect.~\ref{sec:srcdet}). The number of degrees of freedom is two for
  point sources and three for extended sources plus the number of images
  involved in the fit (equalling the number of instruments in each
  observation, for which the mask is valid at the source position, times the
  number of energy bands) and varies from source to source. The decision
  whether a detection enters the final source list is based on the equivalent
  likelihoods. Sections~\ref{sec:processing} and \ref{sec:detqual} describe
  how a large number of input images can affect them and thus the source
  selection in the fitting process. Sources with a minimum equivalent
  likelihood of six in the whole stack or at least one contributing
  observation are included in the stacked catalogue.

  \emph{Source flux.} The fitted count rate per image is converted to flux
  using the energy conversion factors (ECFs) of
  \citetalads[Paper\,][]{2016A&A...590A...1R}.  All-EPIC fluxes are means of
  the fluxes per instrument and observation weighted by their inverse squared
  errors. They are null with undefined flux errors but non-zero count errors
  for an observation if no counts are found within the PSF area of a
  source. The ECFs depend on the instrument, the observing mode, and the
  filter used, and on the spectral shape of the source. Therefore, the
  combined fluxes merging different instruments and setups across the
  observations are affected by cross-calibration uncertainties
  \citepads[see][]{2009A&A...496..879M}. The underlying spectral model of the
  3XMM ECFs is an absorbed power law with a column density of $3\times
  10^{20}\,\textrm{cm}^{-2}$ and a photon index of 1.7.

  \emph{Source extent.} The radial extent and extent likelihood of a source
  are fitted simultaneously in all observations. The $\beta$ model used to
  parameterise the extent is described in Sect.~4.4.4 of
  \citetalads[Paper\,][]{2009A&A...493..339W}. Sources with an extent radius
  below 6\arcsec\ or an extent likelihood below four cannot be resolved and
  are considered point-like. Their extent is set to zero and their extent
  likelihood to null.

  \emph{Mask fraction.} The PSF-weighted detector coverage of a source is
  given for each instrument separately. It is the fraction of the point spread
  function, for extended sources convolved with the $\beta$ extent model,
  falling on valid detector pixels. For one observation, it is conservatively
  defined as the minimum mask fraction of the five energy bands, indicating
  the most restrictive mask. The stacked mask fraction is the largest value of
  the contributing observations, indicating the best one.

  \emph{Source flags.}  A modified version of \texttt{dpssflag}, the task also
  in use for the 3XMM catalogues, is employed for an automated quality
  flagging to warn the user about complexities in the environment of the
  source that might affect the significance of the detection or the source
  parameters and their accuracy. The sources are not visually
  screened. Strings of nine booleans indicate different potential issues of a
  detection in total and for each instrument, described in Sect.~7.3 of
  \citetalads[Paper\,][]{2009A&A...493..339W}. A true EPIC flag means a
  warning for at least one instrument. The nine booleans are converted to a
  single integer summary flag STACK\_FLAG. Sources with a flag value of `0'
  come without any warning. Flag `1' indicates reduced detection quality in at
  least one instrument and observation: low detector coverage or a source
  position close to another source or to bad detector pixels. The list of
  known bad pixels is hard-coded within \texttt{dpssflag}. `2' is attributed
  to potentially spurious sources, for example those found within the PSF
  radius of another source. Flag `3' in the summary row indicates that the
  source has received flag 2 in all contributing observations. The integer
  flags are not directly comparable to the 3XMM SUM\_FLAGs, which have been
  set for individual observations and include additional information from
  visual screening.

  \begin{table}
    \caption{Overview of the catalogue of unique sources in spatially
      overlapping XMM-Newton observations, selected from the 3XMM-DR7
      observations taken between 2000 February 3 and 2016 December 15.}
    \label{tab:catalogue}
    \centering
    \begin{tabular}{l@{~}r}
      \hline\hline\noalign{\smallskip}
      Description                                      & Number        \\
      \hline\noalign{\smallskip}
      Number of stacks                                 &    434        \\
      Number of observations                           &   1\,789      \\
      Time span first to last observation              & Feb 20, 2000  \\
                             \strut\hfill $-$          & Apr 02, 2016  \\
      Approximate sky coverage                         & 150 sq.\ deg. \\
      Approximate multiply observed sky area           & 100 sq.\ deg. \\
      Total number of sources                          &  71\,951      \\
      Sources with several contributing observations   &  57\,665      \\
      Sources with one contributing observation        &  14\,286      \\
      Sources with flag `0' or `1'                     &  69\,526      \\
      Total detection likelihood of at least six       &  64\,221      \\
      Total detection likelihood of at least ten       &  53\,492      \\
      Extended sources (radius $\geq 6\arcsec$)        &   3\,346      \\
      Point sources with VAR\_PROB$\leq$1\,\%          &   5\,607      \\
      Point sources with VAR\_PROB$\leq$10$^{-5}$      &   1\,927       \\
      \hline
    \end{tabular}
  \end{table}

  \emph{Long-term variability between observations.} Three new sets of
  parameters inform about the inter-observation variability of a source, based
  on typical EPIC count numbers in the Gaussian regime: \emph{(i)} the
  $\chi^2$ of the long-term flux changes and the associated probability that
  they are consistent with the flux measurements of a non-variable object,
  \emph{(ii)} the ratio between maximum and minimum flux with its
  1$\sigma$-error, and \emph{(iii)} the maximum flux variation in terms of
  sigma. They are directly derived from the EPIC fluxes and flux errors in all
  contributing observations and in each energy band, resulting in six columns
  per quantity.
  \begin{equation}
    \textrm{VAR\_CHI2} = \frac{1}{n-1}\sum_{k=1}^n
    \left(\frac{F_k-F_\textrm{EPIC}}{\sigma_k}\right)^2
  \end{equation}
  is a reduced $\chi^2$ of flux variability between the mean all-EPIC flux
  $F_\textrm{EPIC}$ over all observations and the individual fluxes $F_k$
  derived for each observation, $k$ running from 1 to number $n$ of
  observations. The associated VAR\_PROB describes the probability that the
  observed flux values are consistent with constant source flux over all
  observations. It is the cumulative chi-square probability
  \begin{equation}
    \textrm{VAR\_PROB} = \int_{\chi^2}^\infty
    \frac{x^{\nu/2-1}e^{-x/2}}{2^{\nu/2}\Gamma(\nu/2)} dx
  \end{equation}
  to reach at least VAR\_CHI2=$\chi^2$ at $\nu=n-1$ degrees of
  freedom. $\Gamma$ denotes the gamma function. A low VAR\_PROB thus indicates
  a high chance that the source shows inter-observation variability.
  \begin{equation}
    \textrm{FRATIO} = F_\textrm{max}/F_\textrm{min}
  \end{equation}
  gives the ratio between the highest and the lowest flux recorded across the
  observations, and
  \begin{equation}
    \textrm{FRATIO\_ERR} =
    \left(\frac{\sigma_{F\textrm{min}}^2}{F_\textrm{min}^2}+
    \frac{\sigma_{F\textrm{max}}^2}%
         {F_\textrm{max}^2}\right)^{0.5}\,\frac{F_\textrm{max}}{F_\textrm{min}}
  \end{equation}
  its 1$\sigma$ error.
  \begin{equation}
    \textrm{FLUXVAR} = \max_{k,l\in[1,n]}\frac{|F_k-F_l|}{\sqrt{\sigma_k^2+\sigma_l^2}}
  \end{equation}
  is the largest difference between pairs of fluxes in terms of sigma, with
  $k$ and $l$ running from 1 to number $n$ of observations.

  \emph{Observation characteristics.} Each row per observation includes the
  modified Julian dates of its start and end time, the filter, the instrument
  mode, and the mean position angle of the spacecraft. In the summary row, the
  beginning of the first and the end of the last contributing observation are
  given.

  \emph{Columns copied from 3XMM-DR7.} For sources with a counterpart in the
  3XMM-DR7 catalogue of sources, information on position, quality flag, and
  intra-observation variability of the 3XMM-DR7 source are copied to the
  summary rows of the stacked catalogue. The observation-specific rows list
  the parameters of the 3XMM-DR7 detection that contributes to the unique
  source, if one is found. Column DIST\_\mbox{3XMMDR7} gives the distance
  between the stacked detection and the 3XMM-DR7 counterpart. More details on
  the matching can be found in Sect.~\ref{sec:3xmmdr7}.

  \subsection{General characteristics}
  \label{sec:generalchar}

  \begin{figure}
    \centering
    \includegraphics[width=79mm]{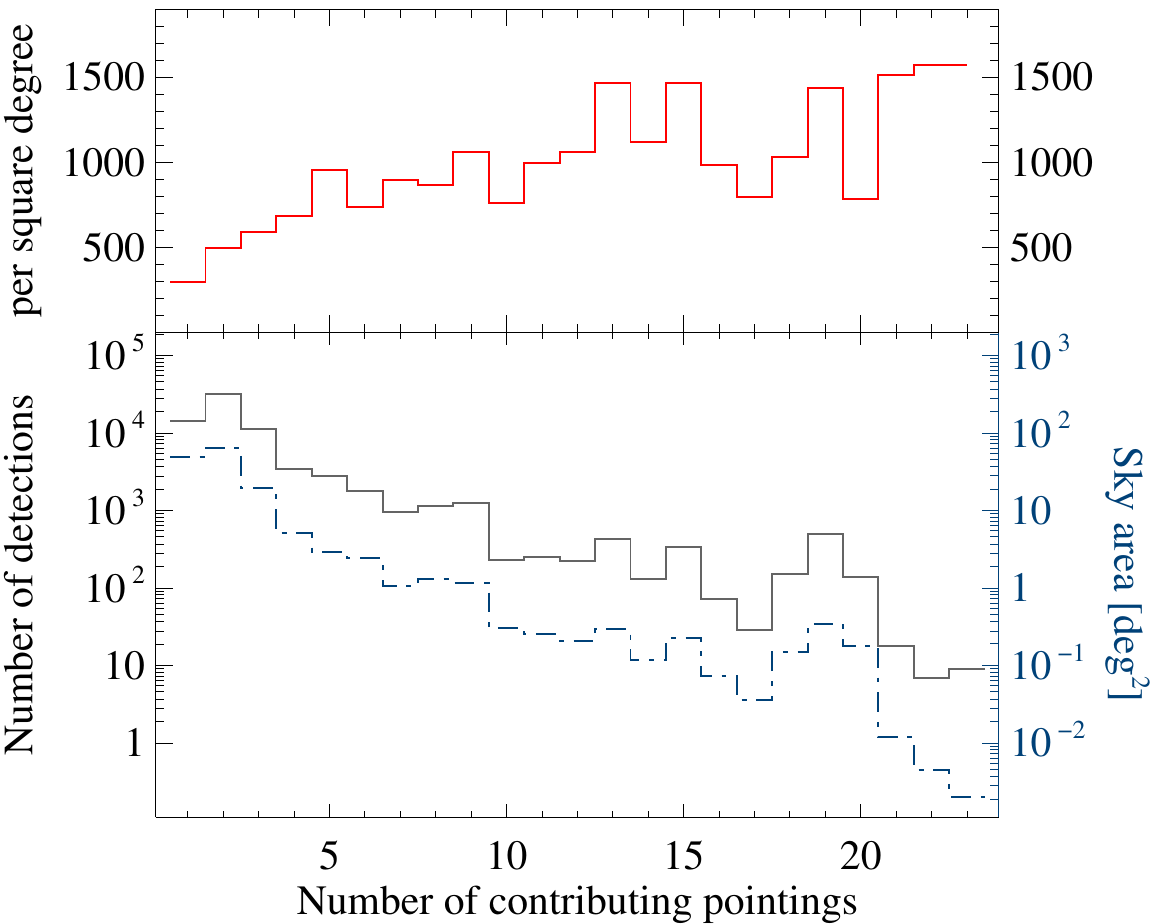}
    \caption{Number of detections (grey solid), detections per square degree
      (red solid), and approximate sky coverage in square degrees (blue
      dash-dotted) per number of contributing observations.}
    \label{fig:ncontrib}
  \end{figure}

  \begin{figure}
    \centering
    \includegraphics[width=72mm]{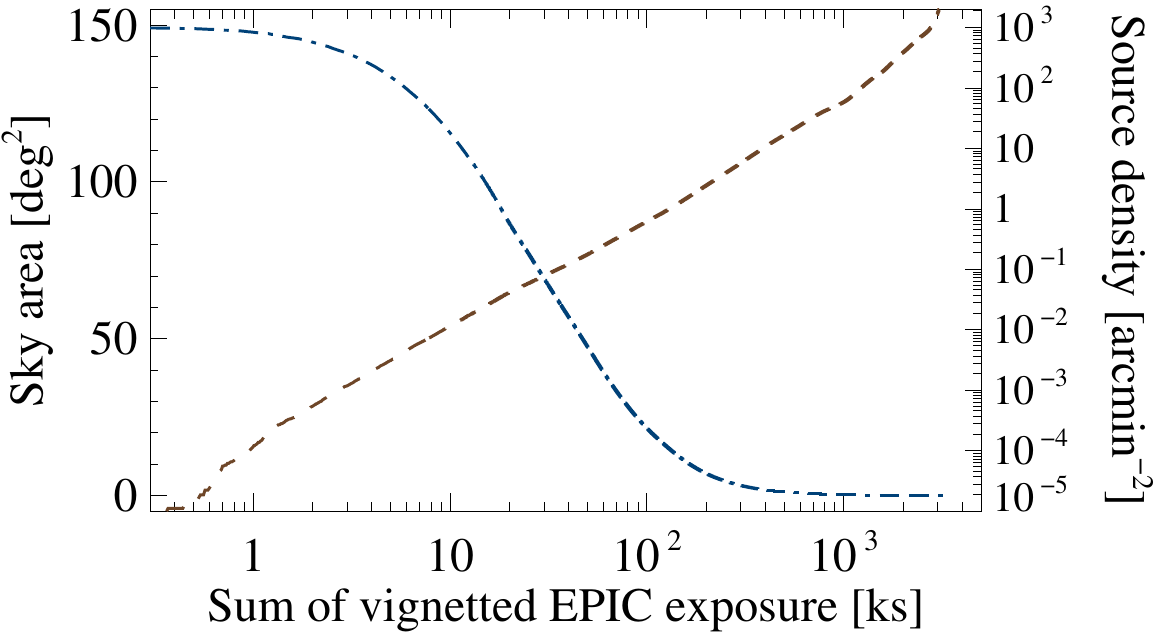}
    \caption{Sky area (blue dash-dotted) and source density (brown dashed) as
      an approximate measure of detection sensitivity over the total EPIC
      exposure times of the catalogue stacks. The plot shows the vignetted
      exposure time averaged over the five energy bands and summed for the
      three EPIC instruments pn, MOS1, and MOS2.}
    \label{fig:exparea}
  \end{figure}

  \begin{figure}
    \centering
    \includegraphics[width=88mm]{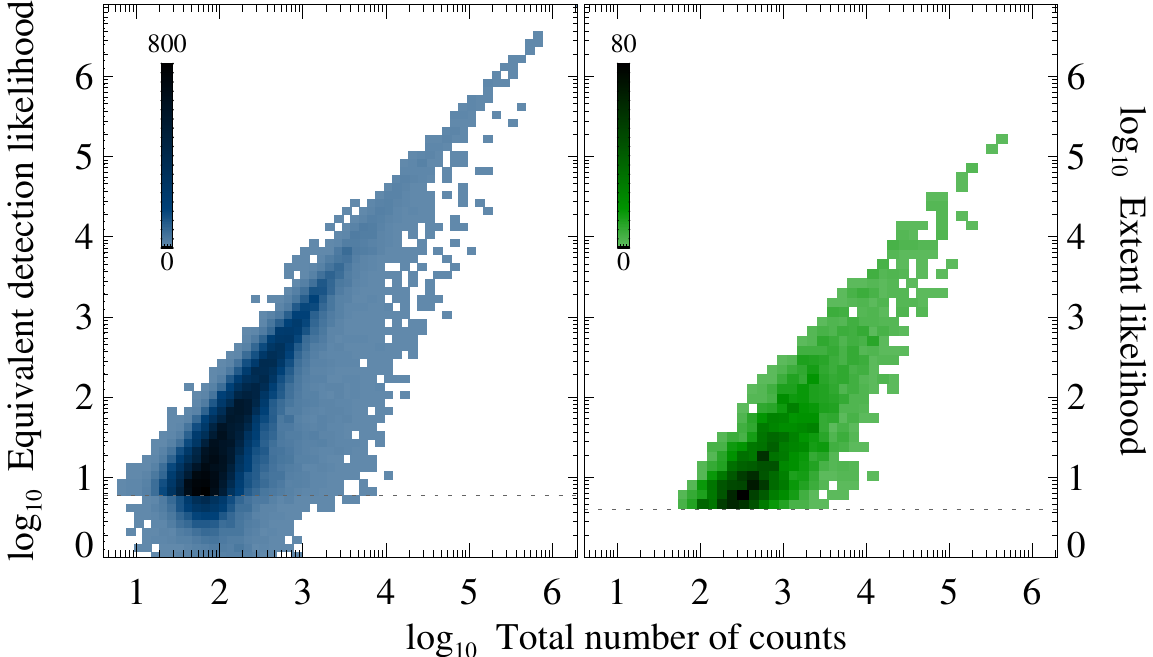}\vskip2pt
    \includegraphics[width=88mm]{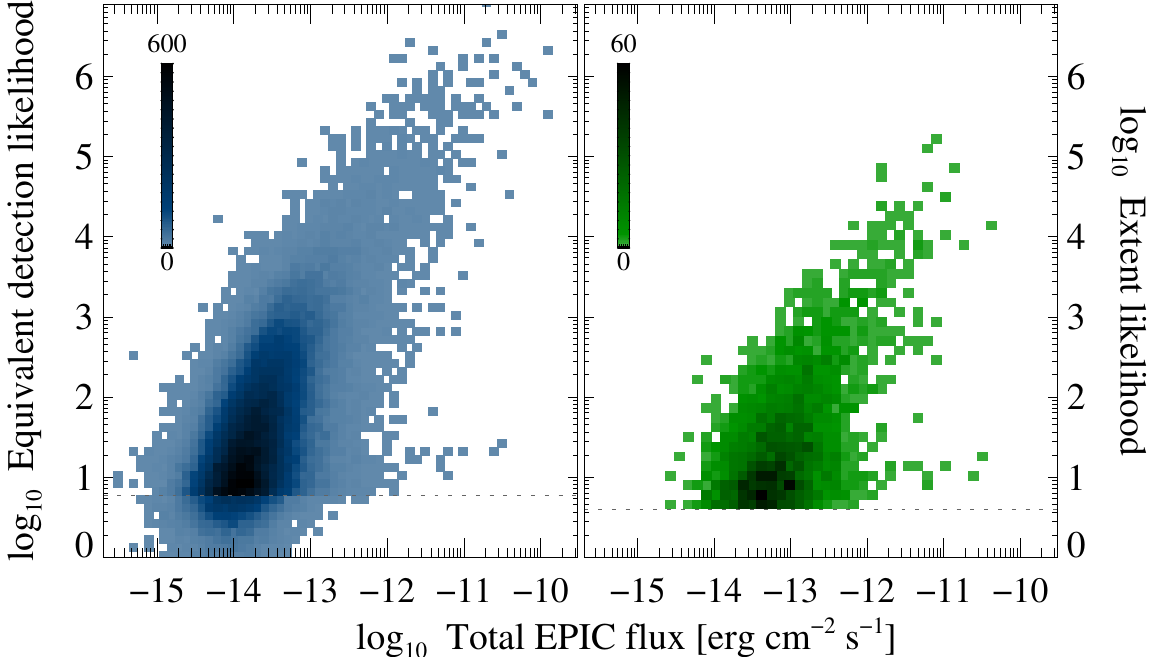}
    \caption{Relation of stacked detection and extent likelihoods to source
      flux and counts in 3XMM-DR7s. Dotted lines mark the lower limits:
      minimum detection likelihood to include a source in the source list and
      minimum extent likelihood per fit to consider a source extended. Colour
      density scales with the source number per plotting bin.}
    \label{fig:likes}
  \end{figure}

  \begin{figure}
    \centering
    \includegraphics[width=72mm]{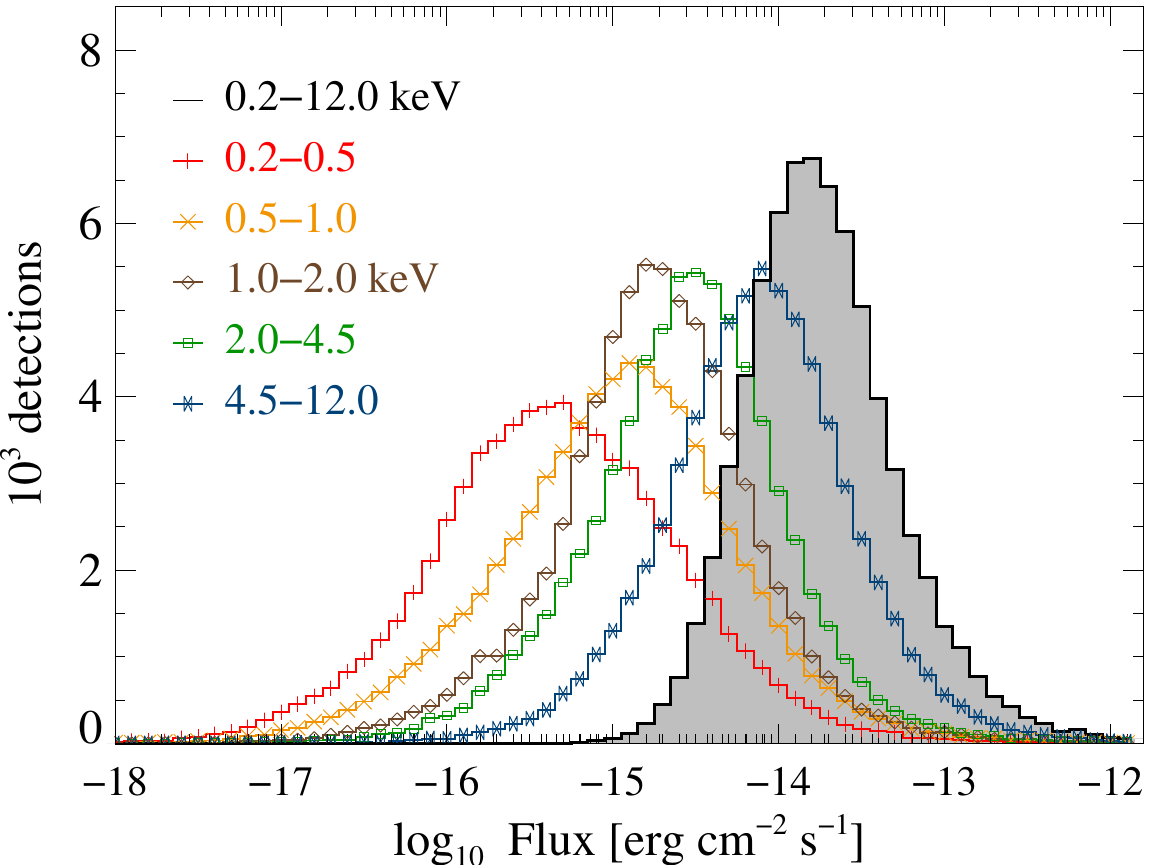}
    \caption{Flux distribution in the catalogue of sources from overlapping
      observations, in total (grey, filled) and for each of the five energy
      bands (energy increases from left to right).}
    \label{fig:fluxes}
  \end{figure}

  \begin{figure}
    \centering
    \includegraphics[width=88mm]{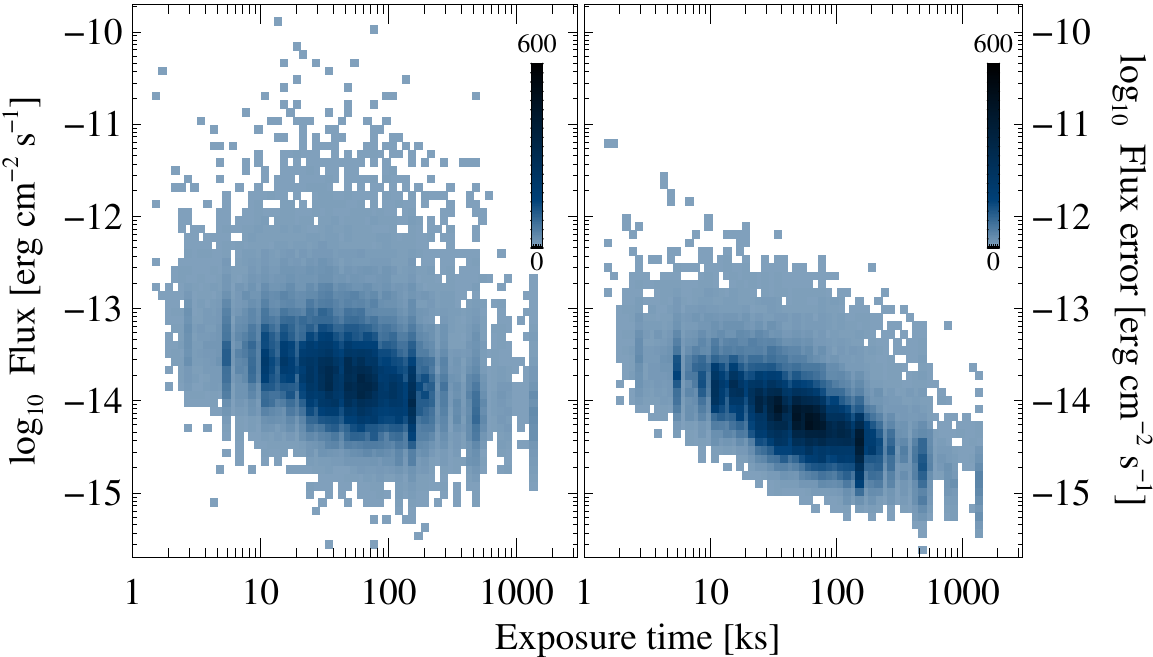}
    \caption{Relation between exposure time EP\_ONTIME and all-EPIC fluxes
      (left panel) and flux errors (right panel) in 3XMM-DR7s. All-EPIC fluxes
      are the weighted means of the fluxes in the individual energy bands. The
      exposure time is summed over all contributing observations. Colour
      density scales with the source number per plotting bin.}
    \label{fig:ontimesscatter}
  \end{figure}

  The 71\,951 unique sources in the stacked catalogue are detected in 1\,789
  observations in 434 stacks, covering more than sixteen years of observations
  in total. The longest time span for a single source is 14.5\,years. 96.6\,\%
  of the sources have been assigned a good automatic quality flag of 0 or 1,
  and 74.3\,\% are detected with a total likelihood of at least ten; a
  somewhat smaller share than in the 3XMM-DR7 catalogue of unique sources
  (80\,\%), where the detection likelihood of repeatedly observed sources is
  given as the highest per-observation likelihood, while the total likelihood
  in the stacked catalogue is calculated using Eq.~\ref{eq:eml_ltot}. 57\,665
  of the sources are covered by more than one observation with a maximum of 23
  visits of a source, and 14\,286 were observed once. An overview of the
  catalogue properties is given in Table~\ref{tab:catalogue}. Since most of
  the stacks comprise two observations, the majority of sources has been
  detected twice (Fig.~\ref{fig:ncontrib}). The absolute number of catalogue
  sources and covered sky area decrease with increasing stack size because few
  large stacks are included in the catalogue. The relative source density per
  unit sky area increases with the stack size thanks to the long total
  exposure (Fig.~\ref{fig:exparea}). The figures include the sources from
  non-overlapping chip areas with one contributing observation.

  With the longer total effective exposure time of the stacks compared to
  individual observations, more counts are collected per source. Hence, the
  sources are measured with higher detection likelihoods than in single
  observations, extended sources additionally with higher extent likelihood,
  and more sources are detected. The likelihood distributions in the stacked
  catalogue over total exposure time per source are shown in
  Fig.~\ref{fig:likes}. In its left panels, the effect of the modified
  likelihood cut becomes obvious. While a hard cut of six has been applied to
  the other 3XMM catalogues, 7,730 sources with a total equivalent detection
  likelihood below six are present in the stacked catalogue: They exceed the
  threshold in at least one contributing observation, not in the whole
  stack. A hard cut of four is applied to the extent likelihood,
  simultaneously determined from all contributing observations.

  The distribution of source fluxes in the stacked catalogue -- in total and
  per energy band -- is shown in Fig.~\ref{fig:fluxes}.  It is similar to the
  distributions determined from the other 3XMM catalogues, in agreement with
  the expectation that the fluxes derived by stacked source detection are
  consistent with those derived from the individual observations, but better
  constrained.

  Almost 4.7\,\% of the catalogue sources are resolved as extended with a core
  radius of the $\beta$-profile extent model of at least 6\arcsec. In general,
  the characterisation of extended sources is affected by larger uncertainties
  than that of point sources: their intensity profile is less sharp, imposing
  larger position errors on extended sources, and the beta function is only an
  approximation to the true extent profile, imposing uncertainties on the
  measured extent radius, which is a free parameter of the fit. For short
  observations and faint extended sources, the measured extent relates to the
  exposure time if insufficient counts are collected to describe them
  reliably. In stacked source detection, the source extent can now be fitted
  simultaneously in all observations irrespective of their individual exposure
  time, making use of the total counts. While uncertainties remain, for
  example owing to deviations from the true extent profile of a source, the
  extent parameters can be determined more precisely, and the risk of fitting
  background fluctuations by spurious extended sources is lower. The
  experiments with artificial stacks (Sect.~\ref{sec:artstack}) confirm that
  extended sources are detected more reliably even if observations of
  different durations are combined. The high percentage of sources with
  quality flag 0 or 1 among all extended sources, similar to the one among the
  point sources, also indicates reasonably low spurious content. Still, large
  position errors and quality flags 2 and 3 should be taken as signs that an
  extended detection is uncertain.

  \subsection{Accuracy of the source parameters}
  \label{sec:simul}

  \begin{figure}
    \centering
    \includegraphics[width=63mm]{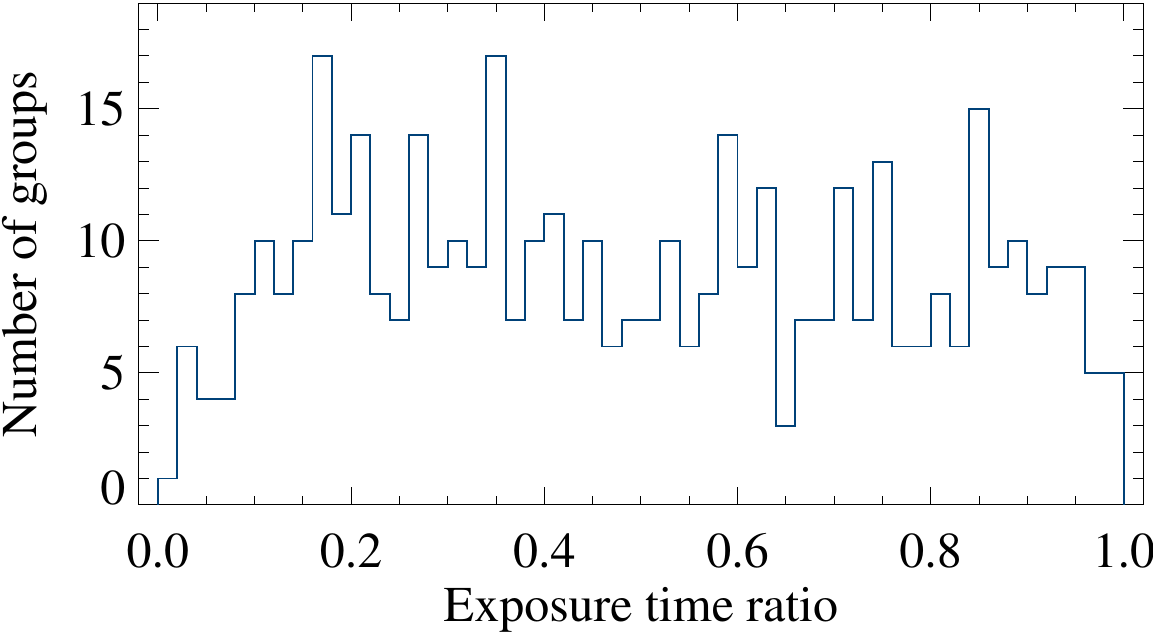}
    \caption{Distribution of exposure time ratios for the 269 stacks
      comprising two observations.}
    \label{fig:expratio}
  \end{figure}

  Owing to the larger exposure time and count number of the stacked
  observations compared to single observations, stacked source detection
  becomes more sensitive to faint sources, and the flux errors decrease
  significantly with exposure time, confirmed by the larger number of
  catalogue sources having low flux and small flux errors at longer EP\_ONTIME
  (Fig.~\ref{fig:ontimesscatter}). The dependence of parameter accuracy on the
  exposure time, shown on the example of the flux errors in the right panel of
  Fig.~\ref{fig:ontimesscatter}, applies to all error columns in the
  catalogue. The smaller errors reflect the smaller scatter of possible
  parameter values and higher fit accuracy in the stacked source detection.
  XMM-Newton source detection employs the C statistic in the
  maximum-likelihood analysis, which is distributed as $\chi^2$ plus an
  additive term proportional to $n^{-0.5}$
  \citepads{1976A&A....52..307C,1979ApJ...228..939C}, negligible for large
  count numbers $n$. The one-dimensional 1$\sigma$ error on a parameter is
  derived by stepping the parameter until $C=C_\textrm{min}+1$ is reached,
  corresponding to the 68\,\% accuracy level of a $\chi^2$ statistic. The
  confidence limits of parameters derived from images with few photons in the
  source-fitting area and of highly coupled parameters may be actually larger
  than those for $C_\textrm{min}+1$, and an additional error component might
  thus be considered when interpreting the statistical errors on the stacked
  parameters, for instance regarding fluxes of sources close to the detection
  limit or position matches in a cross-correlation with other catalogues. For
  the position error, an estimate is derived in Sect.~\ref{sec:astrometry}.

\begin{figure}
    \centering
    \includegraphics[height=35mm]{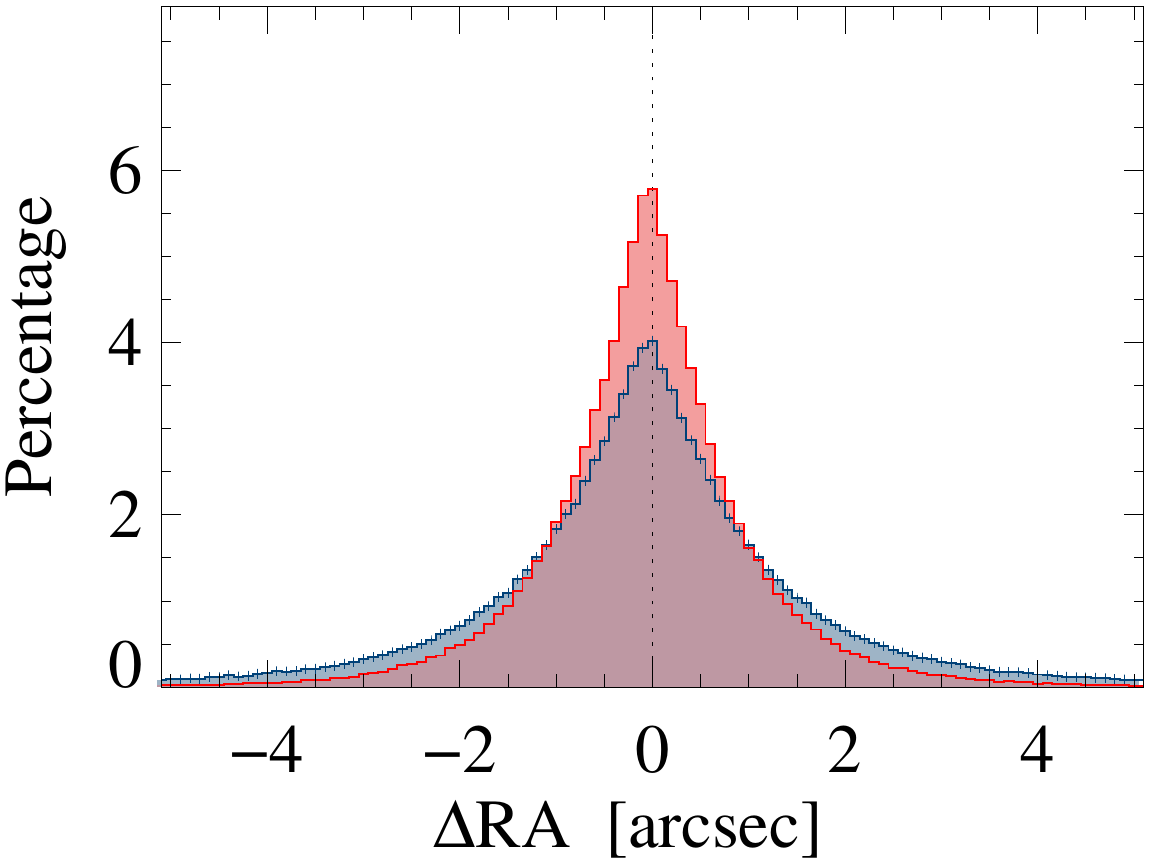}%
    \includegraphics[height=35mm]{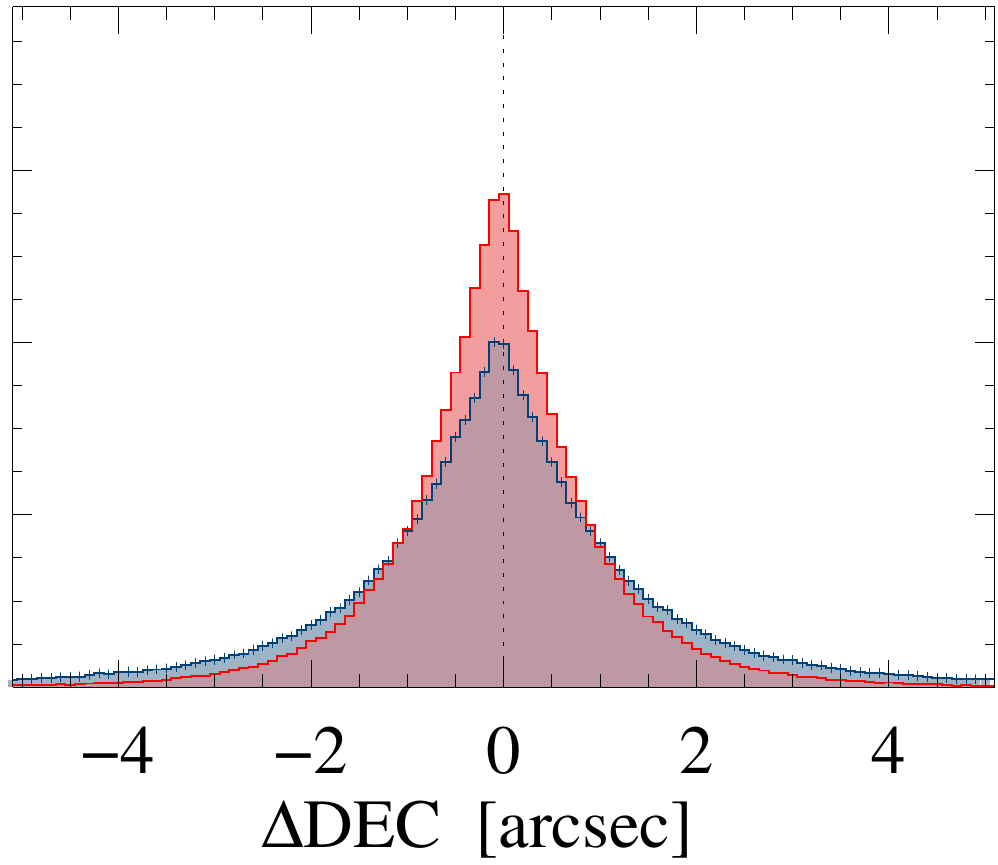}\vskip2pt
    \includegraphics[height=35mm]{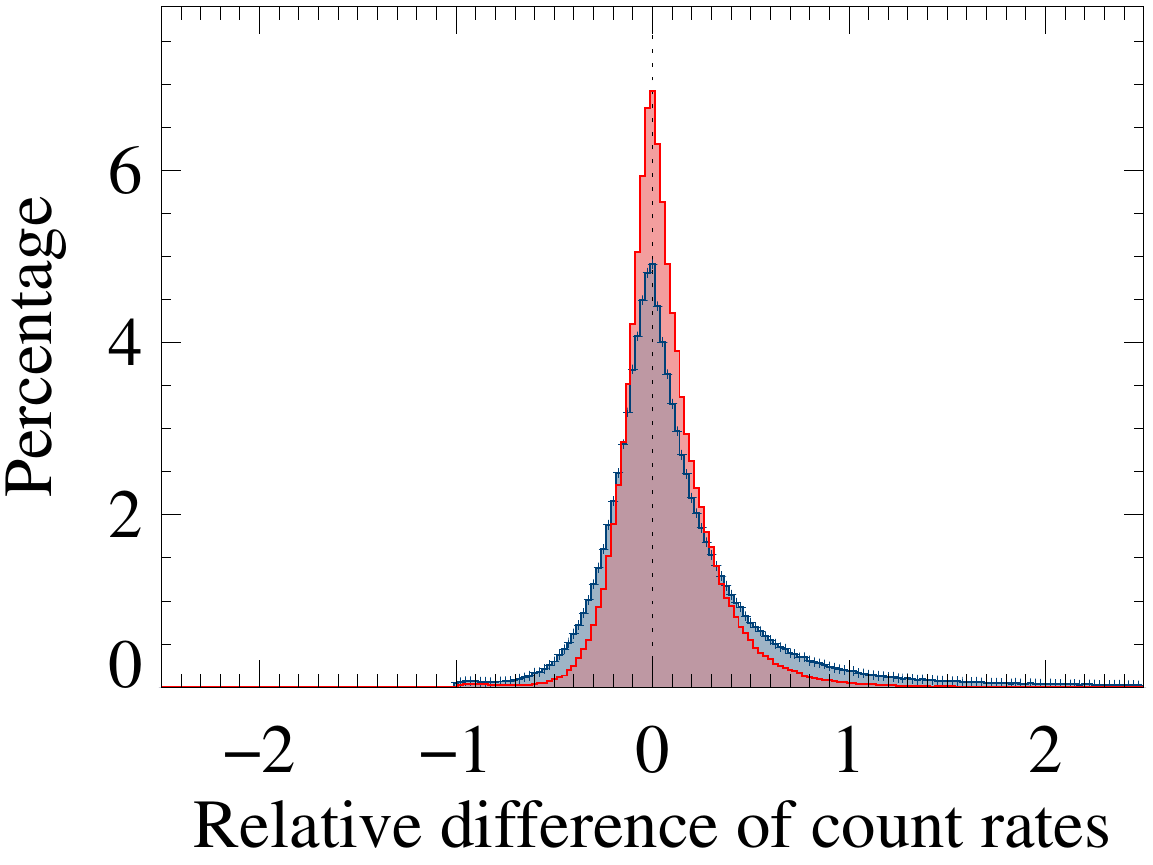}%
    \includegraphics[height=35mm]{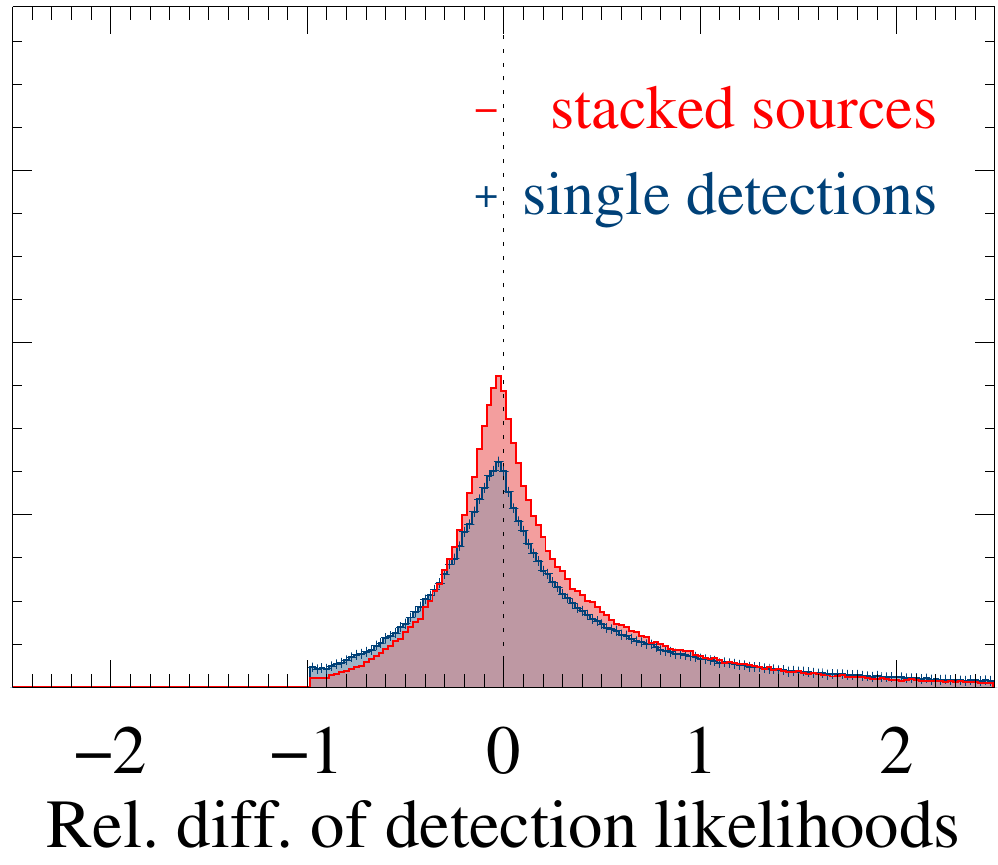}
    \caption{Accuracy of the source parameters of point sources from stacked
      (red) and non-stacked (blue) source detection, derived from simulated
      images of stacks comprising two observations. The coordinates in the
      upper panels are absolute offsets: results of source detection on
      simulated images minus input value. Count rates and equivalent detection
      likelihoods in the lower panels are relative differences: simulations
      minus input divided by the input value.}
    \label{fig:simulation}
  \end{figure}

  As demonstrated in Section~\ref{sec:artstack}, the number of detections in
  two-observation stacks increases reliably compared to a single observation
  for exposure time ratios of more than about 0.4 if not taking the
  likelihoods during the individual observations into account. The
  distribution of exposure time ratios of these stacks is shown in
  Fig.~\ref{fig:expratio}. In order to investigate the accuracy of the stacked
  source parameters quantitatively, the code was applied to simulated images
  and the results compared to the input parameter values. We start from the
  modelled source images of catalogue observations, which were created by the
  task \texttt{emldetect} as by-products of our stacked catalogue
  pipeline. These are the sum of the background maps and the PSF models of all
  sources that passed the likelihood cut. To maximise the multiply covered sky
  area, a subset of 108 stacks of two observations with a maximum offset of
  1\arcmin\ between their respective aim points was selected. They comprise a
  total of 10\,925 catalogue sources. For each of their source images, 25
  images were simulated by drawing random values from a Poisson distribution
  around the input brightness of the source image pixel by pixel. On the
  resulting $108\times 25=2700$ simulated stacks and 5400 observations, source
  detection was performed. The new source parameters derived from the
  simulations were compared to the input values on a per-stack and a
  per-observation basis. The distributions of the offsets from the input
  values are shown in Fig.~\ref{fig:simulation} for the free fit parameters
  coordinates and count rate and for the total equivalent likelihood. They are
  neatly centred at zero, confirming that the true values are reproduced, and
  are narrower for stacked source detection than for the individual
  simulations, confirming that the stacked source parameters have a higher
  precision and accuracy.

  \subsection{Performance of stacked compared to non-stacked source detection}
  \label{sec:detqual}

  To quantify the improvement of the detection sensitivity of stacks over
  individual observations within consistently designed data sets and
  source-detection runs, source detection has been performed separately on
  each catalogue observation, using the same method and parameters as applied
  to the stacks of observations. The 126\,658 individual detections were
  matched into a joint list of 71\,921 tentative unique sources within a
  matching radius of 15\arcsec. We compare the stacked sources first with the
  individual detections and then with the joint sources, again using a radius
  of 15\arcsec. The joint source lists are expected to deviate from 3XMM-DR7
  due to the different background models and image
  creation. Section~\ref{sec:3xmmdr7} includes a comparison with 3XMM-DR7.

\begin{figure}
    \centering
    \includegraphics[height=35mm]{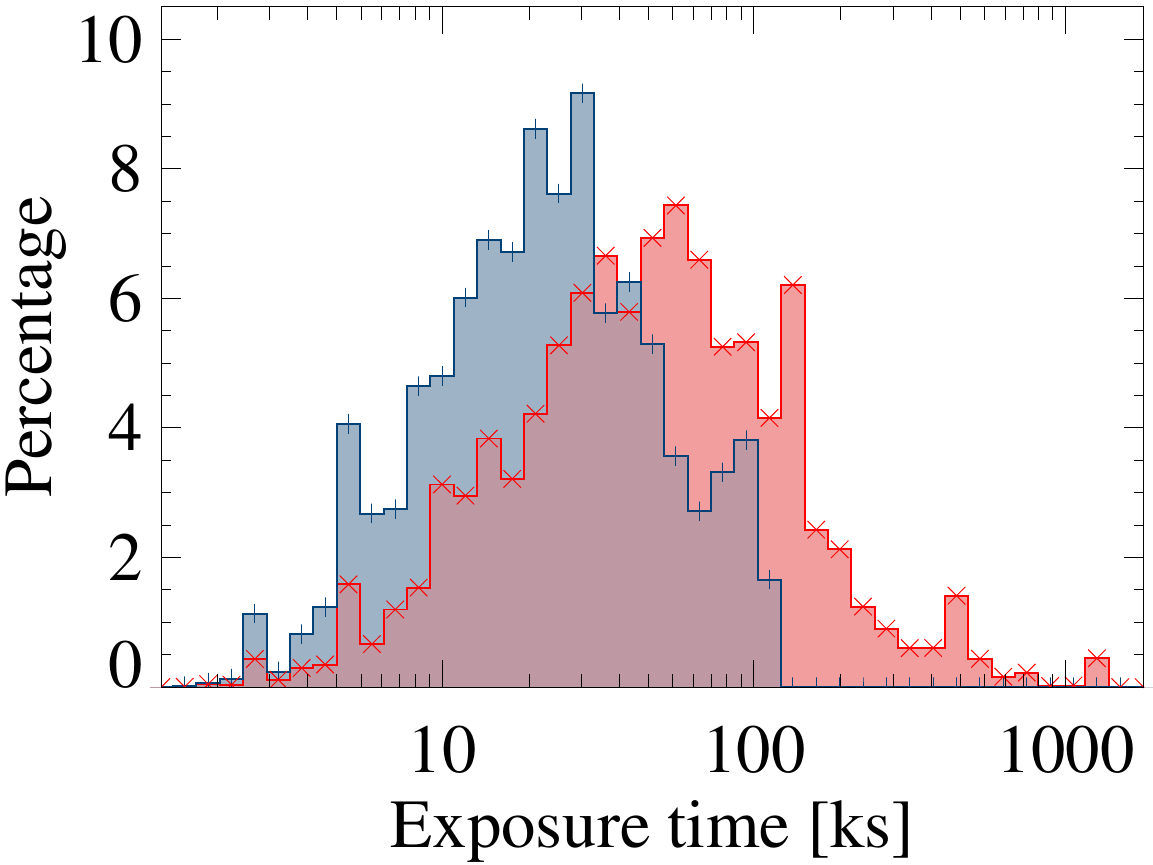}%
    \includegraphics[height=35mm]{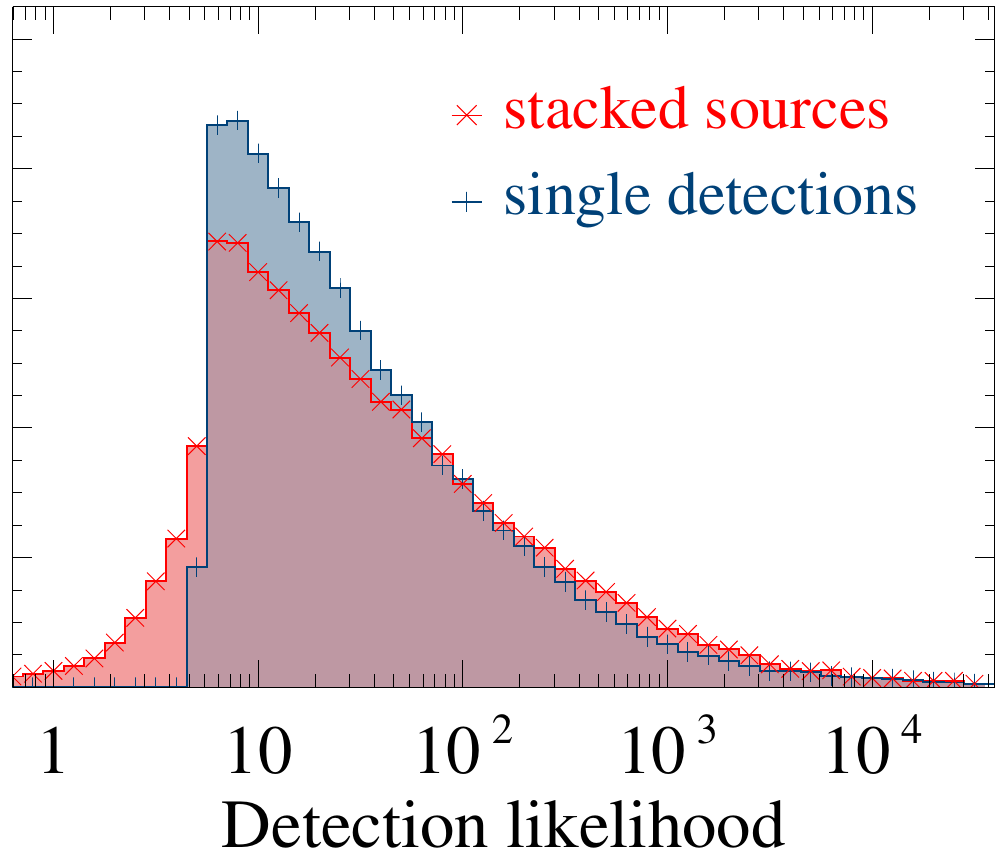}\vskip2pt
    \includegraphics[height=35mm]{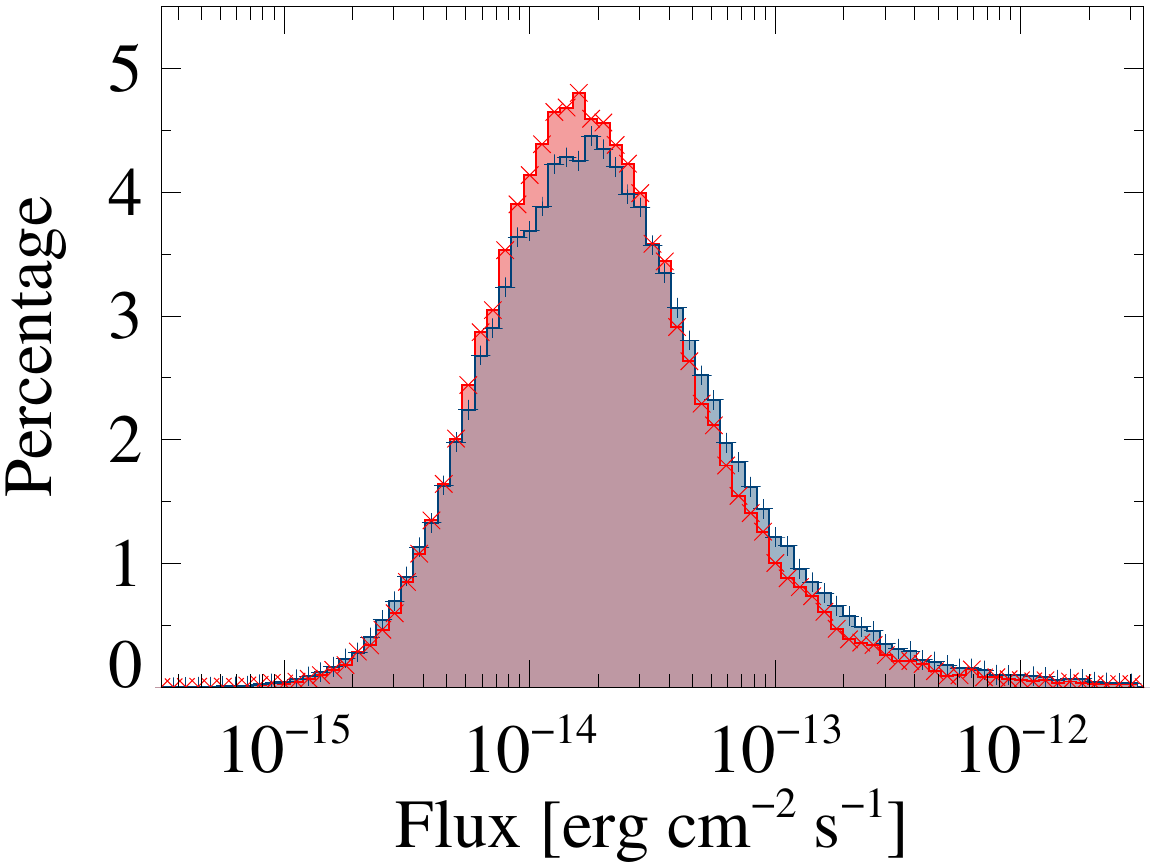}%
    \includegraphics[height=35mm]{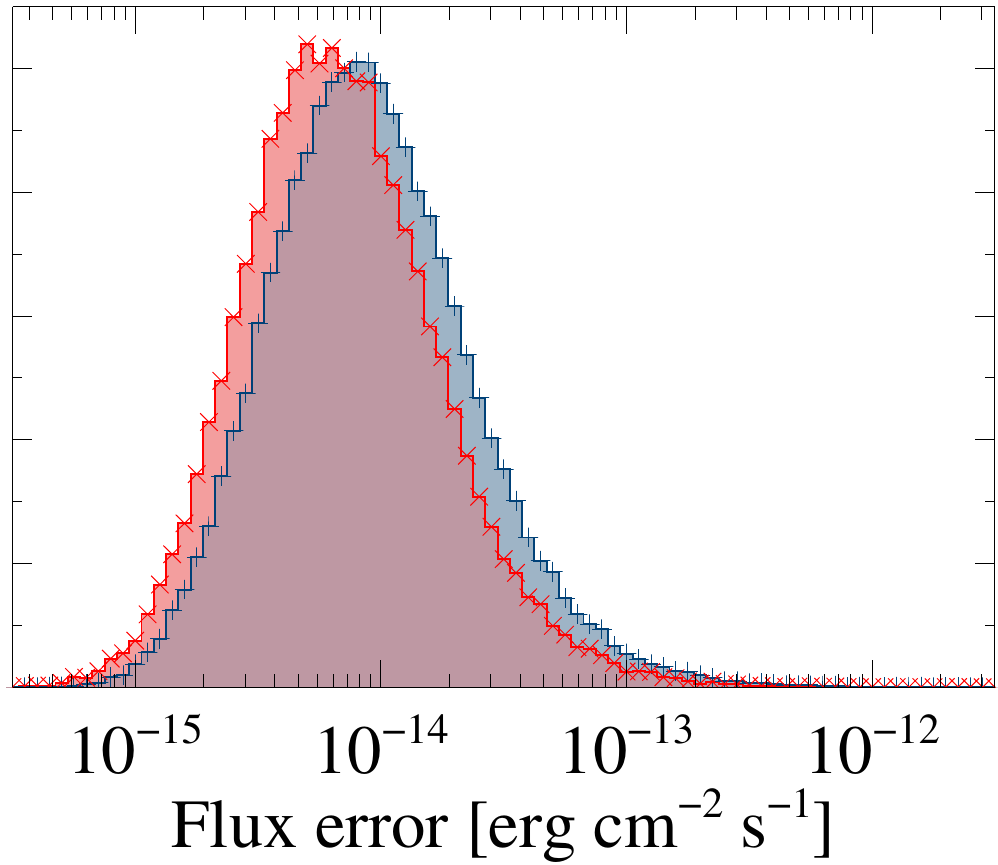}
    \caption{Normalised distribution of all-EPIC good time intervals,
      detection likelihoods, fluxes, and flux errors of the sources from
      stacked source detection (red) and of the individual detections in the
      source lists per single observation (blue).}
    \label{fig:compsingle}
  \end{figure}

  Figure~\ref{fig:compsingle} shows distributions of four main source
  parameters of the stacked catalogue and the individual detections, all
  normalised to their total number. The longer effective exposure times and
  smaller flux errors of the stacked catalogue with respect to all detections
  from the individual observations are clearly visible. The stacked detection
  likelihoods tend to be higher than that of the individual detections, but
  include small values for sources that are significant in only one
  contributing observation. Fluxes are expected to be consistent. Differences
  in their distributions may indicate a larger share of low-flux sources in
  the stacked catalogue and better sensitivity to faint sources.

  To quantify potential gain and loss of sources in stacks compared to
  individual observations, the detections that are not recovered by stacked
  source detection are investigated. 4\,931 are found in the single runs
  only. The vast majority -- over 98\,\% -- are detected in one observation
  with low likelihood without a potential second detection within
  15\arcsec\ although located in overlap areas. About 10\,\% may be subject to
  source confusion, overlapping with neighbouring detections within
  30\arcsec. A large fraction of 40\,\% of the not recovered `single-only'
  detections are extended, 416 even with an extent radius of more than
  1\arcmin. They have large positional uncertainties which may affect the
  matching, and a high chance to be spurious detections.

  For the comparison between the stacked catalogue and the joint source lists,
  the positions of the merged sources are defined as the mean positions of the
  contributing single detections and their extent as the maximum extent among
  them. 4\,347 sources are found by stacked source detection only, meaning
  that they have no counterpart in the joint source list within a
  15\arcsec\ radius. Most of them are located in areas covered by several
  observations. Only 15.7\,\% of them are extended, 121 with an extent radius
  larger than 1\arcmin. The point-like stack-only sources tend to have higher
  detection likelihoods and slightly better constrained fluxes than point-like
  single-only detections. Together with the experiment described in
  Sect.~\ref{sec:artstack} and Figs.~\ref{fig:pseudomosaics} and
  \ref{fig:CDFSmatch}, this clearly indicates that a larger fraction of the
  stack-only than of the single-only sources are reliable detections and that
  the spurious source content is significantly reduced by stacked source
  detection.

\begin{figure}
    \includegraphics[width=62mm]{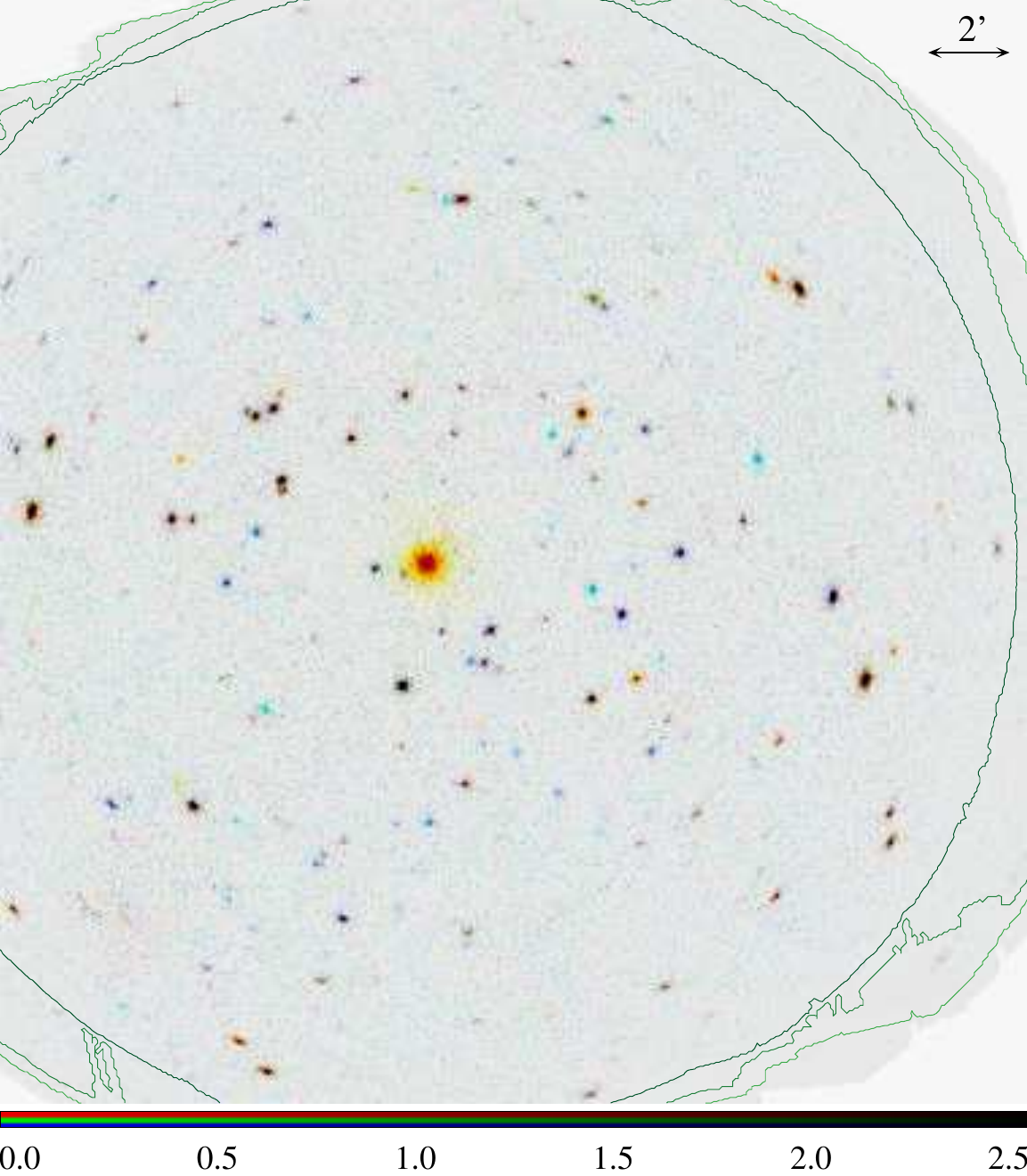}\vskip3pt
    \includegraphics[width=88mm]{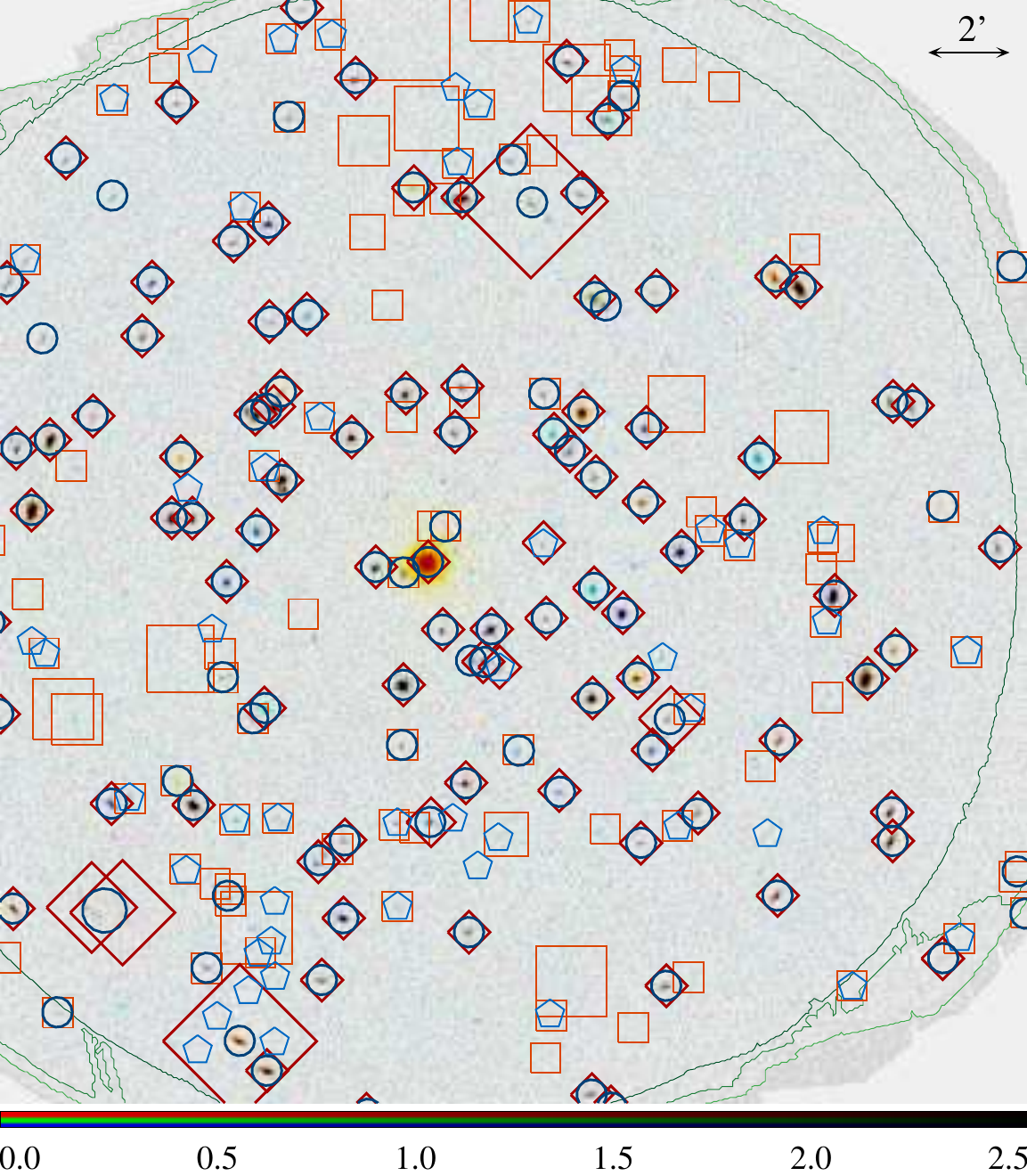}
    \caption{Example with large deviations between stacked and joint source
      list: nineteen observations of \object{HD\,81809}. For clarity, the
      mosaicked image is shown both without and with source
      identifications. \emph{Blue symbols:} Sources detected in the
      stack. Thick circles are used for sources with an equivalent detection
      likelihood above six in total or in at least two observations, thin
      pentagons for the others. \emph{Red symbols:} Joined individual
      detections. Thick diamonds are used for those merged from more than two
      observations, thin squares for the others. The plot symbols have a
      minimum radius of 22\arcsec\ and scale with the source extent if it is
      larger than that. The contours enclose areas within at least two (red),
      seven (orange), and twelve (white) observations overlap.}
    \label{fig:detimaHD}
  \end{figure}

  Figure~\ref{fig:detimaHD} illustrates the differences between stacked and
  non-stacked detections in an example of 19 observations. The images are
  background-subtracted, normalised by their exposure time per pixel, and
  combined into a mosaic for display purposes. Plot symbols indicate the
  significance of the detection, the number of contributing observations, and
  the source extent. Several joint-only detections are very extended, thus
  most likely spurious, and disappear in the stack. Additional example images
  of stacks comprising two to five observations are shown in
  Fig.~\ref{fig:detexamples}.


  \subsection{Astrometry}
  \label{sec:astrometry}

  \begin{figure}
    \centering
    \includegraphics[width=70mm]{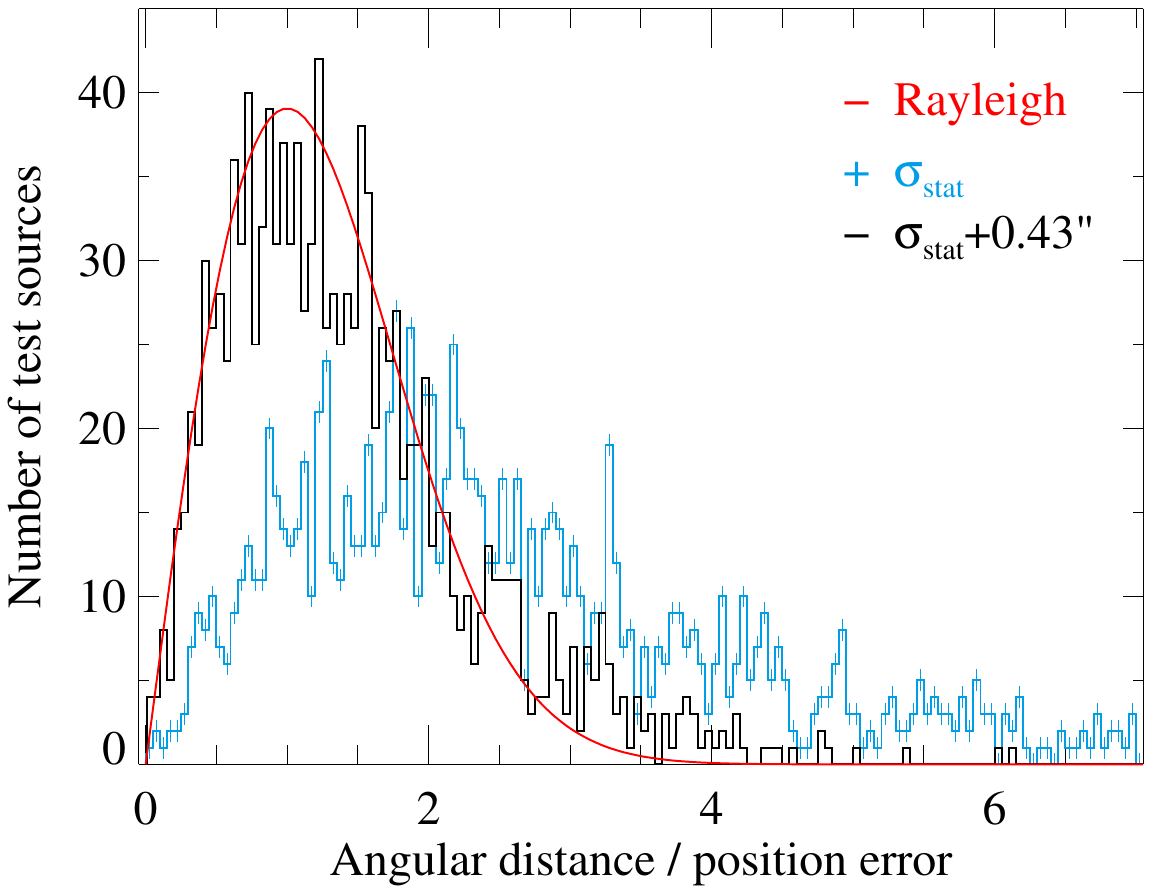}
    \caption{Error-normalised position offsets between sources in the stacked
      catalogue and associated quasars in SDSS-DR12 compared to a Rayleigh
      distribution (red). \emph{Light blue:} Based on the one-dimensional
      purely statistical position errors RADEC\_ERR/$\sqrt{2}$ given in the
      catalogue. \emph{Black:} Using the best-fit additional error component
      0.43\arcsec, linearly added to the statistical error on the X-ray
      position.}
    \label{fig:rayleigh}
  \end{figure}

  The source positions in the stacked catalogue are determined simultaneously
  from all observations using their respective calibration. For the 2XMM and
  3XMM catalogues, the observations are rectified after performing source
  detection by comparing the measured X-ray positions of the brightest sources
  in a field with positions in optical and infra-red catalogues and applying
  the derived coordinate shifts and field rotation to all sources in the
  field. The approach cannot be used for the source lists from which the
  stacked catalogue is compiled, because the different observations per stack
  might be affected by different shifts. New, more detailed PSF models,
  upgrades to the source-detection tasks, and a refined boresight calibration
  have helped to determine the source positions for the 3XMM catalogues more
  precisely than for previous versions even without this field rectification
  \citepalads[see Paper\,][]{2016A&A...590A...1R}. Using them, no additional
  astrometric corrections are applied to the first stacked catalogue. The
  stacked position errors from the joint fit are purely statistical
  uncertainties of the measurements. Systematic uncertainties like the
  inaccuracies of the (positional) cross-calibration of the contributing
  observations are thus not included in the stacked catalogue, but can be
  estimated from the deviations between measured and expected positions of
  point sources with well-defined astrometry.

  For the 2XMM catalogues, the mean additional 1$\sigma$ position error has
  been determined to be about 1\arcsec\ before and 0.35\arcsec\ after
  astrometric correction from a comparison with optical quasar positions in
  the Sloan Digital Sky Survey (SDSS), assuming that the error-normalised
  angular distances are Rayleigh distributed
  \citepalads[Paper\,][]{2009A&A...493..339W}. Following this approach, the
  (uncorrected) X-ray positions of the unique sources of the stacked catalogue
  are matched with the SDSS release DR12 \citepads{2017AJ....154...28B}
  without further restrictions on off-axis angle or quality flags. As for the
  other 3XMM catalogues \citepalads[Paper\,]{2016A&A...590A...1R}, a matching
  radius of 15\arcsec\ is used. The 1\,288 quasars among the best matches are
  selected, and the histogram of their positional offsets $x=\delta/\sigma$ is
  compared with a Rayleigh distribution $xe^{-0.5x^2}$, $\delta$ being the
  angular distance between the positions in SDSS and in 3XMM-DR7s, and
  $\sigma$ the combined circularised one-dimensional position errors, namely
  $(0.5\times(\textrm{errMaj}^2+\textrm{errMin}^2))^{0.5}$ for SDSS and
  RADEC\_ERR/$\sqrt{2}$ for 3XMM-DR7s. An additional error component on the
  X-ray position is varied until best agreement between the measured histogram
  and the Rayleigh distribution is reached. Since the nature of the additional
  error is unknown, the fit is performed for two alternatives, a quadratic sum
  $\sigma=({\sigma_\textrm{stat}^2+\sigma_\textrm{sys}^2})^{0.5}$ and a linear
  $\sigma=\sigma_\textrm{stat}+\sigma_\textrm{sys,lin}$. The best fits are
  achieved with a quadratically added component of
  $\sigma_\textrm{sys}$=0.73\arcsec\ and with a linearly added component of
  $\sigma_\textrm{sys,lin}$=0.43\arcsec, respectively, which can be considered
  the parameter range of the mean systematic error on the stacked source
  positions (not included in the catalogue). Figure~\ref{fig:rayleigh} shows
  the position offsets between stacked sources and SDSS quasars normalised by
  the pure statistical errors, with the linearly added
  0.43\arcsec\ uncertainty on the X-ray positions, and the respective Rayleigh
  distribution.

  For comparison, the same method is applied to the uncorrected positions of
  the individual detections in 3XMM-DR7. Their distribution of offsets from
  associated SDSS quasars is fitted with
  $\sigma_\textrm{sys}$=1.01\arcsec\ and
  $\sigma_\textrm{sys,lin}$=0.59\arcsec. In the 3XMM catalogues, errors on the
  field translation and rotation are determined during the field
  rectification, and their combination is applied as additional error
  component. Its median in DR7, restricted to detections with a quasar
  association, is 0.43\arcsec. Although derived from astrometrically
  uncorrected data, the parameter range of the additional error component for
  the stacked catalogue is far below the pixel size and smaller than for the
  individual DR7 detections in the same sample of observations.

  \subsection{Long-term source variability between observations}
  \label{sec:interobsvar}

  \begin{figure*}
    \centering
    \includegraphics[height=48mm]{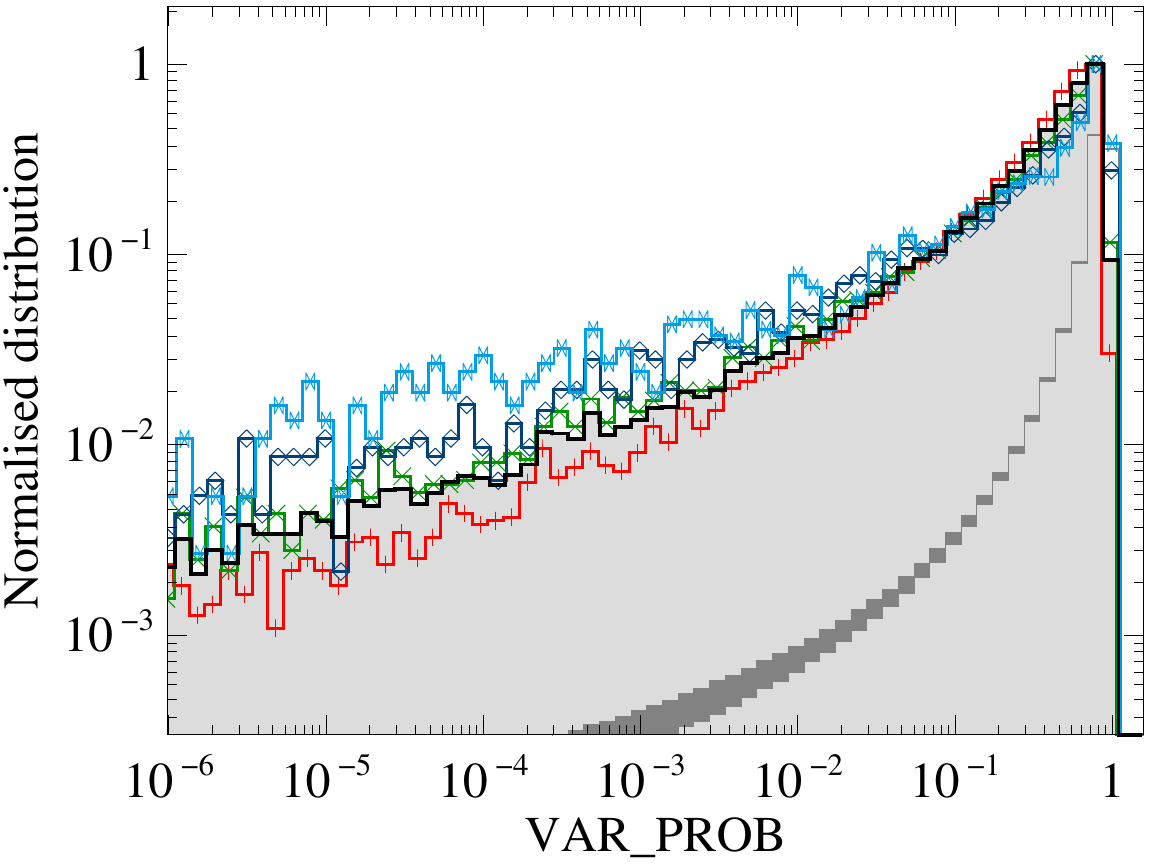}\hskip2mm%
    \includegraphics[height=48mm]{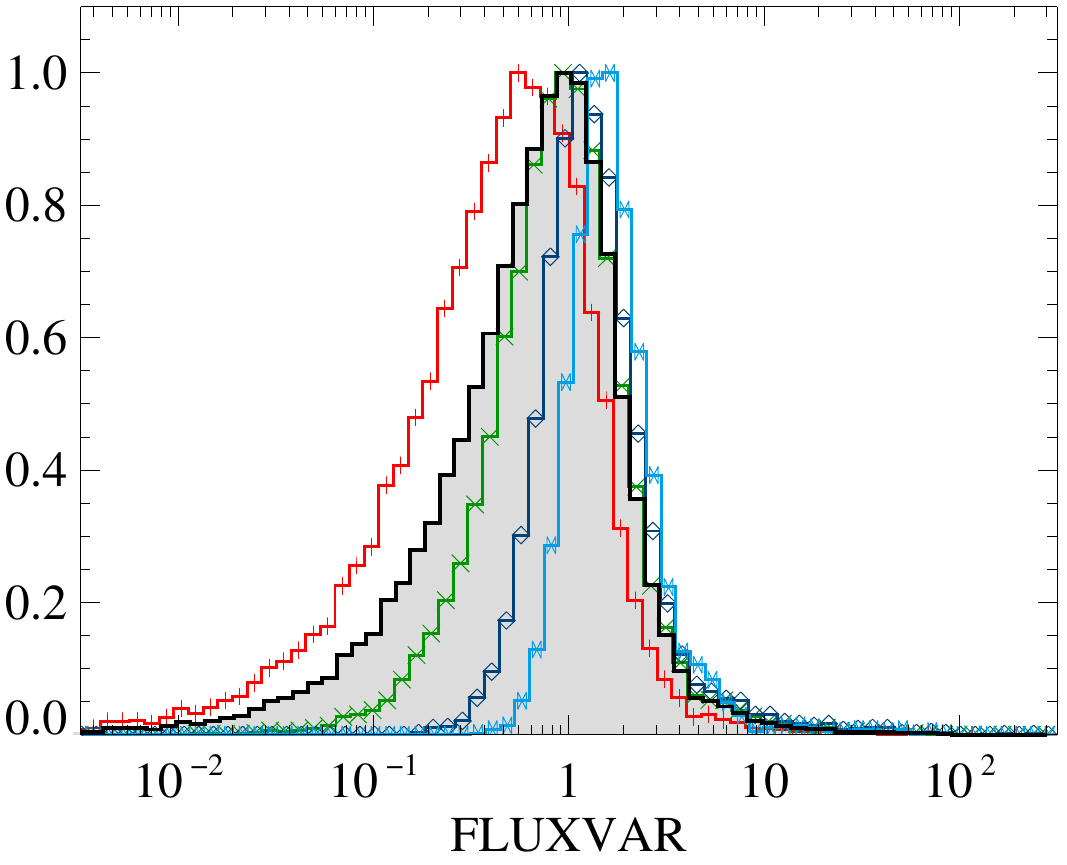}%
    \includegraphics[height=48mm]{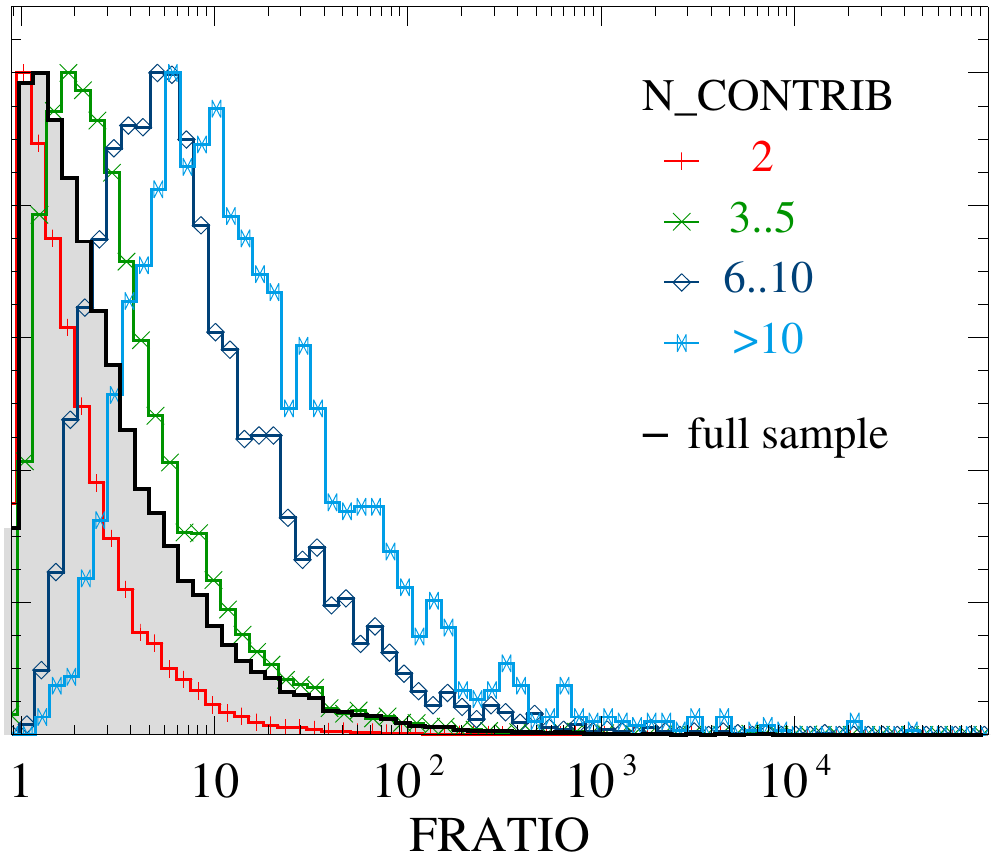}
    \caption{Three of the all-EPIC long-term variability parameters for point
      sources and their different dependence on the number of contributing
      observations. A low value of VAR\_PROB and high values of FLUXVAR and
      FRATIO can indicate a long-term variable source. The dark grey area in
      the left panel indicates the 1$\sigma$ range of the probability estimate
      for constant sources. All histograms are normalised to their maximum.}
    \label{fig:ltv}
  \end{figure*}

  The stacked catalogue can serve as a database for long-term variability of
  serendipitous XMM-Newton sources: Irrespective of the detection probability
  within a single observation, fluxes and flux errors are determined for each
  observation that covers the source of interest without the need to match
  individual detections or to determine upper flux limits, increasing the
  chance to identify transients. Inter-observation variability in XMM-Newton
  data has been explored previously by \citetads{2012ApJ...756...27L} based on
  high signal-to-noise detections in 2XMM-DR3i and through the EXTraS project
  \citepads[Exploring the X-ray Transient and variable
    Sky,][]{2016ASSP...42..291D} based on 3XMM-DR5 and slew observations,
  published as the EXTraS long-term Variability Catalogue
  \citep{2017extras_wp5}. Variability in other missions has been discussed for
  example by \citetads[][Chandra]{2010ApJS..189...37E},
  \citetads[][Swift]{2014ApJS..210....8E}, and
  \citetads[][ROSAT]{2016A&A...588A.103B}.

  For each stacked catalogue source that has been observed at least twice with
  non-zero counts, five quantities describing its inter-observation
  variability are derived from the total flux and the EPIC fluxes of the
  contributing observations (see Sect.~\ref{sec:columns}). Since they are
  based on mean fluxes, they provide information on potential long-term
  variability only and are not probed for intra-observation variability. For
  787 detections, an observation-level EPIC flux has been set to null, because
  no counts were detected during this observation. Null fluxes do not
  contribute to the variability parameters in the present catalogue. Upper
  limits for such cases will be included in future releases.

  The parameters show little dependence on the energy band, with the highest
  values being present in the well-populated bands 2$-$4, but clear dependence
  on the number of contributing observations N\_CONTRIB. VAR\_PROB is least
  dependent on it because it is normalised by the degrees of freedom.
  Distributions of the variability parameters are given in
  Fig.~\ref{fig:ltv}. All histograms peak at higher parameter values for
  larger N\_CONTRIB. This dependence is qualitatively consistent with the
  results of \citet{2017extras_wp5}. They simulate sparsely sampled long-term
  light curves for objects with constant mean fluxes, derive the maximum flux
  variations in terms of sigma, and show their change with the number of
  light-curve points, owing to larger statistical fluctuations for a larger
  number of points. More than half of the repeatedly observed sources in the
  stacked catalogue are covered by only two snapshots. Thus, the distributions
  for low numbers of contributing observations dominate the overall
  result. The catalogue does not include boolean variability flags, since the
  parameter thresholds to consider a source tentatively variable strongly
  relates to the scientific question to be addressed. For example, 5\,607 or
  10.2\,\% of the repeatedly observed point sources in the catalogue have
  VAR\_PROB$\leq$1\,\%. Using a more restrictive probability cut of $10^{-5}$,
  1\,927 or 3.5\,\% point sources can be considered long-term variable. To
  provide a rough estimate of the false-alarm rate among them, we assume
  constant flux for all catalogue sources and randomise the observation-level
  fluxes using Poisson distributed count numbers. This is repeated five
  hundred times, and the resulting distribution of probability values for
  non-variable sources included in Fig.~\ref{fig:ltv}.

  When filtering on high variability, sources with generally unreliable
  variability parameters should be excluded, in particular detections with
  poor quality flags and extended sources. Poorly constrained flux values in
  individual observations and false positives on detector features like bad
  pixels or stray light may also mimic variability. Many of them can be
  identified and removed by applying cuts to the errors on the flux
  ratios. High-proper motion objects and Solar-System bodies cannot be
  uniquely identified by the source-detection process which assumes stable
  source positions in all images. For example, the high-proper motion binary
  \object{61 Cygni} separates into ten individual sources from eighteen
  overlapping observations in the stacked catalogue, recorded at different
  levels of (apparent) variability. Visual inspection of the source images
  which are distributed together with the catalogue (Sects.~\ref{sec:access}
  and \ref{sec:auxprods}) helps to reveal these cases.


  In the 3XMM catalogues, intra-observation variability is investigated for
  all detections with at least 100 counts. We select sources with a
  counterpart in 3XMM-DR7 (see Sect.~\ref{sec:3xmmdr7}) and compare the DR7
  parameters on intra-observation variability with the inter-observation
  variability from this work. Some, but not all of them are expected to be
  identified on all time scales as variable. A long-term variable source may
  be constant over the time span of a single observation, and variability on
  short time scales does not necessarily imply long-term variability of the
  mean fluxes, as for regular periodicity of up to a few hours. Information on
  short-term variability is provided for 11\,172 point-like DR7 counterparts
  to stacked sources. 579 are flagged as short-term variable in at least one
  DR7 observation, and 477 of them have several observations in the stacked
  catalogue. As expected, a considerable number of short-term variable sources
  also show signs of long-term variability: 355 with a probability below 1\,\%
  that the measurements are consistent with constant flux, 282 with a
  probability below $10^{-5}$. Thus, 122 of the sources whose DR7 counterpart
  is flagged as short-term variable are not clearly long-term variable in the
  stacked catalogue. For 29 of them, the DR7 observation that triggered the
  short-term variability flag is not part of the sample selected for the
  stacked catalogue according to the criteria listed in
  Sect.~\ref{sec:observations}.

\begin{figure}
    \centering
    \includegraphics[width=88mm]{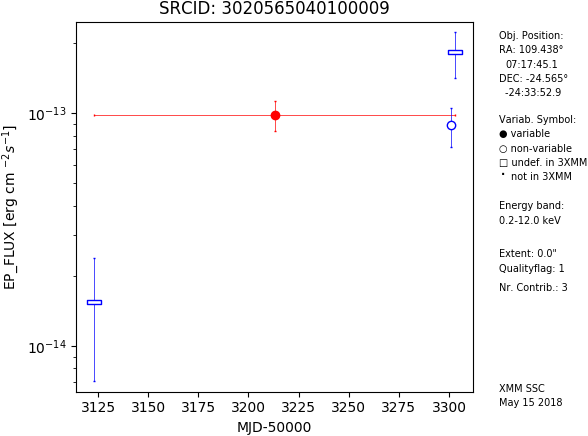}\vskip3mm
    \includegraphics[width=88mm]{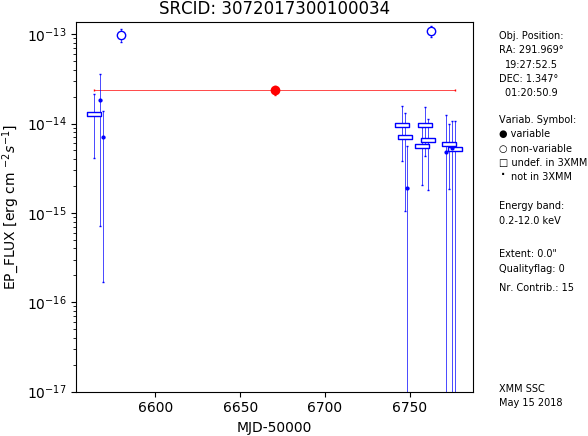}
    \caption{Example light curves of candidates for long-term variability in
      the stacked catalogue, produced as auxiliary catalogue products. The
      objects have no counterpart with SDSS classification within 5\arcsec,
      and their 3XMM-DR7 counterpart is not known to be short-term
      variable. Plot symbols inform about short- and long-term variability and
      non-detections in 3XMM-DR7 and are explained in
      Appendix~\ref{sec:auxprods}.}
    \label{fig:ltvnew}
  \end{figure}

  To demonstrate the potential of the new variability parameters for transient
  detection and the advantage of the combined source fitting, we select
  tentatively variable stacked sources and match them with catalogues from
  surveys at different energies within a radius of 5\arcsec, similar to the
  multi-wavelength cross-matching presented at the end of
  Sect.~\ref{sec:3xmmdr7}. We de-select sources with a matching identification
  in Simbad \citepads{2000A&AS..143....9W}, a counterpart in the pre-release
  version of the second Chandra Source Catalogue CSC2
  \citepads{2010ApJS..189...37E}, or a spectral classification in SDSS-DR12
  \citepads{2017AJ....154...28B}. Two example light curves of the remaining
  candidates for new long-term variable X-ray sources are shown in
  Fig.~\ref{fig:ltvnew}.

  \subsection{Cross-matching with the 3XMM serendipitous source catalogue DR7 and multi-wavelength catalogues}
  \label{sec:3xmmdr7}

  The stacked catalogue is based on a subset of 3XMM-DR7 observations. DR7s
  and DR7 were thus cross-matched to identify new detections from the stacks
  and to transfer DR7-specific information into the new resource, for example
  on short-term variability. To suppress false associations with spurious DR7
  detections, a cleaned version of 3XMM-DR7 was created for this matching
  exercise. It includes all unique sources with at least one detection in an
  observation that was used to create the stacked catalogue and at least one
  detection with a good quality flag (SUM\_FLAG 0 or 1). A source from the
  stacked catalogue and a unique source from the DR7 subset are matched if
  they are separated by less than 2.27 times the sum of their position
  errors. The factor 2.27 converts the errors from a Gaussian 68.30\,\%
  confidence region to the 99.73\,\% confidence region of a Rayleigh
  distribution, which is appropriate for coordinate errors. For the sources in
  the stacked catalogue, the simultaneously determined coordinates RA and DEC
  are used together with the pure statistical error derived from the column
  RADEC\_ERR, and the linearly added 0.43\arcsec\ component derived in
  Sect.~\ref{sec:astrometry}. For the unique 3XMM-DR7 sources, the merged
  astrometrically corrected positions SC\_RA and SC\_DEC are used together
  with the combined statistical and systematic error derived from the
  catalogue column SC\_POSERR. The matching radius per source is thus
  $r_\textrm{match} = 2.27\times\left(\sigma_\textrm{DR7,total} +
  \sigma_\textrm{stack,stat} + \sigma_\textrm{stack,sys.lin}\right)$.

  60\,908 3XMM-DR7 counterparts of stacked sources are found and their
  contributing DR7 detections are identified. The associated DR7 sources are
  included in the stacked catalogue with their identifiers, coordinates, and
  short-term variability information. The combined parameters of the unique
  source are copied from the 3XMM-DR7 catalogue of sources to the DR7s summary
  row. The parameters of each contributing DR7 detection are copied from the
  3XMM-DR7 catalogue of detections to the corresponding observation-level row
  in the stacked catalogue, if the observation was used for DR7s. This applies
  to a total of 114\,200 individual DR7 detections of the 60\,908 unique
  sources. The observation-level values of the DR7 associated columns remain
  undefined in the stacked catalogue if the DR7 source has not been detected
  in the respective observation. The offset between the associated DR7s and
  DR7 sources is given in the column DIST\_\mbox{3XMMDR7} in the summary rows,
  and the offset between the stacked source and the contributing DR7 detection
  in the respective observation-level row if applicable.

  The parameters of associated stacked and 3XMM-DR7 sources are generally
  consistent with each other within a few percent, which is within their
  uncertainties. Sources from stacked source detection with a 3XMM-DR7
  association, for example, have a median flux and median flux error of
  $1.98\pm 0.69\times 10^{-14}\,\textrm{erg\,cm}^{-2}\,\textrm{s}^{-1}$. The
  3XMM-DR7 counterparts (unique sources) have a median flux and median flux
  error of $1.78\pm 0.59\times
  10^{-14}\,\textrm{erg\,cm}^{-2}\,\textrm{s}^{-1}$, including sources with
  different numbers of contributing observations in the stacked catalogue and
  in 3XMM-DR7. For the detections per observation, the median values in the
  stacked catalogue are $2.05\pm0.91 \times
  10^{-14}\,\textrm{erg\,cm}^{-2}\,\textrm{s}^{-1}$, compared to $2.08\pm
  0.86\times 10^{-14}\,\textrm{erg\,cm}^{-2}\,\textrm{s}^{-1}$ in 3XMM-DR7.

  Within the subset of observations that have entered the stacked catalogue,
  128\,509 individual detections are listed in 3XMM-DR7. About 11\,\% are not
  recovered by stacked source detection. The differences mainly lie in a
  higher rejection rate of spurious sources through stacked source detection
  (see Sects.~\ref{sec:artstack} and \ref{sec:detqual}), the
  maximum-likelihood correction scheme (Sect.~\ref{sec:srcdet}), and the
  different background models. The percentage of low-quality detections with
  SUM\_FLAG$>$1 among the missing DR7 sources is 30\,\%, significantly higher
  than in the complete DR7 subset ($\sim16$\%).  The different background
  treatment and other subtleties affect the net counts of tentative sources,
  hence their detection likelihood and inclusion into a catalogue. For
  example, photons are distributed somewhat differently into pixels owing to a
  different spatial binning in the two catalogues compared here. This will
  cause fluctuations of the source content close to the detection likelihood
  limit. A detailed discussion on how different source-detection runs and
  different background values can affect the final source selection is given
  in \citetalads[Paper\,][]{2016A&A...590A...1R}.

  Of the 71\,951 sources in the stacked catalogue, 11\,043 do not have a
  counterpart in the DR7 subset and are thus new findings. The increase of the
  source content compares quite well with a first-order estimate based on area
  overlap, increased exposure time $T$ in the overlap, and an assumed $\log N
  - \log S$. Choosing stacks with two members only (the most abundant
  composition) and using $N\propto S^{-\Gamma+1}$, $S\propto T^{-0.5}$, the
  gain of sources through an additional exposure is
  $(T_\mathrm{total}/T_\mathrm{part})^{(\Gamma-1)/2} - 1$. According to
  \citetads{2008A&A...492...51M}, the power-law index $\Gamma$ above and below
  the flux break ranges from 1.8 to 2.6 in the dominating energy bands. With a
  $\Gamma$ of 1.8 for sources at the sensitivity limit of XMM-Newton, the
  expected gain is 16\,\%.

  When comparing the source parameters in the two catalogues, the differing
  methods used to derive them should be kept in mind. All values in the
  stacked catalogue are fitted simultaneously or directly derived from the
  stacked fit, while individual detections are matched to compile the 3XMM-DR7
  catalogue of unique sources. In particular, the stacked source coordinates
  are fitted simultaneously, while the merged coordinates of the 3XMM-DR7
  unique sources are weighted means of the individually fitted coordinates of
  the contributing detections. 3XMM-DR7 fields are astrometrically rectified
  by comparing them with optical and infra-red catalogues before the
  coordinates are merged. The values given in the RADEC\_ERR columns of the
  stacked and the 3XMM catalogues are the statistical errors of the fit to the
  positions, while the merged DR7 SC\_POSERR position errors include an
  additional component from the astrometric correction. Observations from
  which detections are merged into unique 3XMM-DR7 sources can be missing from
  the stacked catalogue because of the selection criteria of clean
  observations. Conversely, each selected observation in the stack is used to
  derive the source parameters irrespective of the detection likelihood during
  this observation, while a low-likelihood detection is not included in the
  3XMM-DR7 catalogue and does not contribute to the merged unique source. Out
  of the 60\,908 associated sources, only 26\,356 thus have the same number of
  contributing observations in both catalogues, while 26\,395 have more and
  8\,157 have fewer contributions in the stacked catalogue than in 3XMM-DR7.

  The stacked catalogue has been also cross-matched with a selection of
  external optical and infra-red catalogues and the pre-release version of the
  Chandra CSC 2.0 using the X-Match service of the Centre de Donn\'ees
  astronomiques de Strasbourg \citepads{2011ASPC..442...85P}. The best match
  within a radius of 5\arcsec\ is chosen. Table~\ref{tab:xmatch} gives the
  number of matches and their percentage with respect to the stacked
  catalogue. 57\,268 or 80\,\% of the sources have a tentative optical or
  infra-red counterpart, 59\,227 one in any of the selected catalogues
  including CSC~2.0. To estimate the fraction of false associations, a
  histogram of the position offsets between the stacked catalogue and all
  matches in the external catalogue is produced up to 30\arcsec. For a uniform
  local source density, the number of all spurious matches depends linearly on
  the offset. This linear component dominates the offset histogram above
  approximately 5\arcsec\ for the chosen external catalogues. From a linear
  fit, the number of all spurious associations within the matching radius
  5\arcsec\ is derived and subtracted from the number of all associations,
  resulting in the expected number of true associations. Its deviation from
  the number of best matches (first column of Table~\ref{tab:xmatch}) gives an
  estimate of their spurious content and is included in the fourth column of
  Table~\ref{tab:xmatch}.

  \begin{table}
    \caption{Cross-matches of the sources in the stacked catalogue: best
      X-match within a radius of 5\arcsec, given in total numbers in the
      column labelled `Matches' and as percentage of the 71\,951 stacked
      catalogue sources in the column `Share'. The estimated content of false
      associations among the matches is given in the column `False
      pos'[itives].}
    \label{tab:xmatch}
    \centering
  \begin{tabular}{l@{~~}rr@{~~}r@{~~}c}
    \hline\hline\noalign{\smallskip}
    Catalogue & Matches & Share~~\strut & False pos. & Ref. \\
    \hline\noalign{\smallskip}
    2MASS                & 16\,859 & 23.4\,\% & 32\,\%~~\strut & (1) \\
    AllWISE              & 40\,163 & 55.8\,\% &  7\,\%~~\strut & (2) \\
    GALEX GR5 AIS        &  6\,609 &  9.2\,\% &  9\,\%~~\strut & (3) \\
    UKIDSS DR9 LAS       &  8\,208 & 11.4\,\% & 18\,\%~~\strut & (4) \\
    NOMAD                & 31\,628 & 44.0\,\% & 20\,\%~~\strut & (5) \\
    Pan-STARRS1          & 36\,995 & 51.4\,\% & 18\,\%~~\strut & (6) \\
    SDSS DR12            & 21\,887 & 30.4\,\% & 29\,\%~~\strut & (7) \\
    Gaia DR2             & 28\,321 & 39.4\,\% & 32\,\%~~\strut & (8) \\
    Chandra CSC 2.0 pre1 & 13\,771 & 19.1\,\% &  1\,\%~~\strut & (9) \\
  \hline
  \end{tabular}
  \tablebib{
    (1) \citetads{2006AJ....131.1163S};
    (2) \citetads{2014yCat.2328....0C};
    (3) \citetads{2011Ap&SS.335..161B};
    (4) \citetads{2007MNRAS.379.1599L};
    (5) \citetads{2004AAS...205.4815Z};
    (6) \citetads{2016arXiv161205560C};
    (7) \citetads{2017AJ....154...28B};
    (8) \citetads{2018A&A...616A...1G};
    (9) \citetads{2010ApJS..189...37E}.
  }
  \end{table}

  \subsection{Caveats}
  \label{sec:caveats}

  The following limitations to this first stacked catalogue have been
  identified and described throughout the paper. They are summarised in this
  section.

  The catalogue is based on a selection of good-quality observations. In
  particular, repeated observations of a field have not entered the catalogue
  if they have been attributed a 3XMM-DR7 OBS\_CLASS greater than two.

  The detection likelihoods, calculated as the mathematical equivalent of a
  two-parameter fit, can be low if very few source counts are distributed
  across many images, and faint sources may be lost for purely statistical
  reasons. The effect is largely compensated by the refined box-detection
  strategy and source-selection criteria used to construct the stacked
  catalogue.

  Although the number of spurious detections is reduced by stacked source
  detection with respect to the individual observations, the catalogue is not
  free from spurious content, for example along instrumental features, stray
  light, or residuals in the PSF fit to bright sources. Many of them can be
  identified by visual inspection of the images. A filtering expression on the
  total detection likelihood helps to further decrease the potentially
  spurious content at the expense of losing transient sources.

  The source quality flags are purely derived by the automated quality
  assessment of a modified version of \texttt{dpssflag} without visual
  screening. They warn the users about low detector coverage of a source,
  possible source confusion, a source position on known bad pixels, and
  potential extended spurious detections. Source images published together
  with the catalogue offer the opportunity to inspect the detection area (see
  Sects.~\ref{sec:access} and \ref{sec:auxprods}).

  No astrometric correction has been applied to the measured source
  positions. Their mean systematic error is estimated to be 0.43\arcsec\ up to
  0.74\arcsec, depending on its definition. This astrometric accuracy is
  better than that of the uncorrected source positions listed in the 2XMM and
  3XMM catalogues.

  High-proper motion objects are not uniquely recovered by stacked source
  detection, because the algorithm is not designed to follow position changes
  between observations. They show up as several seemingly long-time variable
  objects in the catalogue and need to be identified manually or via
  comparison with astrometric catalogues.

  \subsection{Access to the catalogue and auxiliary products}
  \label{sec:access}

  The catalogue table is compiled as one file in the Flexible Image Transport
  System (FITS) format and can be downloaded directly from the website of the
  XMM-Newton SSC\footnote{\url{http://xmmssc.irap.omp.eu}}. The website also
  provides the catalogue
  documentation\footnote{\url{http://xmmssc.irap.omp.eu/Catalogue/3XMM-DR7s/3XMM_DR7stack.html}}
  and links to the other resources. The list of observations, also delivered
  in FITS format, informs about all selected OBS\_IDs, their assignment to
  stacks, the area covered, the exposure time ratio to the longest observation
  in the stack, and the setup of the observation including the filters chosen
  per instrument. Web-based user interfaces to the catalogue and the
  associated auxiliary products are provided by the
  XCatDB\footnote{\url{https://xcatdb.unistra.fr/3xmmdr7s/}} and ESA's
  XMM-Newton Science Archive
  (XSA\footnote{\url{https://www.cosmos.esa.int/web/xmm-newton/xsa}}). The
  catalogue is also included in the
  VizieR\footnote{\url{https://vizier.u-strasbg.fr/viz-bin/VizieR?-source=IX/56}}
  and
  HEASARC\footnote{\url{https://heasarc.gsfc.nasa.gov/W3Browse/xmm-newton/xmmstack.html}}
  data services.

  For all sources in the catalogue, auxiliary products are created: broad-band
  X-ray images in the $0.2-12.0$\,keV energy band, false-colour RGB images
  within $0.2-1.0$\,keV, $1.0-2.0$\,keV, and $2.0-12.0$\,keV, corresponding to
  the energy bands 1 plus 2, 3, and 4 plus 5, and optical finding charts from
  the highest-quality image out of Pan-STARRS G
  \citepads{2016arXiv161205560C}, skyMapper G \citepads{2018PASA...35...10W},
  and ESO Online Digitized Sky Survey
  DSS2\footnote{\url{https://archive.eso.org/dss/dss}} blue and red band. All
  images are centred on the source position in the stacked catalogue.  The
  X-ray and RGB images show a section of the mosaics, which are created from
  all observations in a stack using the task \texttt{emosaic}, and cover
  10\arcmin\ $\times$ 10\arcmin. Information on source extent and quality flag
  are included. The optical finding charts have a side length of 2\arcmin.
  For all sources that were observed at least twice with non-zero counts,
  long-term light curves are constructed from the mean all-EPIC fluxes in the
  stack and each contributing observation. Short-term variability according to
  3XMM-DR7 is indicated in the plots if a counterpart has been found.  Details
  on the long-term light curves and on the construction of the optical finding
  charts are given in Sect.~\ref{sec:auxprods} of the
  Appendix. Figure~\ref{fig:auxprod} shows a complete set of the auxiliary
  products for an arbitrarily chosen source.

\section{Summary and conclusions}
\label{sec:summary}

  The first serendipitous source catalogue from overlapping XMM-Newton
  observations, named 3XMM-DR7s, contains 71\,951 unique sources in 1\,789
  observations, taken between 2000 and 2016 and grouped into 434 stacks.  Its
  processing is based on a new module, using existing, improved, and new
  source-detection code, which is distributed as part of the XMM-Newton
  Science Analysis System. Stacked source detection proves to be more
  sensitive to faint sources and likely results in a lower false-positive rate
  than source detection on the individual observations. Source parameters are
  determined with higher accuracy, and the catalogue can be used in particular
  to investigate faint sources and potentially variable sources. About 15\% of
  the sources in 3XMM-DR7s are new with respect to 3XMM-DR7. At least 60\,\%
  of them have tentative counterparts in other catalogues within 5\arcsec.

  The stacked catalogue gives information on the parameters of each source in
  the stack of observations as well as in its contributing observations and on
  long-term flux variability directly from the fitting
  process. Post-processing quality assessment is automatically applied to all
  sources. An accompanying list of observations includes their technical
  details like the observation date and the filters used. The auxiliary source
  images can be accessed via the XSA interface to the stacked catalogue.

  Providing information on source detection and catalogue construction, this
  paper is intended to be the reference for 3XMM-DR7s and subsequent releases
  of stacked catalogues. The future releases are envisaged to be based on less
  restrictive selection criteria of observations to be included in the stacks
  than used for this first edition. They are planned to provide upper-limit
  flux estimates at the source positions. Methods to apply astrometric
  corrections to the individual observations before performing stacked source
  detection will be investigated to further improve its sensitivity.


\begin{acknowledgements}
  We thank the anonymous referee for helpful comments which have increased the
  quality of the paper.

  SSC work at AIP has been supported by Deutsches Zentrum f\"ur Luft- und
  Raumfahrt (DLR) through grants 50\,OX\,1401, 50\,OX\,1701, and
  50\,OR\,1604. We greatly appreciate the fruitful collaboration with the
  colleagues at ESA's XMM-Newton Science Operations Centre (SOC) and the kind
  support by the IT service team at the AIP. NW, MC, and FK acknowledge the
  CNES support. FJC acknowledges financial support through grant
  AYA2015-64346-C2-1P (MINECO/FEDER) and MTC through grant ESP2016-76683-C3-1R
  (MINECO/FEDER), both funded by the Agencia Estatal de Investigaci\'on,
  Unidad de Excelencia Mar\'ia de Maeztu.

  This project has made use of CDS services, CDS, Strasbourg, France, of
  FTOOLS by NASA's HEASARC \citepads{1995ASPC...77..367B}, and of
  TOPCAT/stilts \citepads{2005ASPC..347...29T}.
\end{acknowledgements}


\bibliographystyle{aa}
\bibliography{33938}

\begin{thebibliography}{43}
\expandafter\ifx\csname natexlab\endcsname\relax\def\natexlab#1{#1}\fi

\bibitem[{{Bianchi} {et~al.}(2011){Bianchi}, {Herald}, {Efremova}, {Girardi},
  {Zabot}, {Marigo}, {Conti}, \& {Shiao}}]{2011Ap&SS.335..161B}
{Bianchi}, L., {Herald}, J., {Efremova}, B., {et~al.} 2011, \apss, 335, 161

\bibitem[{{Blackburn}(1995)}]{1995ASPC...77..367B}
{Blackburn}, J.~K. 1995, in ASP Conf.\ Ser.\, Vol.~77, Astronomical Data
  Analysis Software and Systems IV, ed. R.~A. {Shaw}, H.~E. {Payne}, \&
  J.~J.~E. {Hayes}, 367

\bibitem[{{Blanton} {et~al.}(2017){Blanton}, {Bershady}, {Abolfathi},
  {Albareti}, {Allende Prieto}, {Almeida}, {Alonso-Garc{\'{\i}}a}, {Anders},
  {Anderson}, {Andrews}, \& et~al.}]{2017AJ....154...28B}
{Blanton}, M.~R., {Bershady}, M.~A., {Abolfathi}, B., {et~al.} 2017, \aj, 154,
  28

\bibitem[{{Boller} {et~al.}(2016){Boller}, {Freyberg}, {Tr{\"u}mper}, {Haberl},
  {Voges}, \& {Nandra}}]{2016A&A...588A.103B}
{Boller}, T., {Freyberg}, M.~J., {Tr{\"u}mper}, J., {et~al.} 2016, \aap, 588,
  A103

\bibitem[{{Cash}(1976)}]{1976A&A....52..307C}
{Cash}, W. 1976, \aap, 52, 307

\bibitem[{{Cash}(1979)}]{1979ApJ...228..939C}
{Cash}, W. 1979, \apj, 228, 939

\bibitem[{{Chambers} {et~al.}(2016){Chambers}, {Magnier}, {Metcalfe},
  {Flewelling}, {Huber}, {Waters}, {Denneau}, {Draper}, {Farrow}, {Finkbeiner},
  {Holmberg}, {Koppenhoefer}, {Price}, {Saglia}, {Schlafly}, {Smartt},
  {Sweeney}, {Wainscoat}, {Burgett}, {Grav}, {Heasley}, {Hodapp}, {Jedicke},
  {Kaiser}, {Kudritzki}, {Luppino}, {Lupton}, {Monet}, {Morgan}, {Onaka},
  {Stubbs}, {Tonry}, {Banados}, {Bell}, {Bender}, {Bernard}, {Botticella},
  {Casertano}, {Chastel}, {Chen}, {Chen}, {Cole}, {Deacon}, {Frenk},
  {Fitzsimmons}, {Gezari}, {Goessl}, {Goggia}, {Goldman}, {Grebel}, {Hambly},
  {Hasinger}, {Heavens}, {Heckman}, {Henderson}, {Henning}, {Holman}, {Hopp},
  {Ip}, {Isani}, {Keyes}, {Koekemoer}, {Kotak}, {Long}, {Lucey}, {Liu},
  {Martin}, {McLean}, {Morganson}, {Murphy}, {Nieto-Santisteban}, {Norberg},
  {Peacock}, {Pier}, {Postman}, {Primak}, {Rae}, {Rest}, {Riess}, {Riffeser},
  {Rix}, {Roser}, {Schilbach}, {Schultz}, {Scolnic}, {Szalay}, {Seitz},
  {Shiao}, {Small}, {Smith}, {Soderblom}, {Taylor}, {Thakar}, {Thiel},
  {Thilker}, {Urata}, {Valenti}, {Walter}, {Watters}, {Werner}, {White},
  {Wood-Vasey}, \& {Wyse}}]{2016arXiv161205560C}
{Chambers}, K.~C., {Magnier}, E.~A., {Metcalfe}, N., {et~al.} 2016, ArXiv
  e-prints [\eprint[arXiv]{1612.05560}]

\bibitem[{{Cutri} \& {et al.}(2014)}]{2014yCat.2328....0C}
{Cutri}, R.~M. \& {et al.} 2014, VizieR Online Data Catalog, 2328

\bibitem[{{De Luca} {et~al.}(2016){De Luca}, {Salvaterra}, {Tiengo},
  {D'Agostino}, {Watson}, {Haberl}, \& {Wilms}}]{2016ASSP...42..291D}
{De Luca}, A., {Salvaterra}, R., {Tiengo}, A., {et~al.} 2016, The Universe of
  Digital Sky Surveys, 42, 291

\bibitem[{{Evans}(2015)}]{2015ASPC..495..297E}
{Evans}, I. 2015, in Astronomical Society of the Pacific Conference Series,
  Vol. 495, Astronomical Data Analysis Software an Systems XXIV (ADASS XXIV),
  ed. A.~R. {Taylor} \& E.~{Rosolowsky}, 297

\bibitem[{{Evans} {et~al.}(2010){Evans}, {Primini}, {Glotfelty}, {Anderson},
  {Bonaventura}, {Chen}, {Davis}, {Doe}, {Evans}, {Fabbiano}, {Galle}, {Gibbs},
  {Grier}, {Hain}, {Hall}, {Harbo}, {He}, {Houck}, {Karovska}, {Kashyap},
  {Lauer}, {McCollough}, {McDowell}, {Miller}, {Mitschang}, {Morgan},
  {Mossman}, {Nichols}, {Nowak}, {Plummer}, {Refsdal}, {Rots}, {Siemiginowska},
  {Sundheim}, {Tibbetts}, {Van Stone}, {Winkelman}, \&
  {Zografou}}]{2010ApJS..189...37E}
{Evans}, I.~N., {Primini}, F.~A., {Glotfelty}, K.~J., {et~al.} 2010, \apjs,
  189, 37

\bibitem[{{Evans} {et~al.}(2014){Evans}, {Osborne}, {Beardmore}, {Page},
  {Willingale}, {Mountford}, {Pagani}, {Burrows}, {Kennea}, {Perri},
  {Tagliaferri}, \& {Gehrels}}]{2014ApJS..210....8E}
{Evans}, P.~A., {Osborne}, J.~P., {Beardmore}, A.~P., {et~al.} 2014, \apjs,
  210, 8

\bibitem[{{Fernique} {et~al.}(2017){Fernique}, {Allen}, {Boch}, {Donaldson},
  {Durand}, {Ebisawa}, {Michel}, {Salgado}, \& {Stoehr}}]{2017ivoa.spec.0519F}
{Fernique}, P., {Allen}, M., {Boch}, T., {et~al.} 2017, {HiPS - Hierarchical
  Progressive Survey Version 1.0}, IVOA Recommendation 19 May 2017

\bibitem[{{Fernique} {et~al.}(2014){Fernique}, {Boch}, {Donaldson}, {Durand},
  {O'Mullane}, {Reinecke}, \& {Taylor}}]{2014ivoa.spec.0602F}
{Fernique}, P., {Boch}, T., {Donaldson}, T., {et~al.} 2014, {MOC - HEALPix
  Multi-Order Coverage map Version 1.0}, IVOA Recommendation 02 June 2014

\bibitem[{{Fernique} {et~al.}(2010){Fernique}, {Oberto}, {Boch}, \&
  {Bonnarel}}]{2010ASPC..434..163F}
{Fernique}, P., {Oberto}, A., {Boch}, T., \& {Bonnarel}, F. 2010, in
  Astronomical Society of the Pacific Conference Series, Vol. 434, Astronomical
  Data Analysis Software and Systems XIX, ed. Y.~{Mizumoto}, K.-I. {Morita}, \&
  M.~{Ohishi}, 163

\bibitem[{{Fisher}(1932)}]{1932fisher}
{Fisher}, R.~A. 1932, {Statistical methods for research workers} (Edinburgh:
  Oliver and Boyd, 4th ed.)

\bibitem[{{Gabriel} {et~al.}(2004){Gabriel}, {Denby}, {Fyfe}, {Hoar}, {Ibarra},
  {Ojero}, {Osborne}, {Saxton}, {Lammers}, \& {Vacanti}}]{2004ASPC..314..759G}
{Gabriel}, C., {Denby}, M., {Fyfe}, D.~J., {et~al.} 2004, in ASP Conf.\ Ser.\,
  Vol. 314, Astronomical Data Analysis Software and Systems (ADASS) XIII, ed.
  F.~{Ochsenbein}, M.~G. {Allen}, \& D.~{Egret}, 759

\bibitem[{{Gaia Collaboration} {et~al.}(2018){Gaia Collaboration}, {Brown},
  {Vallenari}, {Prusti}, {de Bruijne}, {Babusiaux}, {Bailer-Jones}, {Biermann},
  {Evans}, {Eyer}, \& et~al.}]{2018A&A...616A...1G}
{Gaia Collaboration}, {Brown}, A.~G.~A., {Vallenari}, A., {et~al.} 2018, \aap,
  616, A1

\bibitem[{{Jansen} {et~al.}(2001){Jansen}, {Lumb}, {Altieri}, {Clavel}, {Ehle},
  {Erd}, {Gabriel}, {Guainazzi}, {Gondoin}, {Much}, {Munoz}, {Santos},
  {Schartel}, {Texier}, \& {Vacanti}}]{2001A&A...365L...1J}
{Jansen}, F., {Lumb}, D., {Altieri}, B., {et~al.} 2001, \aap, 365, L1

\bibitem[{{Lawrence} {et~al.}(2007){Lawrence}, {Warren}, {Almaini}, {Edge},
  {Hambly}, {Jameson}, {Lucas}, {Casali}, {Adamson}, {Dye}, {Emerson},
  {Foucaud}, {Hewett}, {Hirst}, {Hodgkin}, {Irwin}, {Lodieu}, {McMahon},
  {Simpson}, {Smail}, {Mortlock}, \& {Folger}}]{2007MNRAS.379.1599L}
{Lawrence}, A., {Warren}, S.~J., {Almaini}, O., {et~al.} 2007, \mnras, 379,
  1599

\bibitem[{{Lin} {et~al.}(2012){Lin}, {Webb}, \& {Barret}}]{2012ApJ...756...27L}
{Lin}, D., {Webb}, N.~A., \& {Barret}, D. 2012, \apj, 756, 27

\bibitem[{{Luo} {et~al.}(2017){Luo}, {Brandt}, {Xue}, {Lehmer}, {Alexander},
  {Bauer}, {Vito}, {Yang}, {Basu-Zych}, {Comastri}, {Gilli}, {Gu},
  {Hornschemeier}, {Koekemoer}, {Liu}, {Mainieri}, {Paolillo}, {Ranalli},
  {Rosati}, {Schneider}, {Shemmer}, {Smail}, {Sun}, {Tozzi}, {Vignali}, \&
  {Wang}}]{2017ApJS..228....2L}
{Luo}, B., {Brandt}, W.~N., {Xue}, Y.~Q., {et~al.} 2017, \apjs, 228, 2

\bibitem[{{Mateos} {et~al.}(2009){Mateos}, {Saxton}, {Read}, \&
  {Sembay}}]{2009A&A...496..879M}
{Mateos}, S., {Saxton}, R.~D., {Read}, A.~M., \& {Sembay}, S. 2009, \aap, 496,
  879

\bibitem[{{Mateos} {et~al.}(2008){Mateos}, {Warwick}, {Carrera}, {Stewart},
  {Ebrero}, {Della Ceca}, {Caccianiga}, {Gilli}, {Page}, {Treister}, {Tedds},
  {Watson}, {Lamer}, {Saxton}, {Brunner}, \& {Page}}]{2008A&A...492...51M}
{Mateos}, S., {Warwick}, R.~S., {Carrera}, F.~J., {et~al.} 2008, \aap, 492, 51

\bibitem[{{Motch} {et~al.}(2009){Motch}, {Pires}, {Haberl}, {Schwope}, \&
  {Zavlin}}]{2009A&A...497..423M}
{Motch}, C., {Pires}, A.~M., {Haberl}, F., {Schwope}, A., \& {Zavlin}, V.~E.
  2009, \aap, 497, 423

\bibitem[{{Page} {et~al.}(2012){Page}, {Brindle}, {Talavera}, {Still}, {Rosen},
  {Yershov}, {Ziaeepour}, {Mason}, {Cropper}, {Breeveld}, {Loiseau}, {Mignani},
  {Smith}, \& {Murdin}}]{2012MNRAS.426..903P}
{Page}, M.~J., {Brindle}, C., {Talavera}, A., {et~al.} 2012, \mnras, 426, 903

\bibitem[{{Pineau} {et~al.}(2011){Pineau}, {Boch}, \&
  {Derriere}}]{2011ASPC..442...85P}
{Pineau}, F.-X., {Boch}, T., \& {Derriere}, S. 2011, in Astronomical Society of
  the Pacific Conference Series, Vol. 442, Astronomical Data Analysis Software
  and Systems XX, ed. I.~N. {Evans}, A.~{Accomazzi}, D.~J. {Mink}, \& A.~H.
  {Rots}, 85

\bibitem[{{Puccetti} {et~al.}(2011){Puccetti}, {Capalbi}, {Giommi}, {Perri},
  {Stratta}, {Angelini}, {Burrows}, {Campana}, {Chincarini}, {Cusumano},
  {Gehrels}, {Moretti}, {Nousek}, {Osborne}, \&
  {Tagliaferri}}]{2011A&A...528A.122P}
{Puccetti}, S., {Capalbi}, M., {Giommi}, P., {et~al.} 2011, \aap, 528, A122

\bibitem[{{Read} \& {Ponman}(2003)}]{2003A&A...409..395R}
{Read}, A.~M. \& {Ponman}, T.~J. 2003, \aap, 409, 395

\bibitem[{{Rosen} \& {Read}(2017)}]{2017extras_wp5}
{Rosen}, S.~R. \& {Read}, A.~M. 2017, {The EXTraS long-term Variability
  Catalogue}, EXTraS Public Data Archive Documentation FP7-SPACE-2013-1 GA
  n.607452

\bibitem[{{Rosen} {et~al.}(2016){Rosen}, {Webb}, {Watson}, {Ballet}, {Barret},
  {Braito}, {Carrera}, {Ceballos}, {Coriat}, {Della Ceca}, {Denkinson},
  {Esquej}, {Farrell}, {Freyberg}, {Gris{\'e}}, {Guillout}, {Heil},
  {Koliopanos}, {Law-Green}, {Lamer}, {Lin}, {Martino}, {Michel}, {Motch},
  {Nebot Gomez-Moran}, {Page}, {Page}, {Page}, {Pakull}, {Pye}, {Read},
  {Rodriguez}, {Sakano}, {Saxton}, {Schwope}, {Scott}, {Sturm}, {Traulsen},
  {Yershov}, \& {Zolotukhin}}]{2016A&A...590A...1R}
{Rosen}, S.~R., {Webb}, N.~A., {Watson}, M.~G., {et~al.} 2016, \aap, 590, A1

\bibitem[{{Saxton} {et~al.}(2008){Saxton}, {Read}, {Esquej}, {Freyberg},
  {Altieri}, \& {Bermejo}}]{2008A&A...480..611S}
{Saxton}, R.~D., {Read}, A.~M., {Esquej}, P., {et~al.} 2008, \aap, 480, 611

\bibitem[{{Skrutskie} {et~al.}(2006){Skrutskie}, {Cutri}, {Stiening},
  {Weinberg}, {Schneider}, {Carpenter}, {Beichman}, {Capps}, {Chester},
  {Elias}, {Huchra}, {Liebert}, {Lonsdale}, {Monet}, {Price}, {Seitzer},
  {Jarrett}, {Kirkpatrick}, {Gizis}, {Howard}, {Evans}, {Fowler}, {Fullmer},
  {Hurt}, {Light}, {Kopan}, {Marsh}, {McCallon}, {Tam}, {Van Dyk}, \&
  {Wheelock}}]{2006AJ....131.1163S}
{Skrutskie}, M.~F., {Cutri}, R.~M., {Stiening}, R., {et~al.} 2006, \aj, 131,
  1163

\bibitem[{{Stewart}(2009)}]{2009A&A...495..989S}
{Stewart}, I.~M. 2009, \aap, 495, 989

\bibitem[{{Str{\"u}der} {et~al.}(2001){Str{\"u}der}, {Briel}, {Dennerl},
  {Hartmann}, {Kendziorra}, {Meidinger}, {Pfeffermann}, {Reppin}, {Aschenbach},
  {Bornemann}, {Br{\"a}uninger}, {Burkert}, {Elender}, {Freyberg}, {Haberl},
  {Hartner}, {Heuschmann}, {Hippmann}, {Kastelic}, {Kemmer}, {Kettenring},
  {Kink}, {Krause}, {M{\"u}ller}, {Oppitz}, {Pietsch}, {Popp}, {Predehl},
  {Read}, {Stephan}, {St{\"o}tter}, {Tr{\"u}mper}, {Holl}, {Kemmer}, {Soltau},
  {St{\"o}tter}, {Weber}, {Weichert}, {von Zanthier}, {Carathanassis}, {Lutz},
  {Richter}, {Solc}, {B{\"o}ttcher}, {Kuster}, {Staubert}, {Abbey}, {Holland},
  {Turner}, {Balasini}, {Bignami}, {La Palombara}, {Villa}, {Buttler},
  {Gianini}, {Lain{\'e}}, {Lumb}, \& {Dhez}}]{2001A&A...365L..18S}
{Str{\"u}der}, L., {Briel}, U., {Dennerl}, K., {et~al.} 2001, \aap, 365, L18

\bibitem[{{Taylor}(2005)}]{2005ASPC..347...29T}
{Taylor}, M.~B. 2005, in ASP Conf.\ Ser.\, Vol. 347, Astronomical Data Analysis
  Software and Systems XIV, ed. P.~{Shopbell}, M.~{Britton}, \& R.~{Ebert}, 29

\bibitem[{{Turner} {et~al.}(2001){Turner}, {Abbey}, {Arnaud}, {Balasini},
  {Barbera}, {Belsole}, {Bennie}, {Bernard}, {Bignami}, {Boer}, {Briel},
  {Butler}, {Cara}, {Chabaud}, {Cole}, {Collura}, {Conte}, {Cros}, {Denby},
  {Dhez}, {Di Coco}, {Dowson}, {Ferrando}, {Ghizzardi}, {Gianotti}, {Goodall},
  {Gretton}, {Griffiths}, {Hainaut}, {Hochedez}, {Holland}, {Jourdain},
  {Kendziorra}, {Lagostina}, {Laine}, {La Palombara}, {Lortholary}, {Lumb},
  {Marty}, {Molendi}, {Pigot}, {Poindron}, {Pounds}, {Reeves}, {Reppin},
  {Rothenflug}, {Salvetat}, {Sauvageot}, {Schmitt}, {Sembay}, {Short},
  {Spragg}, {Stephen}, {Str{\"u}der}, {Tiengo}, {Trifoglio}, {Tr{\"u}mper},
  {Vercellone}, {Vigroux}, {Villa}, {Ward}, {Whitehead}, \&
  {Zonca}}]{2001A&A...365L..27T}
{Turner}, M.~J.~L., {Abbey}, A., {Arnaud}, M., {et~al.} 2001, \aap, 365, L27

\bibitem[{{Voges} {et~al.}(1999){Voges}, {Aschenbach}, {Boller},
  {Br{\"a}uninger}, {Briel}, {Burkert}, {Dennerl}, {Englhauser}, {Gruber},
  {Haberl}, {Hartner}, {Hasinger}, {K{\"u}rster}, {Pfeffermann}, {Pietsch},
  {Predehl}, {Rosso}, {Schmitt}, {Tr{\"u}mper}, \&
  {Zimmermann}}]{1999A&A...349..389V}
{Voges}, W., {Aschenbach}, B., {Boller}, T., {et~al.} 1999, \aap, 349, 389

\bibitem[{{Watson} {et~al.}(2001){Watson}, {Augu{\`e}res}, {Ballet}, {Barcons},
  {Barret}, {Boer}, {Boller}, {Bromage}, {Brunner}, {Carrera}, {Cropper},
  {Denby}, {Ehle}, {Elvis}, {Fabian}, {Freyberg}, {Guillout}, {Hameury},
  {Hasinger}, {Hinshaw}, {Maccacaro}, {Mason}, {McMahon}, {Michel}, {Mirioni},
  {Mittaz}, {Motch}, {Olive}, {Osborne}, {Page}, {Pakull}, {Perry}, {Pierre},
  {Pietsch}, {Pye}, {Read}, {Roberts}, {Rosen}, {Sauvageot}, {Schwope},
  {Sekiguchi}, {Stewart}, {Stewart}, {Valtchanov}, {Ward}, {Warwick}, {West},
  {White}, \& {Worrall}}]{2001A&A...365L..51W}
{Watson}, M.~G., {Augu{\`e}res}, J.-L., {Ballet}, J., {et~al.} 2001, \aap, 365,
  L51

\bibitem[{{Watson} {et~al.}(2009){Watson}, {Schr{\"o}der}, {Fyfe}, {Page},
  {Lamer}, {Mateos}, {Pye}, {Sakano}, {Rosen}, {Ballet}, {Barcons}, {Barret},
  {Boller}, {Brunner}, {Brusa}, {Caccianiga}, {Carrera}, {Ceballos}, {Della
  Ceca}, {Denby}, {Denkinson}, {Dupuy}, {Farrell}, {Fraschetti}, {Freyberg},
  {Guillout}, {Hambaryan}, {Maccacaro}, {Mathiesen}, {McMahon}, {Michel},
  {Motch}, {Osborne}, {Page}, {Pakull}, {Pietsch}, {Saxton}, {Schwope},
  {Severgnini}, {Simpson}, {Sironi}, {Stewart}, {Stewart}, {Stobbart}, {Tedds},
  {Warwick}, {Webb}, {West}, {Worrall}, \& {Yuan}}]{2009A&A...493..339W}
{Watson}, M.~G., {Schr{\"o}der}, A.~C., {Fyfe}, D., {et~al.} 2009, \aap, 493,
  339

\bibitem[{{Wenger} {et~al.}(2000){Wenger}, {Ochsenbein}, {Egret}, {Dubois},
  {Bonnarel}, {Borde}, {Genova}, {Jasniewicz}, {Lalo{\"e}}, {Lesteven}, \&
  {Monier}}]{2000A&AS..143....9W}
{Wenger}, M., {Ochsenbein}, F., {Egret}, D., {et~al.} 2000, \aaps, 143, 9

\bibitem[{{Wolf} {et~al.}(2018){Wolf}, {Onken}, {Luvaul}, {Schmidt}, {Bessell},
  {Chang}, {Da Costa}, {Mackey}, {Martin-Jones}, {Murphy}, {Preston}, {Scalzo},
  {Shao}, {Smillie}, {Tisserand}, {White}, \& {Yuan}}]{2018PASA...35...10W}
{Wolf}, C., {Onken}, C.~A., {Luvaul}, L.~C., {et~al.} 2018, \pasa, 35, e010

\bibitem[{{Zacharias} {et~al.}(2004){Zacharias}, {Monet}, {Levine}, {Urban},
  {Gaume}, \& {Wycoff}}]{2004AAS...205.4815Z}
{Zacharias}, N., {Monet}, D.~G., {Levine}, S.~E., {et~al.} 2004, in Bulletin of
  the American Astronomical Society, Vol.~36, American Astronomical Society
  Meeting Abstracts, 1418

\end{thebibliography}

\begin{appendix}

\section{The automated method used to identify high-background fields}
\label{sec:hbmethod}

  To establish a high-background threshold for each EPIC instrument from a
  large sample of exposures, a mean background count rate per unit area
  between 0.2 and 12.0\,keV is determined for each of the about 8\,000
  3XMM-DR7 EPIC observations taken in full-frame or large-window mode in the
  following way. From the event lists pre-filtered with the 3XMM GTIs,
  source-excised images are created per instrument by excluding circular
  regions around known 3XMM-DR7 sources with the radius being the maximum of
  \emph{(i)} 30\arcsec, \emph{(ii)} the square root of the counts as a rough
  approximation to PSF scaling, and \emph{(iii)} -- if the source has a
  summary flag of 0 or 1 in 3XMM-DR7 indicating a good-quality detection --
  the source extent. For bright sources with an EPIC/pn count rate above
  1\,count\,s$^{-1}$, summed over all five energy bands, stripes along the
  readout direction are excluded to get rid of out-of-time events. To simplify
  the procedure, the stripes have a constant width of 40\arcsec\ over the
  whole chip extent. Corresponding source-excised masks are derived from the
  source-excised images, which give the valid pixels per instrument during the
  observation.

  The averaged background count rate per area in units of
  cts\,arcsec$^{-2}$\,s$^{-1}$ in each source-excised image is the total
  number of photons divided by the number of pixels in the source-excised
  mask, the pixel size in square arcseconds, and the net exposure time in
  seconds. For EPIC-pn, the four quadrants are treated separately, because
  they are independent of each other and can have different lifetimes and thus
  background levels, in particular if they are operated in continuous counting
  mode while the telemetry of the instruments is saturated and data are
  transmitted incompletely and unusable for scientific analyses. The maximum
  background value of the quadrants is used as a measure of the whole
  observation.

  The method has two general limitations. Firstly, it does not distinguish
  between high sky background and emission of very extended sources within the
  field of view. Both scenarios are considered problematic for (stacked)
  source detection and treated in the same way. Secondly, background features
  that are prominent on small scales only like stray light are not reliably
  flagged by this method, since the count rate is calculated as an average
  over the chip or chip quadrants. A measure of spatial background variability
  over the field of view can be used to identify these cases and may be
  implemented in the future.

\begin{figure}
    \centering
    \includegraphics[width=88mm]{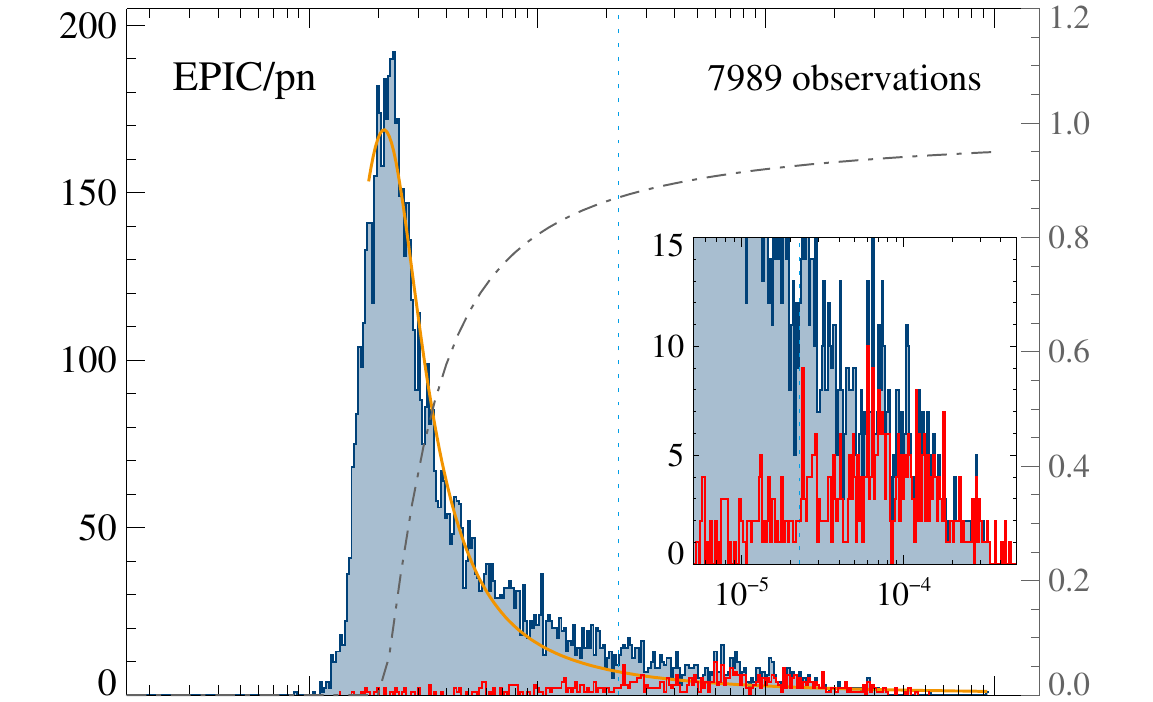}
    \includegraphics[width=88mm]{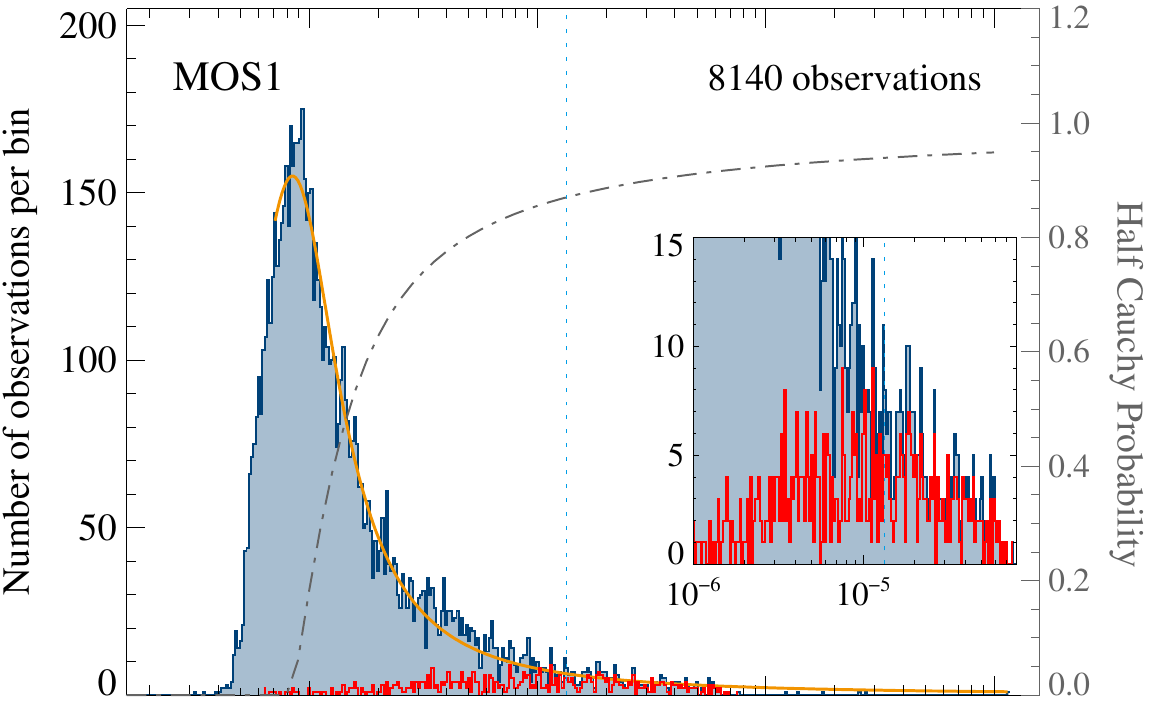}
    \includegraphics[width=88mm]{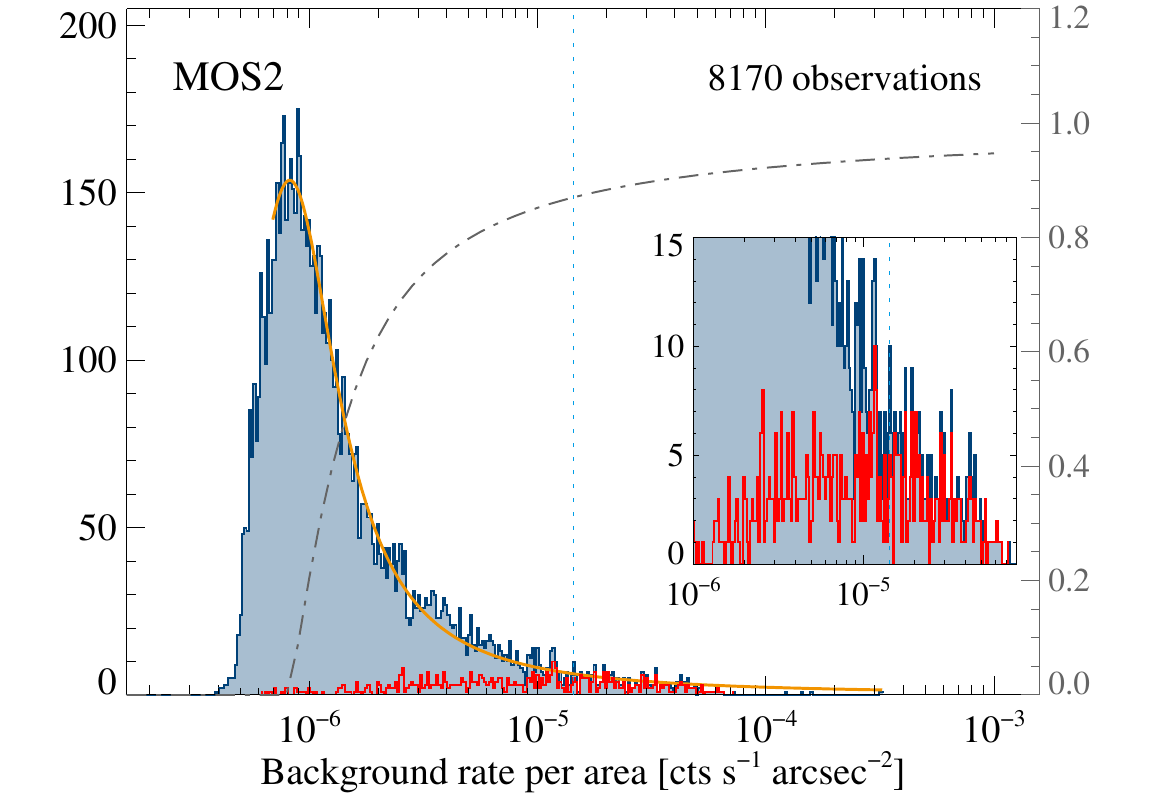}
    \caption{Histograms of the derived background rates per area of all
      considered observations of EPIC-pn, MOS1, and MOS2 (from top to
      bottom). The orange line shows the Lorentz fit to the histogram and the
      grey dash-dotted line the Half Cauchy probability distribution, with the
      scale given in the right axis. The dashed vertical line marks the 87\,\%
      probability cut used to discard observations as high-background
      contaminated. The red histogram shows the distribution of observations
      that have a HIGH\_BACKGROUND flag in 3XMM-DR7 for comparison.
      \emph{Insets:} zoom to the highest background values.}
    \label{fig:bkgrates}
  \end{figure}

  The distributions of mean background values are shown in
  Fig.~\ref{fig:bkgrates} for each EPIC instrument. Observations that have a
  HIGH\_BACKGROUND flag in 3XMM-DR7 are plotted in red, with a zoom to high
  rates in the inset. The 3XMM flag is set for the whole observation if at
  least one instrument experienced increased background. In the plots per
  instrument of Fig.~\ref{fig:bkgrates}, some observations with a low mean
  background level in one instrument are therefore marked in red owing to a
  DR7 background flag triggered by one of the other instruments. The height of
  the peak in the histograms is estimated from a fit with an empirically
  chosen Lorentz function $l(r)=h/((r-c)^2/w^2+1)$ with count rate $r$, peak
  centre $c$, height $h$, and half width at half maximum $w$, omitting the
  left wing of the peak. It translates into a cumulative Half Cauchy
  probability distribution $2\arctan((\log_{10}(r) - c)/w)/\pi$, regarding
  background count rates left of the peak as low background with probability
  zero. For the stacked catalogue, a probability cut of 87\,\% was used to
  exclude high-background observations from stacked source detection
  (Sect.~\ref{sec:highbkg}).

\section{Auxiliary information on the stacked catalogue and its selection of
  observations}

  \begin{figure*}
      \includegraphics[width=.33\linewidth]{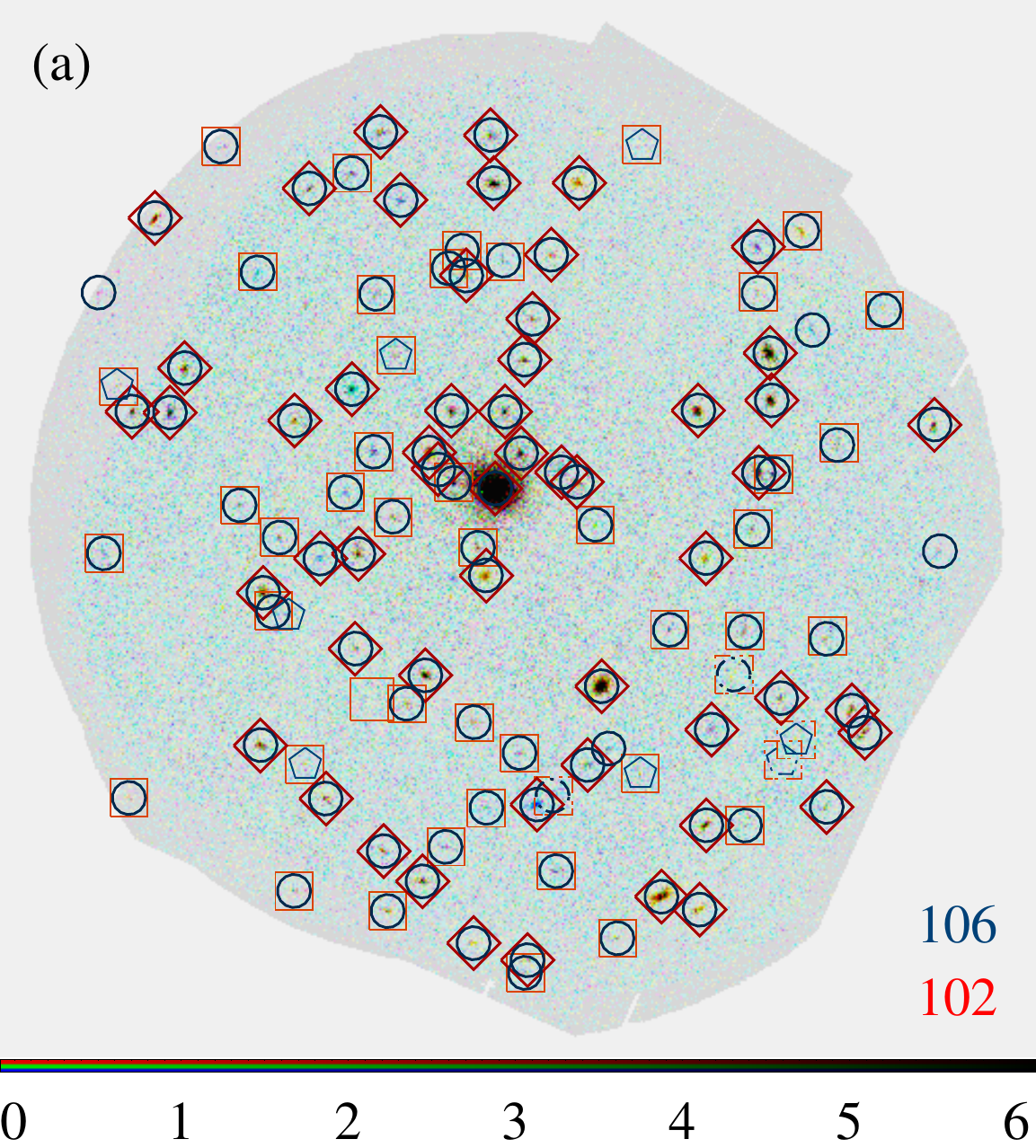}\hfill
      \includegraphics[width=.33\linewidth]{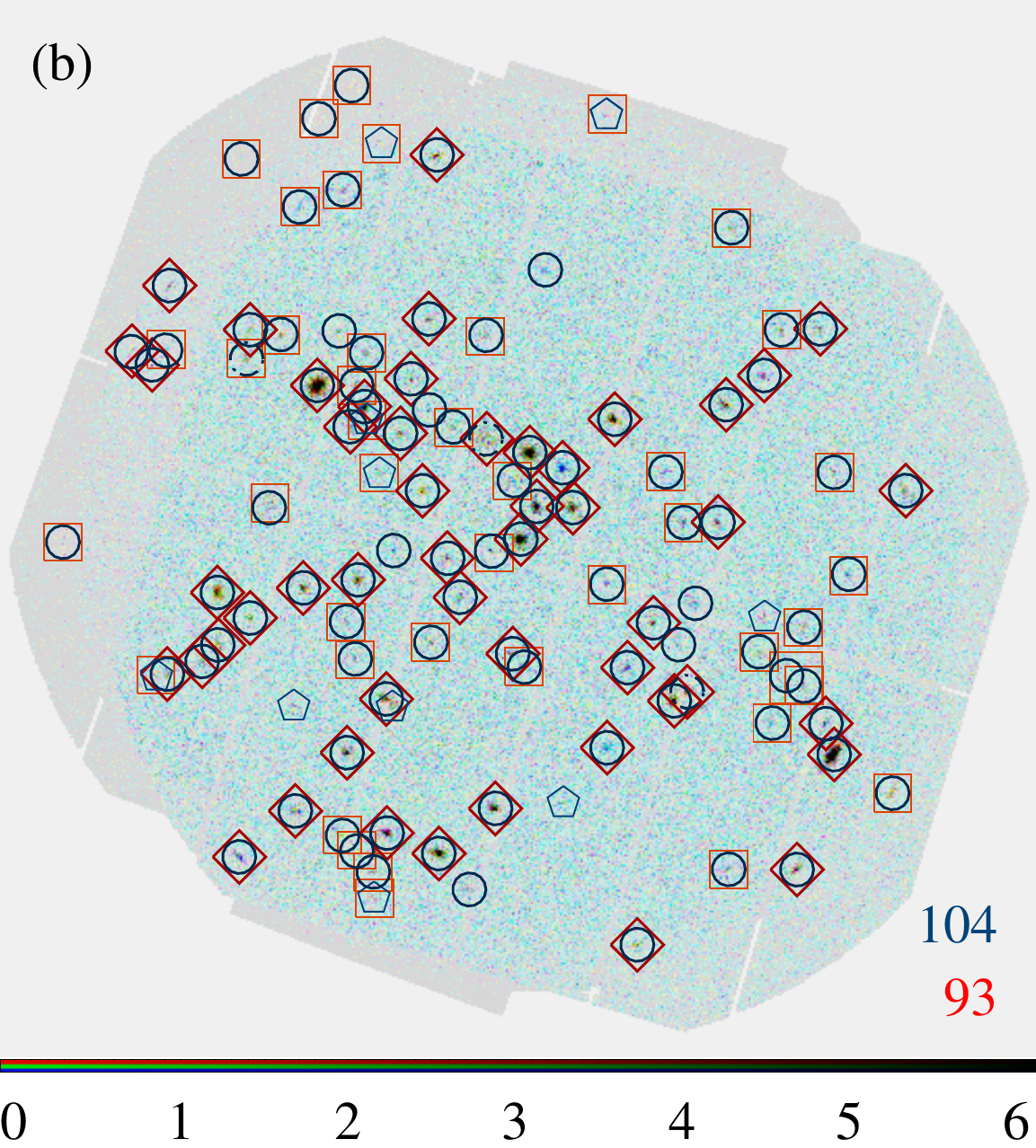}\hfill
      \includegraphics[width=.33\linewidth]{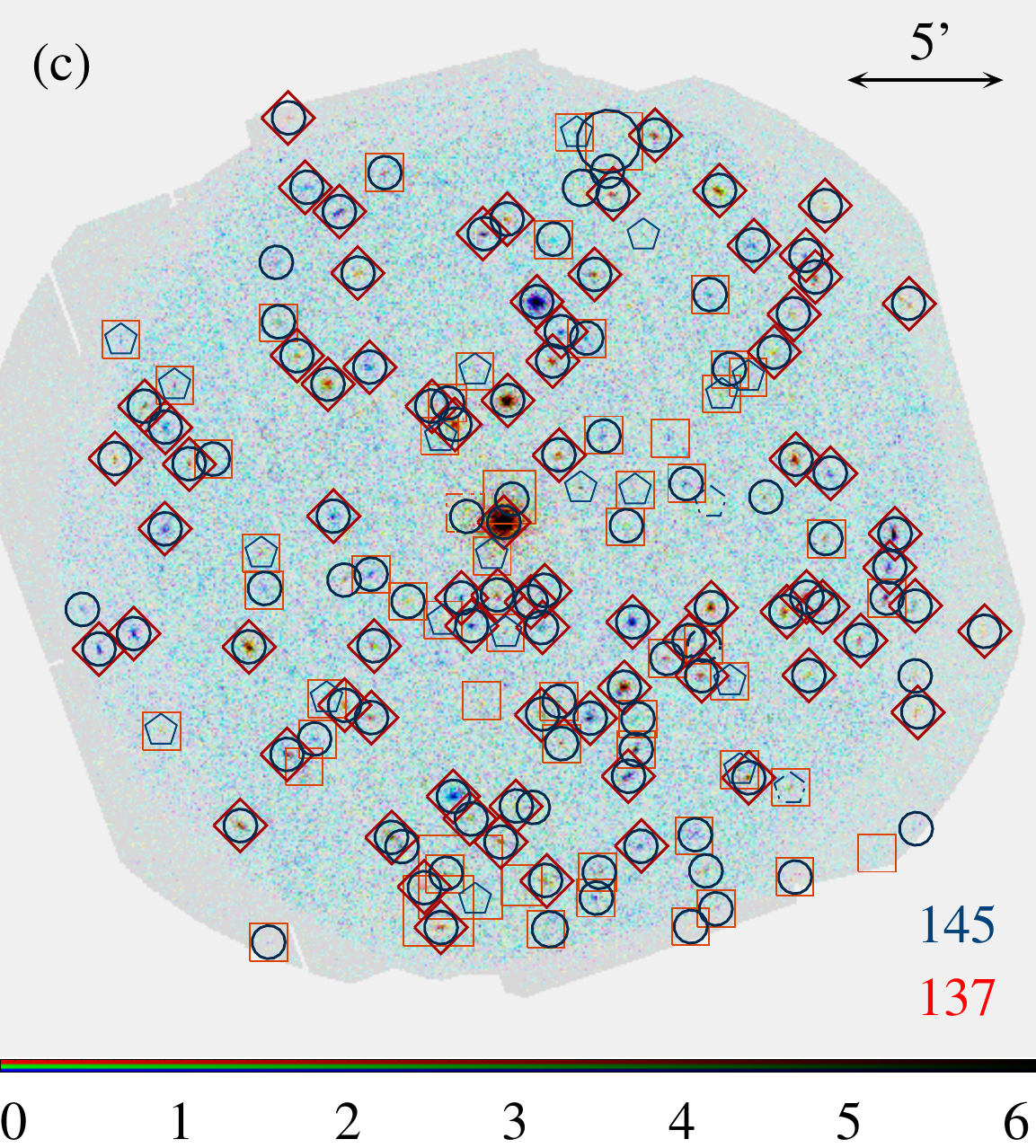}\vskip4pt
      \includegraphics[width=.33\linewidth]{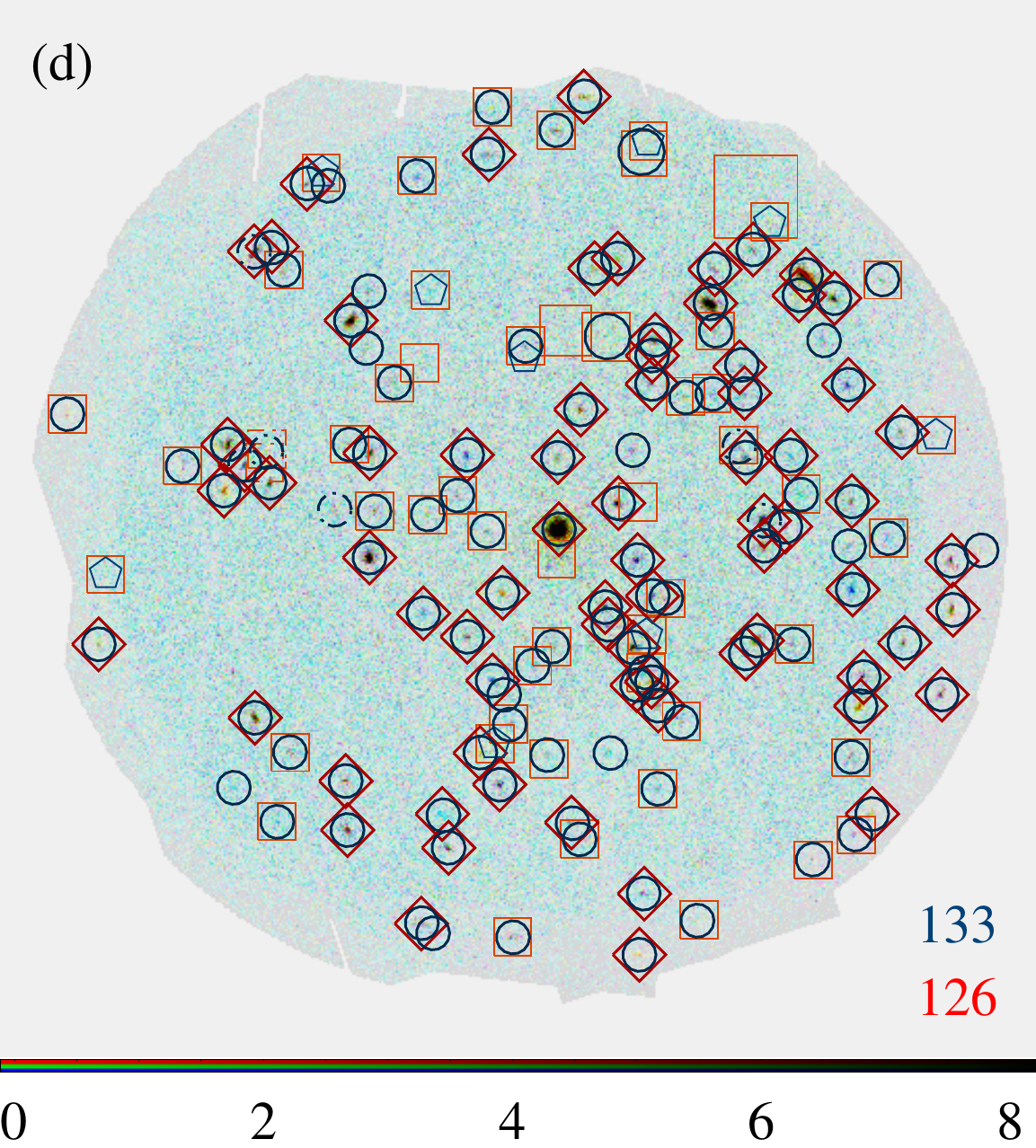}\hfill
      \includegraphics[width=.33\linewidth]{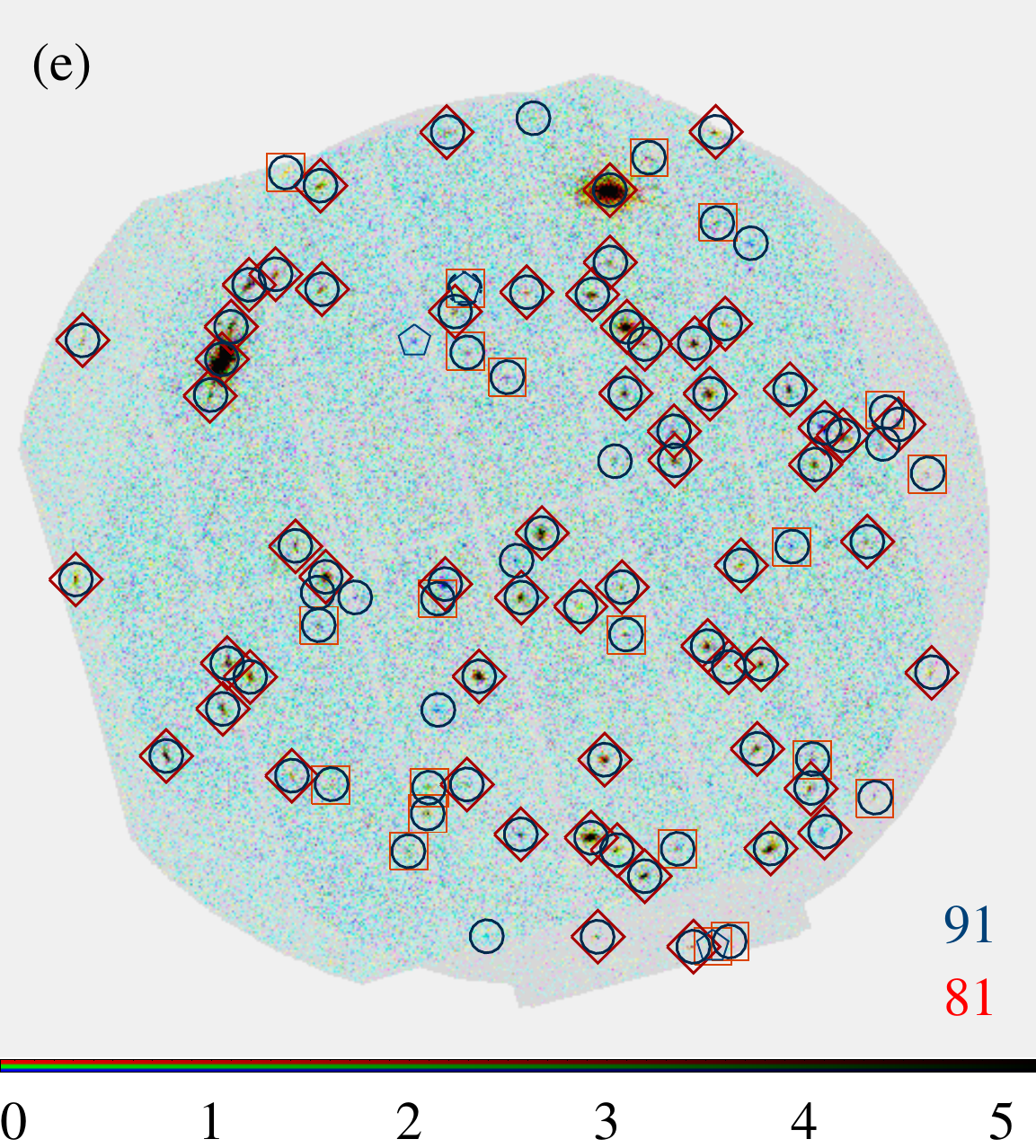}\hfill
      \includegraphics[width=.33\linewidth]{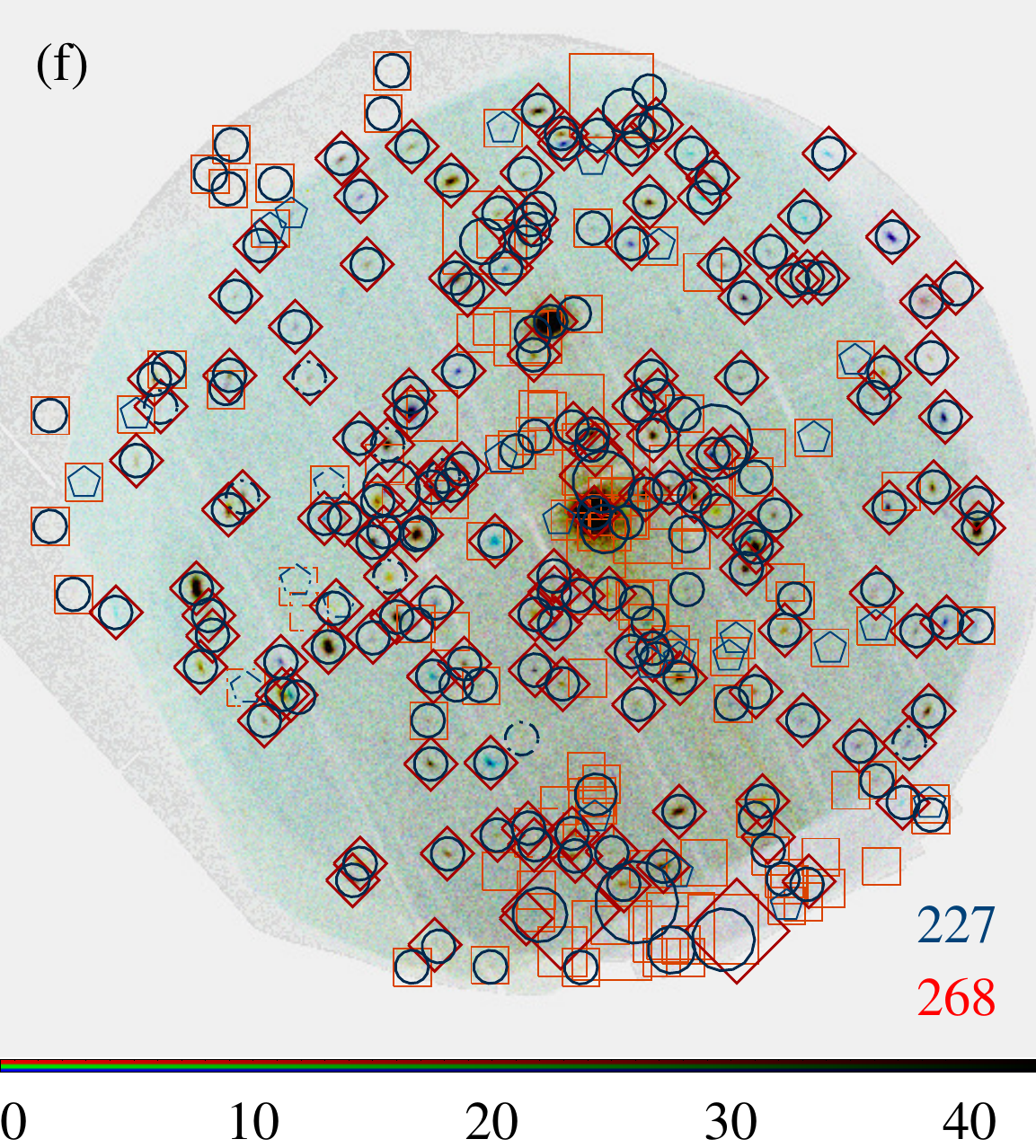}
    \caption{Example detection images of catalogue stacks and detections in
      the respective individual observations, including a field with
      considerable spurious content. As in Fig.~\ref{fig:detimaHD}, stacked
      detections are shown in blue and combined individual detections in
      red. Thick circles and diamonds mark detections that are significant in
      at least two observations, thin pentagons and boxes the others. Dashed
      symbols are used for detections that have been flagged by
      \texttt{dpssflag}. \emph{(a)} Two observations with identifiers
      0693662101 and 0723780201. \emph{(b)} Two observations with identifiers
      0203840101 and 0203840201. \emph{(c)} Three observations with
      identifiers 0205650401, 0205650601, and 0205650701. \emph{(d)} Two
      observations with identifiers 0674320301 and 0674320401. \emph{(e)} Two
      observations with identifiers 0505010501 and 0505011201. \emph{(f)} Five
      observations with identifiers 0124712501, 0204040101, 0204040301,
      0304320201, and 0304320301.}
    \label{fig:detexamples}
  \end{figure*}

  \begin{figure*}
    \centering
    \includegraphics[height=.35\linewidth]{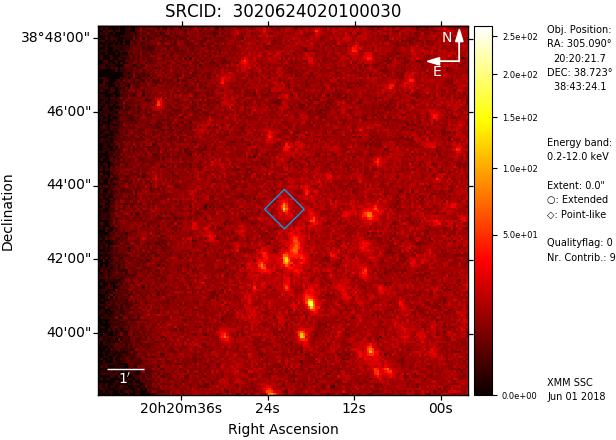}\hfill%
    \includegraphics[height=.35\linewidth]{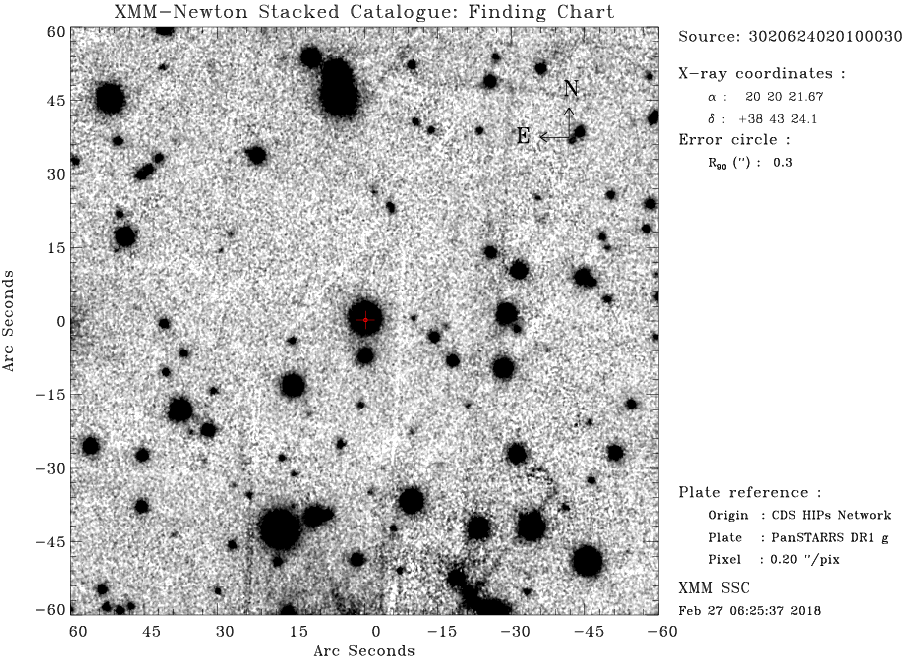}\vskip4pt
    \includegraphics[height=.35\linewidth]{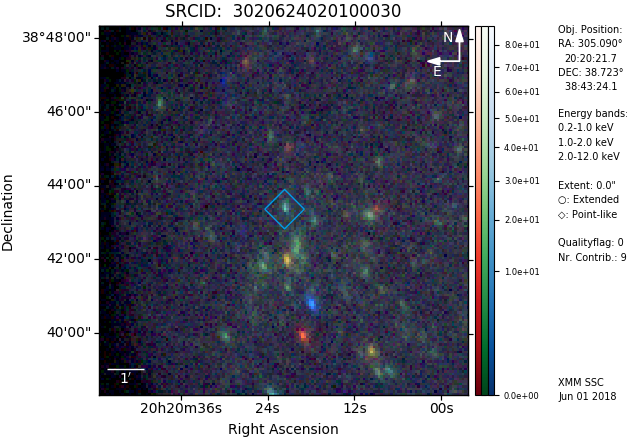}\hfill%
    \includegraphics[height=.35\linewidth]{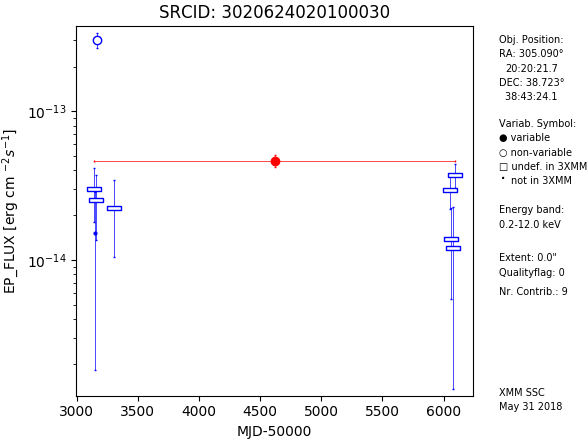}
    \caption{Examples of the auxiliary products accompanying each catalogue
      source: broad-band X-ray image, false-colour RGB image of three X-ray
      energy bands, optical finding chart, and long-term light curve.}
    \label{fig:auxprod}
  \end{figure*}

  \subsection{Proposal categories included in the catalogue}
  \label{sec:propcat}

  Table~\ref{tab:xsa} lists the number of catalogue observations per
  XMM-Newton proposal category. Most of the 3XMM-DR7 observations comprising
  objects with large extent have been de-selected from the first stacked
  catalogue.

  \subsection{Example detection images}
  \label{sec:detimages}

  Figure~\ref{fig:detexamples} shows examples of the differences between
  source detection on two to five stacked observations and on the individual
  observations (see Sect.~\ref{sec:detqual}). As for the 3XMM catalogues of
  unique sources, the individual detections have been joined within a matching
  radius of 15\arcsec.

  \subsection{Catalogue columns}
  \label{sec:columnlist}

  Table~\ref{tab:columns} gives an overview of all columns included in the
  catalogue and a short description of how the stacked parameters and the
  parameters per contributing observation are calculated. Entries centred
  within the two columns for stacked and observation-level values are valid
  for both of them. `Null' stands for undefined values / not-a-number, `zero'
  for 0.0. Weighted means of values $x_i$ with errors $\sigma_i$ are
  calculated as $\bar{x}=(\sum_{i=0}^n x_i/\sigma_i^2)/\sigma_x^2$ with
  $\sigma_x^2=1/\sum_{i=0}^n \sigma_i^{-2}$. Values copied from the nearest
  3XMM-DR7 source within a matching radius of three times the summed position
  errors are labelled by the suffix \_\mbox{3XMMDR7}.

  \subsection{Auxiliary products}
  \label{sec:auxprods}

  The optical finding charts have been generated in three steps using tools
  based on the HiPS standard \citepads{2017ivoa.spec.0519F} initially designed
  by the Astronomical Data Center (CDS) of the Observatoire de Strasbourg and
  adopted by the Virtual Observatory. The procedure uses a large collection of
  multi-order coverage maps \citepads{2014ivoa.spec.0602F} describing the sky
  coverage of many surveys and catalogues, which is operated by the CDS. The
  list of the HiPS surveys covering the position is requested from this
  database. The optical survey having the highest priority is selected and
  transmitted to an Aladin instance \citepads{2010ASPC..434..163F} running
  behind a Tomcat server. This service gets the HiPS tiles covering the
  requested region from a CDS server and converts them into a FITS image. The
  operations are controlled by a Java client which runs an IDL\footnote{Based
    on proprietary Interactive Data Language software,
    \url{https://www.harrisgeospatial.com/SoftwareTechnology/IDL.aspx}.} task
  producing the PDF file lastly. The image cuts are tuned to highlight the
  fainter features. The IDL code is derived from the Astronomical Catalogue
  Data Subsystem (ACDS) task of the XMM-Newton pipeline.

  The long-term light curves are created for sources with non-zero counts
  during at least two observations and include the stacked EPIC flux value and
  the EPIC fluxes during the contributing observations. Different plot symbols
  are used to indicate tentative short- and long-term variability. The stacked
  flux is plotted with a filled circle, if the variability VAR\_PROB of the
  source fluxes to be consistent with constant flux is $1\,\%$ or lower.
  Probabilities of short-term variability are included in 3XMM-DR7 for
  detections with at least 100\,counts and indicated in the long-term light
  curves by the plot symbols of the observation-level fluxes in the stacked
  catalogue. The flux is plotted with a filled circle, if a 3XMM-DR7
  observation has been associated with the source in the stacked catalogue
  (cf.\ Sect.~\ref{sec:3xmmdr7}) and if its short-term variability flag
  VAR\_FLAG\_\mbox{3XMMDR7} is true. Open circles are used in the opposite
  cases for tentatively non-variable sources. An open box of arbitrary size
  means that too few counts were collected during the observation to derive
  information on short-term variability in 3XMM-DR7, and a small dot that no
  DR7 detection has been associated with the source.

  Figure~\ref{fig:auxprod} shows them for an example source detected in nine
  stacked observations.

  \begin{table}
    \caption{XSA proposal categories of the selected observations.}
    \label{tab:xsa}
    \begin{tabular}{lp{.74\linewidth}@{~~}r}
      \hline\hline\noalign{\smallskip}
      \multicolumn{2}{l}{XSA proposal category} & number \\
      \hline\noalign{\smallskip}
      A & Stars, White Dwarfs and Solar System
        & 382 \\
      B &  White Dwarf Binaries, Neutron Star Binaries, Cataclysmic Variables, ULX and Black Holes
        & 192 \\
      C & Supernovae, Supernova Remnants, Diffuse Emission, Diffuse galactic Emission and Isolated Neutron Stars
        & 119 \\
      D & Galaxies, Galactic Surveys and X-ray Background
        & 111 \\
      E & Galaxies, Groups of Galaxies, Clusters of Galaxies and Superclusters
        & 229 \\
      F & Active Galactic Nuclei, Quasars, BL-Lac Objects and Tidal Disruption Events
        & 291 \\
      G & Groups of Galaxies, Clusters of Galaxies and Superclusters
        & 162 \\
      H & Cosmology, Extragalactic Deep Fields and Large Extragalactic Areas
        & 303 \\
      \hline
      \end{tabular}
    \end{table}

  \longtab[2]{
  \begin{landscape}
  \begin{longtable}{l@{~~}c@{\quad}cp{.325\linewidth}p{.16\linewidth}p{.16\linewidth}}
  \caption{Description of all catalogue columns. The table column `Stack'
    comprises a short summary of how the parameter value in the summary row
    per stack is calculated, `Observation' of the value derived for the
    individual contributing observation. Values per instrument are labelled by
    $CC$, which stands for one of PN, M1, M2. The energy bands, defined as for
    3XMM, are abbreviated by $1\leq n\leq 5$ and for the hardness ratios by
    $1\leq i\leq 4$. Broad-band EPIC values are derived in the 0.2$-$12.0\,keV
    range.}
    \label{tab:columns}\\
  \hline\hline
  \noalign{\smallskip} Column name &
  Format & Unit & Description & Observation & Stack \\ \hline \endfirsthead
  \caption{Continued.} \\
  \hline\noalign{\smallskip}
    Column name & Format & Unit  & Description & Observation & Stack \\
  \hline\strut
  \endhead
  \hline
  \endfoot
  \hline
  \endlastfoot
  \noalign{\smallskip}\multicolumn{6}{l}{Columns 1$-$4: Detection characteristics\medskip} \\
   \mbox{SRCID}             & long    &
       & Identifier of the detection.
       & \multicolumn{2}{c}{Assigned to all rows per source} \\
   OBS\_ID           & string  &
       & XMM-Newton observation identifier.
       & Per observation & \strut\hfill null \hfill\strut \\
   N\_OBS            & integer &
       & Number of observations involved in the stack.
       & \strut\hfill null \hfill\strut & Per stack \\
   N\_CONTRIB        & integer &
       & Number of observations in which the source was fitted.
       & \strut\hfill null \hfill\strut & Per stack \\

  \noalign{\bigskip}\multicolumn{6}{l}{Columns 5$-$15: Coordinates and
      associated identifier\medskip} \\
   RA                & double  & degrees
       & Right ascension (J2000).
       & \multicolumn{2}{c}{Combined fit} \\
   DEC               & double  & degrees
       & Declination (J2000).
       & \multicolumn{2}{c}{Combined fit} \\
   RADEC\_ERR        & float  & arcsec
       & Square root of the sum of the squared 1$\sigma$ errors in X\_IMA and
         Y\_IMA $\sqrt{\sigma_\textrm{X\_IMA}^2+\sigma_\textrm{Y\_IMA}^2}$,
         converted to arcsec.
       & \multicolumn{2}{c}{Combined fit} \\
   LII               & double  & degrees
       & Galactic longitude.
       & \multicolumn{2}{c}{Combined fit} \\
   BII               & double  & degrees
       & Galactic latitude.
       & \multicolumn{2}{c}{Combined fit} \\
   X\_IMA(\_ERR)     & float   & pixel
       & X coordinate within the re-binned image and its 1$\sigma$ error. Image
         pixels are binned to 4\arcsec\ $\times$ 4\arcsec.
       & \multicolumn{2}{c}{Combined fit} \\
   Y\_IMA(\_ERR)     & float   & pixel
       & Y coordinate within the re-binned image and its 1$\sigma$ error. Image
         pixels are binned to 4\arcsec\ $\times$ 4\arcsec.
       & \multicolumn{2}{c}{Combined fit} \\
   DIST\_NN          & float   & arcsec
       & Distance to the nearest neighbouring detection.
       & \multicolumn{2}{c}{Combined fit} \\
   IAUNAME           & string  &
       & IAU name `3XMMs\,J\emph{hhmmss.s}$\pm$\emph{ddmmss}' of the source.
       & \multicolumn{2}{c}{Assigned to all rows per source} \\

  \noalign{\bigskip}\multicolumn{6}{l}{Columns 16$-$23: 3XMM counterpart:
      values copied from 3XMM-DR7\medskip} \\
   IAUNAME\_\mbox{3XMMDR7} & string &
       & IAU name assigned to the nearest unique source in 3XMM-DR7.
       & \multicolumn{2}{p{.3\linewidth}}{\strut\hfill Assigned to all rows per source\hfill\strut\protect\newline \strut\hfill with a 3XMM-DR7 counterpart\hfill\strut} \\
   \mbox{SRCID}\_\mbox{3XMMDR7} & long &
       & Unique source identifier of the nearest source in 3XMM-DR7 within a
         correlation radius of three times the sum of the position
         errors. Null (undefined) if no counterpart is found.
       & \multicolumn{2}{p{.3\linewidth}}{\strut\hfill Assigned to all rows per source\hfill\strut\protect\newline \strut\hfill with a 3XMM-DR7 counterpart\hfill\strut} \\
   \mbox{DETID}\_\mbox{3XMMDR7} & long &
       & Identifier of the 3XMM-DR7 per-observation detection that contributes
         to the unique source \mbox{SRCID}\_\mbox{3XMMDR7}, if existing.
       & Per OBS\_ID & null \\
   RA\_\mbox{3XMMDR7} & double & degrees
       & Mean right ascension of the unique 3XMM-DR7 source and its contributing
         detections after field rectifications.
       & RA of\protect\newline \mbox{DETID}\_\mbox{3XMMDR7} & SC\_RA of\protect\newline \mbox{SRCID}\_\mbox{3XMMDR7} \\
   DEC\_\mbox{3XMMDR7} & double & degrees
       & Mean declination of the unique 3XMM-DR7 source and its contributing
         detections after field rectifications.
       & DEC of\protect\newline \mbox{DETID}\_\mbox{3XMMDR7} & SC\_DEC of\protect\newline \mbox{SRCID}\_\mbox{3XMMDR7} \\
   POSERR\_\mbox{3XMMDR7} & float & arcsec
       & Position error of the unique 3XMM-DR7 source and its contributing
         detections, including the error derived from the field rectification.
       & POSERR of\protect\newline \mbox{DETID}\_\mbox{3XMMDR7} & SC\_POSERR of\protect\newline \mbox{SRCID}\_\mbox{3XMMDR7} \\
   DIST\_\mbox{3XMMDR7} & double & arcsec
       & Distance between the source and the nearest unique source / its
         contributing detection in 3XMM-DR7.
       & Distance to\protect\newline \mbox{DETID}\_\mbox{3XMMDR7} & Distance to\protect\newline \mbox{SRCID}\_\mbox{3XMMDR7} \\
   NDETECT\_\mbox{3XMMDR7} & integer &
       & Number of DR7 detections N\_DETECTIONS of the nearest unique source in
         3XMM-DR7.
       & null & Of\protect\newline \mbox{SRCID}\_\mbox{3XMMDR7}\\

  \noalign{\smallskip}\multicolumn{6}{l}{Columns 24$-$71: Fluxes\medskip} \\
   EP\_FLUX          & float   & erg\,cm$^{-2}$\,s$^{-1}$
       & All-EPIC flux between 0.2 and 12\,keV: dead-time corrected count
         rates in the entire PSF, multiplied with the respective energy
         conversion factors. Null (undefined) if the exposure is zero, like on
         chip gaps, or if the covered PSF fraction is below $15\,\%$.
       & \multicolumn{2}{@{\hskip.03\linewidth}p{.25\linewidth}@{\hskip.03\linewidth}}{\centering Mean of the instrument fluxes weighted by the inverse squared errors} \\
   EP\_FLUX\_ERR     & float   & erg\,cm$^{-2}$\,s$^{-1}$
       & 1$\sigma$ error on EP\_FLUX.
       & \multicolumn{2}{@{\hskip.03\linewidth}p{.25\linewidth}@{\hskip.03\linewidth}}{\centering Inverse square root of the
         sum of inverse squared errors per instrument} \\
   EP\_$n$\_FLUX(\_ERR)    & float   & erg\,cm$^{-2}$\,s$^{-1}$
       & All-EPIC flux and 1$\sigma$ flux error in energy band $n$. Null
         (undefined) if the exposure is zero or if the covered PSF fraction is
         below $15\,\%$. Zero with non-zero errors, if no counts are detected in
         spite of sufficient PSF coverage.
       & \multicolumn{2}{@{\hskip.03\linewidth}p{.25\linewidth}@{\hskip.03\linewidth}}{\centering Mean of the instrument values weighted by the inverse squared errors} \\
   $CC$\_FLUX(\_ERR)       & float   & erg\,cm$^{-2}$\,s$^{-1}$
       & Total flux and 1$\sigma$ flux error in instrument $CC$. Null / zero as before.
       & \multicolumn{2}{@{\hskip.03\linewidth}p{.25\linewidth}@{\hskip.03\linewidth}}{\centering Sum of fluxes and
         flux errors per energy band} \\
   $CC$\_$n$\_FLUX(\_ERR)   & float   & erg\,cm$^{-2}$\,s$^{-1}$
       & Flux and 1$\sigma$ flux error per instrument $CC$ and energy band
         $n$. Null / zero as before.
       & Fitted per input image & Mean of the observation-level values
         weighted by the exposure time  \\

  \noalign{\bigskip}\multicolumn{6}{l}{Columns 72$-$117: Count rates and count numbers\medskip} \\
   EP\_RATE          & float   & counts\,s$^{-1}$
       & All-EPIC count rate between 0.2 and 12.0\,keV: background-subtracted
         and vignetting corrected count rate in the entire PSF. Null
         (undefined) if the exposure is zero or if the covered PSF fraction
         is below $15\,\%$. Zero with non-zero errors if no counts are
         detected in spite of sufficient PSF coverage.
       & Sum of the count rates per instrument & Mean of the observation-level
         rates weighted by the exposure time \\
   EP\_RATE\_ERR     & float   & counts\,s$^{-1}$
       & 1$\sigma$ error on the count rate. Null (undefined) if the exposure
         is zero or if the covered PSF fraction is below $15\,\%$.
       & Square root of the quadratic sum of the errors per instrument &
         Combined observation-level errors weighted by the exposure time  \\
   $CC$\_RATE(\_ERR)     & float   & counts\,s$^{-1}$
       & Total count rate and count-rate error in instrument $CC$. Null /
         zero as before.
       & Sum of the values per energy band & Mean of the observation-level
         values weighted by the exposure time\\
   $CC$\_$n$\_RATE(\_ERR)  & float   & counts\,s$^{-1}$
       & Count rate and count-rate error per instrument $CC$ and energy band
         $n$. Null / zero as before.
       & Fitted per input image & Mean of the observation-level values
         weighted by the exposure time \\
   EP\_CTS            & float   & counts\,s$^{-1}$

       & All-EPIC counts: background-subtracted source counts in the entire
         PSF. Null (undefined) if the exposure is zero or if the covered PSF
         fraction is below $15\,\%$.
       & \multicolumn{2}{@{\hskip.03\linewidth}p{.25\linewidth}@{\hskip.03\linewidth}}{\centering Sum of the source counts per instrument} \\
   EP\_CTS\_ERR     & float   & counts\,s$^{-1}$
       & 1$\sigma$ error on the counts. Null (undefined) if the exposure is
         zero or if the covered PSF fraction is below $15\,\%$.
       & \multicolumn{2}{@{\hskip.03\linewidth}p{.25\linewidth}@{\hskip.03\linewidth}}{\centering Square root of the quadratic sum of
           the errors per instrument} \\
   $CC$\_CTS(\_ERR)     & float   & counts\,s$^{-1}$
       & Total counts and error in instrument $CC$.
       & \multicolumn{2}{@{\hskip.03\linewidth}p{.25\linewidth}@{\hskip.03\linewidth}}{\centering Sum of the values per energy band \medskip} \\

  \noalign{\smallskip}\multicolumn{6}{l}{Columns 118$-$139: Detection and extent likelihoods\medskip} \\
   EP\_DET\_ML           & float &
       & Total equivalent maximum detection likelihood, normalised to two
         degrees of freedom.
       & \multicolumn{2}{@{\hskip.03\linewidth}p{.25\linewidth}@{\hskip.03\linewidth}}{\centering Calculated from all valid
           contributing images} \\
   $CC$\_DET\_ML           & float &
       & Equivalent maximum detection likelihood for instrument $CC$, normalised
         to two degrees of freedom.
       &
         \multicolumn{2}{@{\hskip.03\linewidth}p{.25\linewidth}@{\hskip.03\linewidth}}{\centering Calculated from all valid contributing images
           per instrument} \\
   $CC$\_$n$\_DET\_ML       & float &
       & Equivalent maximum detection likelihood per instrument $CC$ and energy
         band $n$, normalised to two degrees of freedom.
       & Value per input image & Calculated from all valid
           contributing images per instrument and energy band \\
   EXTENT                 & float & arcsec
       & Source extent radius, derived from the model source PSF convolved with
         a beta profile. Below an extent of 6\arcsec, it is considered
         unresolved and set to zero, the source is `point-like'.
       & \multicolumn{2}{c}{Combined fit} \\
   EXTENT\_ERR            & float & arcsec
       & $1\sigma$ error on the source extent radius. Null (undefined) if the
         extent radius is below 6\arcsec.
       & \multicolumn{2}{c}{Combined fit} \\
   EXTENT\_ML             & float &
       & Likelihood that the source is extended with radius EXTENT. Null
         (undefined) if the source is considered point-like.
       & \multicolumn{2}{c}{Combined fit} \\

  \noalign{\bigskip}\multicolumn{6}{l}{Columns 140$-$171: Hardness ratios\medskip} \\
   EP\_HR$i$              & float &
       & Equivalent all-EPIC hardness ratios
         $(r_{i+1}-r_i)/(r_{i+1}+r_i)$
         between the count rates $r$ in energy bands $i$ and $i+1$.
       & \multicolumn{2}{@{\hskip.03\linewidth}p{.25\linewidth}@{\hskip.03\linewidth}}{\centering Mean of all active instruments} \\
   EP\_HR$i$\_ERR         & float &
       & $1\sigma$ error on hardness ratio EP\_HR$i$.
       & \multicolumn{2}{@{\hskip.03\linewidth}p{.25\linewidth}@{\hskip.03\linewidth}}{\centering Error propagation of the contributing
         values} \\
   $CC$\_HR$i$(\_ERR)        & float &
       & Hardness ratios
         $(r_{i+1}-r_i)/(r_{i+1}+r_i)$
         between the count rates $r$ in energy bands $i$ and $i+1$ per instrument.
       & Per input images & Mean of the observations \\

  \noalign{\bigskip}\multicolumn{6}{l}{Columns 172$-$233: Fit characteristics\medskip} \\
   $CC$\_EXP                 & float & seconds
       & Exposure map values per instrument $CC$, including vignetting
         effects. The effective good exposure time is given in the ONTIME
         columns, described below.
       & \multicolumn{2}{@{\hskip.03\linewidth}p{.25\linewidth}@{\hskip.03\linewidth}}{\centering Sum of the contributing images} \\
   $CC$\_$n$\_EXP            & float & seconds
       & Exposure map values per instrument $CC$ and energy band.
       & Fitted per input image & Sum of the contributing images \\
   $CC$\_BG                  & float & counts\,px$^{-1}$
       & Background model at the central position of the source on the
         CCD. Null (undefined) if the exposure map is zero and if the centre
         of the source is located on a bad chip area.
       & \multicolumn{2}{@{\hskip.03\linewidth}p{.25\linewidth}@{\hskip.03\linewidth}}{\centering Sum of all contributing images} \\
   $CC$\_$n$\_BG             & float & counts\,px$^{-1}$
       & Background model in energy band $n$ at the CCD position of the source.
       & \multicolumn{2}{@{\hskip.03\linewidth}p{.25\linewidth}@{\hskip.03\linewidth}}{\centering Sum of all valid images in energy band $n$} \\
   EP\_ONTIME                & float & seconds
       & Total good exposure time at the central position of the source on the
         CCD. Zero if the centre of the source is located on a bad chip area.
       & Maximum of the good exposure time per instrument & Sum of the maximum
         on-times of all observations \\
   $CC$\_ONTIME              & float & seconds
       & Total good exposure time of instrument $CC$ at the CCD position of the
         source. Times are calculated by the task \texttt{evselect} and not
         vignetting corrected.
       & Read from the headers of the input files. & Sum of the
         observation-level values. \\
   $CC$\_MASKFRAC            & float &
       & Mean chip coverage in the detection mask per instrument, weighted by
         the point spread function of the source.
       & Minimum of the slightly differing values per energy band & Maximum of
         the observation-level values \\
   EP\_OFFAX                 & float & arcmin
       & Angular distance of the source to the boresight.
       & Minimum of the values per instrument. & \strut\hfill null \hfill\strut \\
   $CC$\_OFFAX               & float & arcmin
       & Angular distance of the source to the boresight in instrument $CC$.
       & Combined fit & \strut\hfill null \hfill\strut \\
   $CC$\_$n$\_VIG            & float &
       & Vignetting fraction in instrument $CC$ and energy band $n$ at the
         central position of the source.
       & Read from the calibration data base & \strut\hfill null \hfill\strut \\

  \noalign{\bigskip}\multicolumn{6}{l}{Columns 234$-$238: Quality flags\medskip} \\
   STACK\_FLAG               & integer &
       & Summarised quality flag of the detection. `0' if all flags are
         false. `1' if at least one of flags 1, 2, 3, 9 is true. `2' if at
         least one of flags $4-8$ is true. `3' if STACK\_FLAG is 2 for all
         contributing observations.
       & \multicolumn{2}{@{\hskip.03\linewidth}p{.25\linewidth}@{\hskip.03\linewidth}}{\centering Numeric version of EP\_FLAG} \\
   EP\_FLAG                  & string &
       & Quality flags of the detection, automatically set and combined into a
         nine-characters string. A flag `true' means a warning on a detection
         condition.
       & Worst flag of all instruments & Worst flag of all observations \\
   $CC$\_FLAG                & string &
       & Quality flags in instrument $CC$.
       & Set per input image & Worst flag of all observations \\

  \noalign{\bigskip}\multicolumn{6}{l}{Columns 239$-$268:
      Inter-observation variability of sources with at least two contributing
      observations\medskip} \\
   VAR\_CHI2                  & float &
       & Reduced $\chi^2$ of inter-observation flux variability in all
         contributing observations with valid non-zero fluxes.
       & \strut\hfill null \hfill\strut & $\frac{1}{n-1}\sum_{k=1}^n \left(\frac{F_k-F_\textrm{EPIC}}{\sigma_k}\right)^2$ \\
   VAR\_CHI2\_$n$             & float &
       & Reduced $\chi^2$ of inter-observation flux variability in energy band $n$.
       & \strut\hfill null \hfill\strut & \\
   VAR\_PROB                  & double &

       & Probability that the measured flux values are consistent with
         constant inter-observation flux, derived from VAR\_CHI2. The smaller
         VAR\_PROB, the more likely the source shows inter-observation flux
         variability.
       & \strut\hfill null \hfill\strut 
       & $\int_{\chi^2}^\infty
         \frac{x^{\nu/2-1}e^{-x/2}}{2^{\nu/2}\Gamma(\nu/2)} dx$ \\

   VAR\_PROB\_$n$             & double &
       & Probability that the measured flux values are consistent with constant
         inter-observation flux in energy band $n$.
       & \strut\hfill null \hfill\strut & Derived from VAR\_CHI2\_$n$ \\
   FRATIO                    & float &
       & Ratio between highest and lowest (non-zero, non-null) mean flux in
         the contributing observations.
       & \strut\hfill null \hfill\strut & $F_\textrm{max}/F_\textrm{min}$ \\
   FRATIO\_ERR               & float &
       & $1\sigma$ error on the flux ratio.
       & \strut\hfill null \hfill\strut & $\left(\frac{\sigma_{F\textrm{min}}^2}{F_\textrm{min}^2}+\frac{\sigma_{F\textrm{max}}^2}{F_\textrm{max}^2}\right)^{0.5}\,\frac{F_\textrm{max}}{F_\textrm{min}}$ \\
   FRATIO\_$n$(\_ERR)        & float &
       & Ratio between highest and lowest mean flux in energy band $n$ and its
         $1\sigma$ error.
       & \strut\hfill null \hfill\strut & \\
   FLUXVAR                   & float &
       & Largest difference between mean all-EPIC fluxes in terms of $\sigma$.
       & \strut\hfill null \hfill\strut &
   $\max_{k,l\in[1,n]}\frac{|F_k-F_||}{\sqrt{\sigma_k^2+\sigma_l^2}}$ \\
   FLUXVAR\_$n$              & float &
       & Largest difference between mean all-EPIC fluxes in terms of $\sigma$ in
         energy band $n$.
       & \strut\hfill null \hfill\strut & \\

  \noalign{\smallskip}\multicolumn{6}{l}{Columns 269$-$273:
      3XMM-DR7 intra-observation variability of sources with a DR7 light curve\medskip} \\
   CHI2PROB\_\mbox{3XMMDR7} & double &
       & EPIC $\chi^2$ probability that the time series of the
         nearest unique source in 3XMM-DR7 is consistent with the source having
         constant flux during the observation.
       & EP\_CHI2PROB in 3XMM-DR7 & SC\_CHI2PROB in 3XMM-DR7: Minimum of all contributing 3XMM-DR7 observations \\
   FVAR\_\mbox{3XMMDR7} & double &
       & Fractional variance of the nearest unique source in 3XMM-DR7.
       & FVAR in 3XMM-DR7 of the instrument with minimum CHI2PROB & SC\_FVAR in
         3XMM-DR7: Observation with minimum CHI2PROB \\
   FVARERR\_\mbox{3XMMDR7} & double &
       & $1\sigma$ error on FVAR\_\mbox{3XMMDR7}.
       & FVAR\_ERR in 3XMM-DR7 of the instrument with minimum CHI2PROB  &
         SC\_FVAR\_ERR in 3XMM-DR7: Observation with minimum  CHI2PROB \\
   VAR\_FLAG\_\mbox{3XMMDR7} & boolean &
       & Variability flag of the nearest unique
           source in 3XMM-DR7. True if at least one exposure has CHI2PROB below
         $10^{-5}$.
       & EP\_VAR\_FLAG in 3XMM-DR7 & SC\_VAR\_FLAG in 3XMM-DR7: Observation with
         minimum CHI2PROB \\
   SUM\_FLAG\_\mbox{3XMMDR7} & integer &
       & Integer representation of the quality flags of the nearest 3XMM-DR7
         source from automatic and visual screening.
       & SUM\_FLAG in\protect\newline 3XMM-DR7 & SC\_SUM\_FLAG\protect\newline in 3XMM-DR7  \\

  \noalign{\bigskip}\multicolumn{6}{l}{Columns 274$-$284: Observation characteristics\medskip} \\
   MJD\_FIRST      & double  &  days
       & Modified Julian Date of the start of the observation (JD$-$2\,400\,000.5).
       & Read from the headers of the input files & First contributing observation \\
   MJD\_LAST       & double  &  days
       & Modified Julian Date of the end of the observation (JD$-$2\,400\,000.5).
       & Read from the headers of the input files & Last contributing observation \\
   PA\_PNT         & float   &  degrees
       & Mean position angle of the spacecraft during the observation.
       & File headers & \strut\hfill null \hfill\strut \\
   REVOLUT         & short   &
       & XMM-Newton orbit of the observation.
       & File headers & \strut\hfill null \hfill\strut \\
   $CC$\_SUBMODE   & string &
       & Observing mode of instrument $CC$.
       & File headers & \strut\hfill null \hfill\strut \\
   $CC$\_FILTER    & string &
       & Filter used during the observation.
       & File headers & \strut\hfill null \hfill\strut \\
   URL\_\mbox{3XMMDR7} & string &
       & Web page of the nearest unique source in 3XMM-DR7.
       &
   \multicolumn{2}{@{\hskip.03\linewidth}p{.25\linewidth}@{\hskip.03\linewidth}}{\centering
     Copied from 3XMM-DR7} \\
  \end{longtable}
  \end{landscape}
  }

\end{appendix}

\end{document}